%% file: main.tex
\documentclass[preprint,3p]{elsarticle}

\usepackage[utf8]{inputenc}
\usepackage{geometry,physics,color,booktabs}
\usepackage{amsmath} 
\usepackage{amsfonts} 
\usepackage{amssymb}
\usepackage{amsthm} 
\usepackage{mathtools}
\usepackage{float}
\usepackage{pdfpages}
\usepackage[ruled,vlined]{algorithm2e}
\usepackage{algpseudocode}
\usepackage{subfigure}
\usepackage{marginnote}
\usepackage{bm}
\usepackage{array}

\newcommand{\REV}[1]{#1}

\newtheorem{remark}{Remark}
\newtheorem{lem}{Lemma}
\newtheorem{thm}{Theorem}
\newtheorem{defn}{Definition}

\journal{Computer Methods in Applied Mechanics and Engineering}
\begin{document}
\begin{frontmatter}

\title{Dynamic flux surrogate-based partitioned methods for interface problems}
\cortext[cor1]{Corresponding author}  
\fntext[CCR]{Center for Computing Research, Sandia National Laboratories, Albuquerque, NM 87185, USA}
\fntext[AIS]{Applied Information Sciences, Sandia National Laboratories, Albuquerque, NM 87185, USA}
\fntext[UCON]{University of Connecticut, 341 Mansfield Road U1009, Storrs, CT 06269, USA}
\fntext[sand-blurb]{Sandia National Laboratories is a multimission laboratory managed and operated by National Technology and Engineering Solutions of Sandia, LLC., a wholly owned subsidiary of Honeywell International, Inc., for the U.S. Department of Energy's National Nuclear Security Administration under contract DE-NA-0003525. SAND2024-01215O}

\author[CCR]{Pavel Bochev\corref{cor1}}
\ead{pbboche@sandia.gov}

\author[AIS]{Justin Owen}
\ead{jowen@sandia.gov}

\author[CCR]{Paul Kuberry}
\ead{pakuber@sandia.gov}

\author[UCON]{Jeffrey Connors}
\ead{jeffrey.connors@uconn.edu}


\input{Abstract}
\end{frontmatter}

\section{Introduction}\label{sec:intro}
\input{Introduction}

\section{Notation and background}\label{sec:background}
\input{Notation}

\section{\REV{Model problem and its data-driven partitioned solution}}\label{sec:model}
\input{FOM}

\section{Construction of the dynamic flux surrogate}\label{sec:DMD}
\input{DMD}

\section{\REV{Numerical stability}}\label{sec:stability} 
\input{Stability}

\section{\REV{Generation of training data for the dynamic flux surrogates}}\label{sec:trainDMD}
\input{Training}

\section{Numerical results}\label{sec:num}
\input{Numerics}

\section{Conclusions}\label{sec:concl}
\input{Conclusion}

\section*{Acknowledgments}
This material is based upon work supported by the U.S. Department of Energy, Office of Science, Office of Advanced Scientific Computing Research, Mathematical Multifaceted Integrated Capability Centers (MMICCs) program, under Field Work Proposal 22-025291 (Multifaceted Mathematics for Predictive Digital Twins (M2dt)), Field Work Proposals 23-020467 and 19-020315, and the Laboratory Directed Research and Development program at Sandia National Laboratories.  This paper describes objective technical results and analysis. Any subjective views or opinions that might be expressed in the paper do not necessarily represent the views of the U.S. Department of Energy or the United States Government. 
 
This article has been authored by an employee of National Technology \& Engineering Solutions of Sandia, LLC under Contract No. DE-NA0003525 with the U.S. Department of Energy (DOE). The employee owns all right, title and interest in and to the article and is solely responsible for its contents. The United States Government retains and the publisher, by accepting the article for publication, acknowledges that the United States Government retains a non-exclusive, paid-up, irrevocable, world-wide license to publish or reproduce the published form of this article or allow others to do so, for United States Government purposes. The DOE will provide public access to these results of federally sponsored research in accordance with the DOE Public Access Plan https://www.energy.gov/downloads/doe-public-access-plan.

\bibliographystyle{elsarticle-num}
\bibliography{refs}

\end{document}

%% file: Abstract.tex
\begin{abstract}
\REV{Loosely coupled partitioned methods for multiphysics problems treat each subproblem as a separate entity and advance them independently  in time. In so doing these methods enable code reuse, increase concurrency and provide a convenient framework for plug-and-play multiphysics simulations.
However, mathematically loosely coupled schemes are equivalent to a single step of an iterative solution method, which can compromise their accuracy and stability. 
We present a new data-driven partitioned method for coupled parametric PDEs that can improve upon the accuracy of traditional loosely coupled methods without incurring a  performance penalty. To that end, we replace conventional field transfers across the interface by a surrogate for the dynamics of the interface flux exchanged between the subdomains. To develop this surrogate we apply dynamic mode decomposition to a non-standard staggered-in-time state, comprising the interface flux and small solution patches near the interface. 
The new approach shifts the main computational burden to an offline training phase, whereas application of the surrogate in the online phase amounts to a single matrix-vector multiplication.  We provide stability analysis of the surrogate-based partitioned scheme and include  numerical results  that demonstrate its potential.}
\end{abstract}

%% file: Introduction.tex
Numerical solution of coupled multi-physics problems  can be approached in two distinct ways \cite{Felippa_01_CMAME}. 
\REV{\emph{Tightly coupled} or} monolithic methods treat the coupled system as a a single entity and advance all of its constituent physics components in time simultaneously  \cite{Felippa_01_CMAME}. 
%
As a result, monolithic methods possess excellent stability and accuracy properties but \REV{are} computationally expensive \REV{and require} sophisticated, \emph{physics-based} preconditioners \cite{Newman_13_SISC} \REV{and} solvers \cite{Knoll_04_JCP}. 
In terms of software development and reuse, these methods are also rather inflexible since their modification to include and/or exclude a particular physics component is a non-trivial task that may require significant code refactoring up to reimplementation from scratch.

In contrast, \REV{\emph{loosely coupled}} partitioned methods treat the subproblems comprising the multi-physics system as separate entities and advance them in time independently \cite{Felippa_01_CMAME}. Interactions between the subproblems and their synchronization is performed via data transfers over coupling windows; see, e.g., \cite{Connors_22_SINUM}. 
\REV{As a result, loosely coupled methods are} inherently more flexible \REV{and computationally efficient} than their monolithic counterparts, \REV{and provide a convenient framework for plug-and-play multi-physics simulation capabilities in applications such as  earth system models \cite{Caldwell_19_JAMES,Lemarie_15_PCS}.} 
\REV{However, mathematically, loosely coupled methods are equivalent to a single step of an iterative solution scheme \cite{Lemarie_15_PCS}, which can compromise their accuracy and stability; see, e.g., \cite{Foerster_07_CMAME}.}

\REV{In this paper we formulate and demonstrate numerically a new, data-driven partitioned method for coupled parametric PDEs ($\mu$PDEs) whose cost is comparable to that of traditional loosely coupled schemes, but whose accuracy and stability is closer to that of monolithic methods. The core idea of the approach is to replace conventional remap-based 
\cite{Jiao_04_IJNME,Ullrich_15_MWR,Gatzhammer_14_THESIS,Slattery_16_JCP,Bungartz_16_CF,Kuberry_19_MISC}
field transfers between the subdomains by data-driven surrogates that are learned offline. In particular, in this work we extend the dynamic mode decomposition approach  \cite{Rowley_09_JFM,Mezic_05_ND} to develop computationally efficient  surrogates for the dynamics of the interface flux exchanged between the subdomains.}

\REV{This new class of \emph{dynamic flux surrogate-based} partitioned schemes shifts} the principal computational burden to an \emph{offline} phase, leaving the application of the surrogate as the sole additional cost during the \emph{online} simulation phase. This cost  amounts to a single matrix-vector multiplication and is comparable to the cost of, e.g., remap via linear maps \cite{Ullrich_15_MWR}. Furthermore, learning the surrogate  does not require access to the discretized equations, which makes it minimally intrusive.

We first consider the case of a coupled $\mu$PDE problem with a fixed parameter and show that, with proper training data, a DMD flux surrogate (DMD-FS) can be constructed offline to accurately represent the dynamics of the interface flux for initial conditions not included in the training data. 
We then extend the DMD-FS  to the parametric setting and develop a parametric DMD-FS ($\mu$DMD-FS)  to predict the flux for coupled $\mu$PDEs. We show that $\mu$DMD-FS \REV{ (i)  can handle both  parameters and initial conditions outside the training data, (ii) improves upon the accuracy of conventional loosely coupled schemes, and (iii) has computational cost that is comparable to or even lower than that of the latter.}

\REV{To the best of our knowledge, this work is the first to exploit the concept of \emph{dynamic} surrogates as a way to estimate interface fields in loosely coupled schemes.} 
While other efforts \REV{to develop coupling surrogates exist, they have focused primarily on \emph{static} surrogates for the Poincar\'e-Steklov operators that give the boundary response of the subdomain problems in alternating Schwarz methods for multi-physics problems.} 
For example, \cite{Aletti_17_IJNME} considers a ``single simulation'' scenario in which only one component of a coupled system has to be approximated accurately. To perform such simulations efficiently via Schwarz, the Poincar\'e-Steklov operators  expressing the interactions between the subdomain of interest and the remaining subdomains are replaced by low-rank approximations of the Neumann-to-Dirichlet maps based on the first few eigenfunctions of the Laplace-Beltrami operators on the interfaces.
Similarly,  \cite{Chen_22_arXiv} develops an accelerated Schwarz framework for multiscale PDEs by training a neural network surrogate for the Dirichlet-to-Dirichlet map between the subdomains. Another example is  \cite{Discacciati_23_UNPUB}, which considers a non-overlapping Schwarz method to solve coupled multi-physics problems. The Schwarz iteration is first expressed in terms of  two Poincar\'e-Steklov operators realizing the Dirichlet-to-Neumann and the Neumann-to-Dirichlet maps. Then, each one of these maps is replaced by a surrogate based on kernel interpolation or neural net regression.  \REV{Finally, \cite{Parish_23_arXiv} considers machine-learned surrogates for solid mechanics applications that perform static condensation of the interior degrees of freedom by learning a Dirichlet-to-Neumann map.} 

\REV{Compared to static surrogates our approach offers greater generalizability as it is not tied to a specific instance of a Poincar\'e-Steklov operator. In particular, it can be extended to general imperfect interfaces \cite{Javili_14_CMAME} that are governed by separate evolution equations whose behavior cannot be captured by a static surrogate, but can be learned by a DMD surrogate; see Remark \ref{rem:imperfect}.}

%
%
%
%
The rest of this article is organized as follows. \REV{Section \ref{sec:background} introduces the relevant notation and background, including the DMD approach \cite{Mezic_05_ND} and its parametric reduced Koopman operator interpolation (rKOI)  \cite{Huhn_23_JCP} version.
Section \ref{sec:model} describes the coupled model $\mu$PDE problem, outlines the data-driven loosely coupled approach, and introduces its accuracy and efficiency benchmarks.   
Sections  \ref{sec:DMD}--\ref{sec:trainDMD}  are the core of this paper. We develop the DMD flux surrogates for the interface flux dynamics in Section \ref{sec:DMD}. In Section \ref{sec:stability}, we provide stability analysis of the surrogate-based scheme and a scheme that solves a hybrid monolithic formulation of the model coupled problem. The latter retains the energy bounds of the monolithic problem and is used as a stability benchmark for the data-driven scheme. Next, in Section \ref{sec:trainDMD}, we formulate the training approach for DMD flux surrogate identification.
Section \ref{sec:num} illustrates numerically the performance of the new $\mu$DMD-FS partitioned scheme by comparing its accuracy and efficiency with the benchmark schemes described in Section \ref{sec:model}. 
Section \ref{sec:concl} summarizes our findings, outlines future research directions and 
offers some conclusions.
}

%% file: Notation.tex
Section \ref{sec:notation} summarizes the notation that will be used throughout the paper. Section \ref{sec:DMD-review} reviews the original Dynamic Mode Decomposition (DMD) method, while Section \ref{sec:DMD-param} describes its parametric version.

\subsection{Notation}\label{sec:notation}
\begin{figure}[t!]
  \begin{center}
    \includegraphics[width=0.4\textwidth]{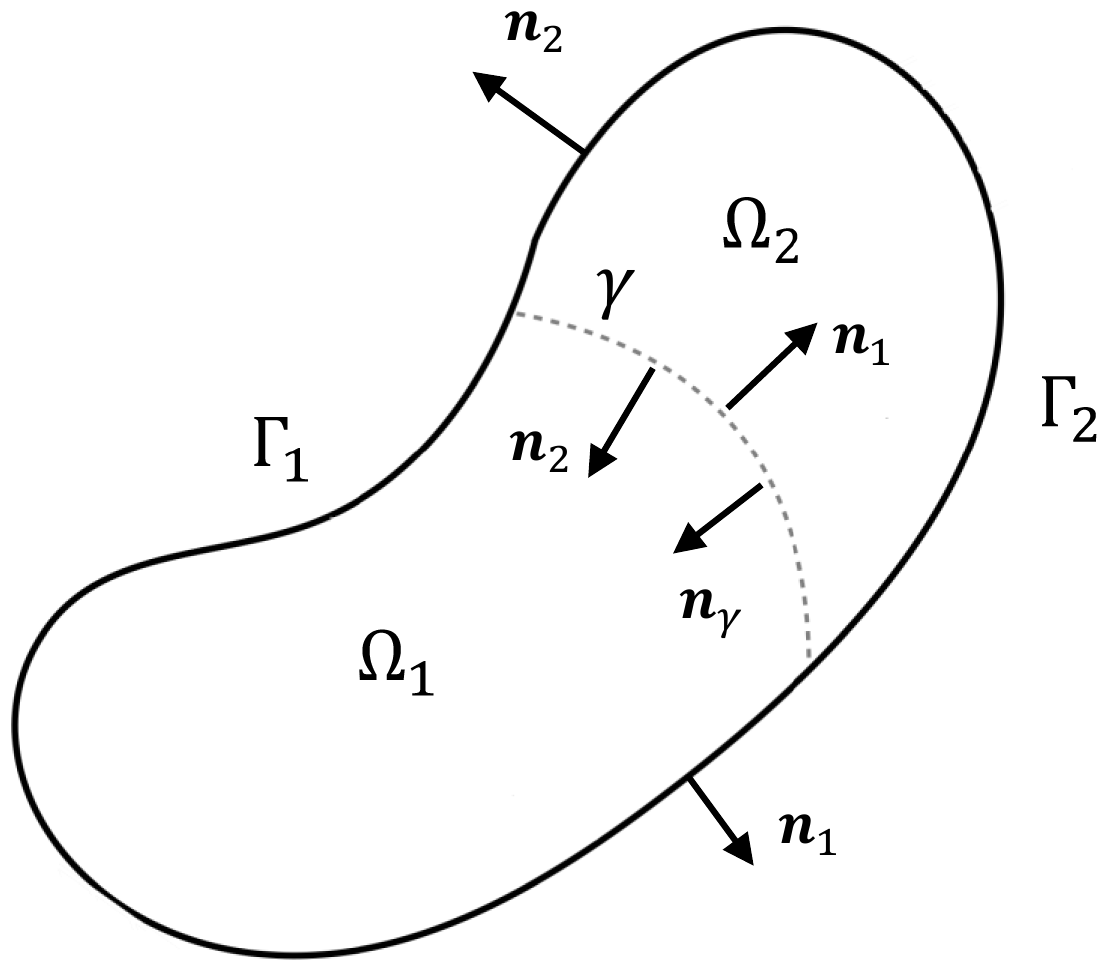}
  \end{center}
\vspace{-3ex}
\caption{A two-dimensional domain  $\Omega$ partitioned into two non-overlapping subdomains, $\Omega_1$ and 
$\Omega_2$, with interface $\gamma$.} \label{fig:dd}
\end{figure}

Let $\Omega \in \mathbb{R}^\nu$, $\nu = 2,3$ be a bounded region with Lipschitz continuous boundary $\Gamma$. We assume that $\Omega$ is divided into two non-overlapping subdomains $\Omega_1$ and $\Omega_2$ by an interface $\gamma$, as shown in Figure \ref{fig:dd}. This configuration is sufficient for our purposes, however the techniques developed in this paper can be extended to interface problems with more than two subdomains. 
We orient the interface by its unit normal $\bm{n}_\gamma$. Without a loss of generality we assume that $\bm{n}_\gamma$ points towards $\Omega_1$ and set $\Gamma_i := \partial \Omega_i  \backslash \gamma$, $i=1,2$. We denote the unit outer normal to $\partial\Omega_i$ as $\bm{n}_i$. Note that $\bm{n}_2$ coincides with $\bm{n}_{\gamma}$ along the interface.

In what follows $L^2(\Omega_i)$ will denote the space of all square integrable functions in $\Omega_i$ with norm and inner product denoted by $\|\cdot\|_{0,\Omega_i}$ and $(\cdot,\cdot)_{0,\Omega_i}$, respectively. Similarly, ${H}^1(\Omega_i)$ is the Sobolev space of order one on  $\Omega_i$ with norm $\|\cdot\|_{1,\Omega_i}$ and ${H}^1_{D}(\Omega_i)$ will denote the subspace of  ${H}^1(\Omega_i)$ whose elements  vanish on  $\Gamma_i$. 

The symbol $\Omega^h_i$ will stand for a conforming quasi-uniform \cite{Ciarlet_02_BOOK} partition of $\Omega_i$ into finite elements $K_{i,s}$ with vertices $\bm{x}_{i,r}$ and mesh parameter $h_i$. We denote the total number of nodes in $\Omega^h_i$ by $n_i$. Each $\Omega^h_i$ induces a conforming mesh $\Gamma^h_i$ on the Dirichlet boundary $\Gamma_i$ with $n_{i,\Gamma}$ nodes. 
In this paper we shall assume that $\Omega_1$ and $\Omega_2$ are meshed independently. As a result, their finite element partitions $\Omega^h_1$ and $\Omega^h_2$ induce two independent finite element meshes on the interface $\gamma$ denoted by $\gamma^h_1$ and $\gamma^h_2$, respectively, with $n_{i,\gamma}$ nodes each. For simplicity we restrict attention to spatially coincident discrete interfaces, however, the nodes on $\gamma^h_1$ and $\gamma^h_2$ are not required to match. 

\REV{In what follows $S_{i}^h$ will denote the lowest-order nodal $C^0$ finite element space  defined with respect to $\Omega^h_i$.} We recall that $S_{i}^h$ is a conforming subspace of ${H}^1(\Omega_i)$; see, e.g., \cite{Ciarlet_02_BOOK}. The elements of $S_{i}^h$ that vanish on the Dirichlet boundary $\Gamma_i$ form the subspace $S_{i,D}^h$, which is a conforming approximation of ${H}^1_{D}(\Omega_i)$. Restriction of $S_{i}^h$ to $\Gamma_i$ and $\gamma$ induce the boundary space $S^h_{i,\Gamma}$ and the interface space $S^h_{i,\gamma}$, respectively. We endow $S_{i}^h$ with a standard Lagrangian basis $\{\ell_{i,r}\}$ such that $\ell_{i,r}(\bm{x}_{i,s}) = \delta_{rs}$, where $\delta_{rs}$ is the Kronecker $\delta$-symbol. 

\REV{Dimensions of these finite element spaces} equal the number of free nodes in their respective finite element partitions. Thus, 
$\text{dim}\, S_{i}^h = n_i$,
$\text{dim}\, S_{i,D}^h = n_{i,D}:=n_i - n_{i,\Gamma}$, 
$\text{dim}\, S_{i,\Gamma}^h = n_{i,\Gamma}$, and 
$\text{dim}\, S^h_{i,\gamma} =  n_{i,\gamma}$. 
Let $\mathbf{u}_i\in\mathbb{R}^{n_i}$ denote the coefficient vector of the finite element function $u^h_i\in S_{i}^h$. 
Without a loss of generality we shall assume that the nodes of $\Omega^h_i$ are numbered in such a way that $\mathbf{u}_i$ is partitioned as
$\mathbf{u}_i = (\mathbf{u}_{i,\gamma},\mathbf{u}_{i,\circ},\mathbf{u}_{i,\Gamma})$, where 
$\mathbf{u}_{i,\gamma}\in\mathbb{R}^{n_{i,\gamma}}$, 
$\mathbf{u}_{i,\circ}\in\mathbb{R}^{n_{i,\circ}}$, and 
$\mathbf{u}_{i,\Gamma}\in\mathbb{R}^{n_{i,\Gamma}}$
are vectors corresponding to the  interface, interior, and Dirichlet coefficients of  $u^h_i$, respectively. Note that with this convention the coefficient vector of $u^h_{i,D}\in S^h_{i,D}$ can be partitioned as $\mathbf{u}_{i,D}=(\mathbf{u}_{i,\gamma},\mathbf{u}_{i,\circ})$.

\subsection{Dynamic Mode Decomposition}\label{sec:DMD-review}
Dynamic Mode Decomposition (DMD) \cite{Mezic_05_ND,Rowley_09_JFM} is a data driven algorithm 
that infers an approximation for the flow map of a dynamical system from snapshots $\mathbf{y}(t)$ of its solutions.
Specifically, given $S+1$  equally spaced in time solution snapshots $\REV{\mathbf{y}_i:=\mathbf{y}(t_i)\in\mathbb{R}^{N}}$, \REV{$i=0,\ldots,S$}, DMD seeks an $\REV{N\times N}$ operator $\mathbf{A}$ such that 
\begin{equation}\label{eq:DMD-single}
\mathbf{y}_{i+1} \approx \mathbf{A} \mathbf{y}_{i} \,.
\end{equation}
To approximate the discrete flow map $\mathbf{A}$ we arrange the snapshots into \REV{$N\times S$}  matrices $\mathbf{Y} := [\mathbf{y}_i]_{i=0}^{S-1}$ and $\mathbf{Y}' := [\mathbf{y}_i]_{i=1}^{S}$. Then, equation \eqref{eq:DMD-single} implies the relationship
\begin{equation}
    \mathbf{Y}' \approx \mathbf{A}\mathbf{Y}.
    \label{operator_equation}
\end{equation}
DMD treats \eqref{operator_equation} as a linear system for the unknown operator $\mathbf{A}$. ``Solving'' this linear system yields $A = \mathbf{Y}' \mathbf{Y}^{+}$ where $\mathbf{Y}^{+}$ is the Moore-Penrose pseudo-inverse.  An effective way to estimate the pseudo-inverse is provided by its truncated singular value decomposition (SVD). 
To that end, one computes the SVD $\mathbf{Y} = \mathbf{U}\mathbf{\Sigma} \mathbf{V}^T$ of the snapshot matrix and retains the first $k$ left singular vectors (``modes'') corresponding to the $k$ leading singular values. This yields the approximation
\begin{equation}\label{eq:tSVD}
\mathbf{Y}^+ = \mathbf{V} \mathbf{\Sigma}^{+} \mathbf{U}^T \approx  \mathbf{V}_k\mathbf{\Sigma}_k^{+}\mathbf{U}_k^T =: \mathbf{Y}_k^+ \,,
\end{equation}
where $\mathbf{V}_k$, $\mathbf{\Sigma}_k^{+}$, and $\mathbf{U}_k$ are the truncated SVD matrices. Using $\mathbf{Y}_k^{+}$ in lieu of $\mathbf{Y}^{+}$ then results in the following DMD approximation of the flow map:
\begin{equation}\label{eq:DMD}
    \mathbf{A}\approx \mathbf{A}_k := \mathbf{Y}' \mathbf{Y}_k^+ = \mathbf{Y}' \mathbf{V}_k \mathbf{\Sigma}_k^{+}\mathbf{U}_k^T \,.
\end{equation}
We refer to  $\mathbf{A}_k$ as the DMD operator. The accuracy of the DMD operator depends on the number $k$ of modes retained in \eqref{eq:tSVD}. This number is usually selected to be the minimum positive integer for which 
\begin{equation}\label{eq:POD-energy}
1-E_k(\mathbf{Y}) \le \epsilon \,,
\end{equation}
where $0<\epsilon <1$ is a given tolerance and 
\begin{align}\label{eq:snapEnergy}
	E_k(\mathbf{Y}):=\frac{\sum_{i=1}^{k} \sigma_i^2}{\sum_{i=1}^n \sigma_i^2}.
\end{align} 
In \eqref{eq:snapEnergy} $\sigma_i$ is the $i^{th}$ singular value of $\mathbf{Y}$, $n$ is the total number of singular values, and $E_k(\mathbf{Y})$ is the relative snapshot energy captured by the first $k$ modes.   Algorithm \ref{dmd_algorithm} summarizes the computation of the DMD operator.
%

\begin{algorithm}
\caption{DMD Algorithm}
\smallskip
\begin{enumerate}
\item Simulate the dynamical system of interest to generate a collection of $S+1$ equally spaced in time solution snapshots $\mathbf{y}_i$, $i=0,\ldots,S$, and arrange them into matrices $\mathbf{Y}$ and $\mathbf{Y}'$.
\smallskip
\item Compute the SVD of the snapshot matrix: $\mathbf{Y}= \mathbf{U}\mathbf{\Sigma} \mathbf{V}^T$.
\smallskip
\item Select a tolerance $0<\epsilon<1$ and find the smallest positive integer $k$ for which \eqref{eq:POD-energy} holds. 
\smallskip
\item Retain the first $k$ left singular vectors and compute the approximate pseudo-inverse $ \mathbf{Y}_k^+$ as in \eqref{eq:tSVD}.
\smallskip
\item Compute the DMD operator $\mathbf{A}_k$ as in \eqref{eq:DMD}.
\end{enumerate}
\label{dmd_algorithm}
\end{algorithm}

\begin{remark}\label{rem:POD-DMD}
In typical modeling situations the DMD operator is applied repeatedly to a given initial state $y_0$ to simulate the evolution of a dynamical system over a sequence of time steps $t_1, t_2,\ldots t_k,\ldots$. In this case simulation efficiency can be significantly improved by replacing $\mathbf{A}_k$ with a reduced order DMD operator $\widetilde{\mathbf{A}}_k := \mathbf{U}_k^T \mathbf{A}_k \mathbf{U}_k \in \mathbb{R}^{k\times k}$ which acts on a reduced state $\widetilde{\mathbf{y}}_i = \mathbf{U}^T_k \mathbf{y}_i$.
In contrast, here we shall always use the DMD operator over a single time interval to estimate the full order interface flux 
\REV{at the current time from the available full order states and flux.} In this context, switching to a reduced order operator $\widetilde{\mathbf{A}}_k$ is not justified \REV{because of the added cost of projecting a full state onto a reduced state and back to a full state at \emph{every time step}.}
\end{remark}

\subsubsection{Parametric Dynamic Mode Decomposition}\label{sec:DMD-param}
\REV{For dynamical systems corresponding to parameterized PDEs ($\mu$PDEs), the standard DMD Algorithm \ref{dmd_algorithm} can provide accurate approximation of the flow map only for a fixed  parameter $\bm{\mu}$.} 
Development of the dynamic flux surrogate-based partitioned method for coupled $\mu$PDEs requires extension of DMD to the parametric context. \REV{Typically this involves} sampling of the parameter space $\mathcal{M}\subset \mathbb{R}^M$ to define a representative set of parameter values $\mathcal{M}_m:=\{\bm{\mu}_i\}_{i=1}^m$. One then simulates the dynamical system of interest for all \REV{$\bm{\mu}_i\in\mathcal{M}_m$} to generate parameterized snapshot matrices $\mathbf{Y}(\bm{\mu}_i)$ and $\mathbf{Y}'(\bm{\mu}_i)$.  

The stacked DMD technique \cite{Taraneh_15_PF} assumes that the states for all parameter values were sampled at the same time instances. This allows one to ``stack'' the snapshots  $\mathbf{Y}(\bm{\mu}_i)$ and $\mathbf{Y}'(\bm{\mu}_i)$ into composite snapshot matrices 
 $[\mathbf{Y}(\bm{\mu}_i)]_{i=1}^m$ and $[\mathbf{Y}'(\bm{\mu}_i)]_{i=1}^m$ containing the known state data for all  
 $\bm{\mu}_i\in\mathcal{M}_m$.
Then, one applies the standard DMD algorithm (Algorithm \ref{dmd_algorithm})  to the composite snapshot matrices. The estimated state for a new parameter value $\bm{\mu}\notin \mathcal{M}_m$ is then computed by interpolating between the output state values. One drawback of the stacked DMD technique is that all possible parameter values, including those far from the current parameter value of interest, contribute to the singular value decomposition of the composite snapshot matrix $[\mathbf{Y}(\bm{\mu}_i)]_{i=1}^m$. 

Reduced Koopman operator inference (rKOI) \cite{Huhn_23_JCP} is an alternative parametric approach to DMD that localizes the interpolation process.  
Given a new parameter value $\bm{\mu}\notin \mathcal{M}_m$ this approach starts by choosing a ball 
$B(\bm{\mu}, r)\subset \mathcal{M}$ of radius $r$, centered at $\bm{\mu}$. Then, rKOI uses Algorithm \ref{dmd_algorithm} to compute the DMD operators $\mathbf{A}_{k_j}(\bm{\mu}_j)$ for all 
\begin{equation}\label{eq:intersect}
\bm{\mu}_j\in \mathcal{M}_m(\bm{\mu},r):=\mathcal{M}_m \cap B(\bm{\mu}, r) \,.
\end{equation}
A Lagrange interpolant $\REV{\mathbf{A}^{I}(\cdot)}$ is then constructed from the set of DMD operators $\{\mathbf{A}_{k_j}(\bm{\mu}_j)\,|\, \bm{\mu}_j\in \mathcal{M}_m(\bm{\mu},r)\}$ and queried at $\bm{\mu}$ to produce $\REV{\mathbf{A}^I}({\bm{\mu}})$. Unlike the stacked DMD, rKOI creates a DMD operator for each parameter $\bm{\mu}_i\in\mathcal{M}_m$, and only operators corresponding to parameters close to the parameter of interest $\bm{\mu}$ contribute to the approximation of the flow map at this parameter. \REV{For this reason we choose to work with rKOI in this paper.}

%% file: FOM.tex
\REV{To develop and demonstrate the dynamic flux surrogate-based partitioned method, we consider the problem of diffusive transport of a scalar quantity across an interface separating two different materials. In this context a loosely coupled partitioned scheme is an attractive and cost-effective alternative to a monolithic solver when codes implementing the two different materials already exist, and/or simulation scenarios exploring designs involving multiple material combinations.}

\REV{To model the dynamics of the scalar quantity we consider a system of two conservation laws}
\begin{equation}\label{eq:TP}
\left\{
 \begin{aligned}
  \dot{u}_i - \nabla \cdot F_i(u_i) &= f_i  \quad \mbox{in} \quad \Omega_i\times[0,T] \\[0.5ex]
                                          {u}_i  &= {g}_i  \quad \mbox{on} \quad \Gamma_i\times[0,T] \\[0.5ex]
                            u_i(\bm{x}, 0)  &= u_{i,0}(\bm{x}) \quad \mbox{in} \quad \Omega_i
  \end{aligned}
  \right.\quad i=1,2,
\end{equation}
coupled by standard interface compatibility conditions
\begin{equation} \label{interface_conditions}
u_1(\bm{x}, t) = u_2(\bm{x}, t) \quad\text{and}\quad F_1(u_1)\cdot\bm{n}_\gamma= F_2(u_2)\cdot\bm{n}_\gamma
 \quad \mbox{on} \quad \gamma\times[0,T]\,,
\end{equation}
that prescribe continuity of the states and the fluxes on the interface, respectively.  In \eqref{eq:TP}--\eqref{interface_conditions}, $f_i$, $g_i$ and $u_{i,0}$ are a given source term, Dirichlet boundary data, and initial condition, respectively.
\REV{For simplicity we restrict attention to fluxes that are linear functions of the states $u_i$ and comprise a sum of diffusive and advective terms, i.e.,}
\begin{equation*}
    F_i(u_i) = \kappa_{i}\nabla u_i - \bm{v}u_i,\quad  i=1,2.
\end{equation*}
\REV{Here  $\kappa_{i}>0$ describes a material property such as diffusivity, permittivity, or thermal conductivity in} $\Omega_i$ and $\bm{v}$ is a velocity field. \REV{We shall refer to the coupled system \eqref{eq:TP}--\eqref{interface_conditions} as the \emph{transmission problem} (TP). We parameterize this problem by the subdomain coefficients $\kappa_i$, i.e., we set $\bm{\mu} = \{\kappa_1,\kappa_2\}$.  For brevity we shall refer to $\kappa_i$ as the diffusion parameter.}

\REV{We choose the  coupled $\mu$PDE problem \eqref{eq:TP}--\eqref{interface_conditions} as a basis for the development and demonstration of our approach because it possesses all of the characteristics relevant to the formulation of the dynamic flux surrogates, yet it is simple enough to avoid non-essential technicalities. However, the data-driven partitioned solution approach in this paper can be applied to more general versions of the model TP, such as problems with non-linear flux functions and/or imperfect interfaces  \cite{Javili_14_CMAME}. Such interfaces allow for jumps in the state and/or the flux due to, e.g., Kapitza thermal resistance \cite{Mahan_09_PRB} and appear in, e.g., climate models \cite{Lemarie_15_PCS}. We briefly discuss some possible extensions of our approach in Section \ref{sec:ext}.}

\subsection{\REV{Conceptualization of the dynamic flux surrogate-based partitioned approach}}\label{sec:closure}

\REV{In this section we describe the key junctures of our data-driven partitioned solution approach for \eqref{eq:TP}--\eqref{interface_conditions} using a synchronous\footnote{\REV{In such a setting the sub-models exchange instantaneous fluxes as opposed to the asynchronous setting where flux exchange happens over time windows.}} explicit time integration setting.  Construction of the dynamic flux surrogate underpinning the approach is discussed in detail in  Section \ref{sec:DMD}, while an extension to more general time integration schemes is sketched in Section \ref{sec:ext}.}

\REV{We start by letting $\lambda = F_1(u_1)\cdot\bm{n}_\gamma= F_2(u_2)\cdot\bm{n}_\gamma$ be the unknown flux across the interface and treating $\lambda$ as Neumann boundary data. This yields the following, formally complete versions of the subdomain $\mu$PDEs:}
\begin{equation}\label{eq:strong_subd}
\left\{
 \begin{aligned}
  \dot{u}_i - \nabla \cdot F_i(u_i) &= f_i                    &\text{in}\ &\Omega_i\times(0,T] \\[0.5ex]
                                          {u}_i  &= {g}_i                   &\text{on}\ & \Gamma_i\times(0,T] \\[0.5ex]
		F_i(u_i)\cdot\bm{n}_i &= (-1)^{i}\lambda          & \text{on}\ & \gamma\times(0,T] \\[0.5ex]
                            u_i(\bm{x}, 0)  &= u_{i,0}(\bm{x}) & \text{in}\ &\quad \Omega_i
  \end{aligned}
  \right.\,;\quad i=1,2.
\end{equation}
\REV{Next, we discretize \eqref{eq:strong_subd} in space to obtain two separate systems of ODEs
\begin{equation}\label{eq:ODE-subd}
\left\{
\begin{aligned}
\mathbf{M}_1\dot{\mathbf{u}}_1  + \mathbf{K}_1 \mathbf{u}_1 &=\mathbf{f}_1 - \mathbf{G}_1^T\bm{\lambda}\,, \ \  t\in (0,T] \\[1ex]
\mathbf{u}_{1}(0) = \mathbf{u}_{1,0} 
\end{aligned}
\right.
\quad\mbox{and}\quad
\left\{
\begin{aligned}
\mathbf{M}_2\dot{\mathbf{u}}_2  + \mathbf{K}_2 \mathbf{u}_2 &=\mathbf{f}_2 + \mathbf{G}_2^T\bm{\lambda}\,, \ \   t\in (0,T] \\[1ex]
\mathbf{u}_2(0)  & = \mathbf{u}_{2,0} 
\end{aligned}
\right.
\end{equation}
where $\mathbf{u}_i(t)$, $\bm{\lambda}(t)$, and $\mathbf{f}_i(t)$ are coefficient vectors of the subdomain states, the Neumann data, and the source terms respectively, $\mathbf{u}_{i,0}$ are initial condition vectors, $\mathbf{M}_i$ are mass matrices, $\mathbf{K}_i$ are stiffness matrices, and $\mathbf{G}_i$ are matrices implementing the Neumann boundary condition on the interface.}

\REV{Construction of the dynamic flux surrogate does not depend on the spatial discretization scheme employed to obtain the subdomain ODEs \eqref{eq:ODE-subd}. In this paper we will discretize \eqref{eq:strong_subd} by a standard Galerkin method implemented with the lowest order $C^0$ finite element spaces $S^h_{i}$; $i=1,2$, introduced in Section \ref{sec:notation}. We will treat the Dirichlet boundary data as a type of system forcing, decomposing the finite element solution as 
\begin{equation}\label{eq:dirichlet-data}
u^h_i(x,t) = u^h_{i,D} + g^h_{i}(x,t);\quad
 u^h_{i,D}(\cdot,t) \in S^h_{i,D}\ \  \mbox{and} \ \ g^h_{i}(\cdot,t)\in S^h_{i,\Gamma}\,,
\end{equation}
where $g^h_{i}$ is the finite element interpolant of the Dirichlet data $g_i$. As a result, the coefficient vector of $u^h_i(x,t)$ is  $\mathbf{u}_i(t) = (\mathbf{u}_{i,D}(t),\mathbf{g}_i(t))$ where $\mathbf{g}_i(t)\in\mathbb{R}^{n_{i,\Gamma}}$ contains the nodal values of the function $g_i(x,t)$ on the Dirichlet boundary $\Gamma_i$ and 
$\mathbf{u}_{i,D}(t) = (\mathbf{u}_{i,\gamma}(t),\mathbf{u}_{i,\circ}(t))$; see Section \ref{sec:notation}. Thus, the subdomain problems \eqref{eq:ODE-subd} are systems of $n_{i,D}$ ODEs for the unknown finite element coefficients $\mathbf{u}_{i,D}(t)$. However, for notational clarity we will denote these coefficients simply by $\mathbf{u}_{i}(t)$. Similarly, $\mathbf{u}_{i,0}$ is the coefficient vector of the approximation of the initial data out of the space $S^h_{i,D}$ and $\mathbf{f}_i(t)$ is a forcing term which combines the contributions from the source term $f_i(x,t)$ and the Dirichlet data $g_i(x,t)$ in \eqref{eq:strong_subd}.
We will approximate the interface flux by a finite element function $\lambda^h\in\Lambda^h$ with coefficient vector $\bm{\lambda}(t)$ where $\Lambda^h = S^h_{1,\gamma}$ or $\Lambda^h=S^h_{2,\gamma}$.}

\REV{Finally, we discretize \eqref{eq:ODE-subd} in time by applying the forward Euler scheme on a uniform partition $0=t_0<t_1<\ldots <t_K = T$ of the simulation time interval $[0,T]$. This yields the fully discrete system 
\begin{equation}\label{eq:FE}
\left\{
\begin{aligned}
\mathbf{u}_{1,k+1} = \mathbf{u}_{1,k} +
\Delta t \mathbf{M}_1^{-1} \left( \mathbf{f}_{1,k} - \mathbf{K}_1\mathbf{u}_{1,k} -	\mathbf{G}_1^T\bm{\lambda}_{k}\right) \\[1ex]
\mathbf{u}_{2,k+1} = \mathbf{u}_{2,k} +
\Delta t \mathbf{M}_2^{-1} \left( \mathbf{f}_{2,k} - \mathbf{K}_2\mathbf{u}_{2,k} + \mathbf{G}_2^T\bm{\lambda}_{k}\right) 
\end{aligned}
\right.\,,
\quad k=0,\ldots,K-1\,,
\end{equation}
where $\mathbf{u}_{i,k}$ is approximation of $\mathbf{u}_{i}(t_k)$, $\mathbf{f}_{i,k} = \mathbf{f}_{i}(t_k)$,  $\bm{\lambda}_{k} = \bm{\lambda}(t_k)$ and $\Delta t$ is the uniform time step.
 The coefficient vector $\mathbf{u}_{i,k}$ corresponds to a finite element function $u^h_{i,k}\in S^h_{i,D}$, which approximates the semi-discrete solution at time $t_k$, i.e., $u^h_{i,k}\approx u^h_{i,D}(x,t_k)$.
}

\REV{Note that the system \eqref{eq:FE} is incomplete because the Neumann data $\bm{\lambda}_{k}$ is unknown at the current time step $t_k$. The basic idea of the dynamic surrogate-based approach is to estimate the flux at this time step as 
\begin{equation}\label{eq:DMD-prelim}
\bm{\lambda}_{k} = \mathbf{A}_\lambda \mathbf{y}_{k-1}\,,
\end{equation}
where $\mathbf{A}_\lambda$ approximates the dynamics of the interface flux of  \eqref{eq:TP}--\eqref{interface_conditions}, and $\mathbf{y}_{k-1}$ is a suitable state vector at the previous time step $t_{k-1}$.
To construct the operator $\mathbf{A}_\lambda$ we will use a Dynamic Mode Decomposition (DMD) approach adapted to our needs. Details of this construction are discussed in the next section.}

\REV{
\begin{remark}\label{rem:IVR}
An alternative way to estimate the Neumann condition $\bm{\lambda}_{k}$ is as follows. We first augment  \eqref{eq:ODE-subd} with the constraint equation 
\begin{equation}\label{eq:constr-dis}
\mathbf{G}_1\dot{\mathbf{u}}_1 - \mathbf{G}_2 \dot{\mathbf{u}}_2 = 0 \,,
\end{equation}
which is equivalent to the first condition in \eqref{interface_conditions} as long as the initial data is continuous across the interface. The resulting coupled system, comprising the subdomain ODEs \eqref{eq:ODE-subd} and the constraint \eqref{eq:constr-dis}, is a hybrid monolithic discretization of the model problem \eqref{eq:TP}--\eqref{interface_conditions} in which $\bm{\lambda}$ is the Lagrange multiplier enforcing the first condition in \eqref{interface_conditions}. 
Elimination of the subdomain states from the hybrid monolithic problem then yields a linear system for $\bm{\lambda}$
\begin{equation}  \label{eq:schur}
    \mathbf{S}\bm{\lambda} = 
    \mathbf{G}_1\mathbf{M}_1^{-1}\mathbf{b}_1 - 
    \mathbf{G}_2\mathbf{M}_2^{-1}\mathbf{b}_2 \,,
\end{equation}
where $\mathbf{b}_i = \mathbf{f}_i - \mathbf{K}_i\mathbf{u}_i$ and 
\begin{equation}\label{eq:schur-matrix}
    \mathbf{S} = \mathbf{G}_1\mathbf{M}_1^{-1}\mathbf{G}_1^T + \mathbf{G}_2\mathbf{M}_2^{-1}\mathbf{G}_2^T \,.
\end{equation}
One can show \cite{Bochev_19_CAMWA}  that the hybrid monolithic system \eqref{eq:ODE-subd}, \eqref{eq:constr-dis} is a well-posed Hessenberg Index-1 Differential Algebraic Equation (DAE)  \cite{Ascher_98_BOOK} and so the Schur complement \eqref{eq:schur-matrix} is a symmetric and positive definite matrix.  We can use this fact to estimate $\bm{\lambda}_{k}$ by setting $t=t_k$ in \eqref{eq:schur} and solving this system to obtain
\begin{equation}\label{eq:schur-estimate}
\bm{\lambda}_k = \mathbf{S}^{-1}\left(
    \mathbf{G}_1\mathbf{M}_1^{-1}\mathbf{b}_{1,k} - 
    \mathbf{G}_2\mathbf{M}_2^{-1}\mathbf{b}_{2,k}\right) \,,
\end{equation}
where $\mathbf{b}_{i,k} = \mathbf{f}_{i,k} - \mathbf{K}_i\mathbf{u}_{i,k}$. The Schur complement estimate \eqref{eq:schur-estimate} of the flux is the basis of the Implicit Value Recovery (IVR) scheme \cite{Bochev_19_CAMWA}, which is a hybridized solution method \cite{Brezzi_91_BOOK} for the transmission problem \eqref{eq:TP}--\eqref{interface_conditions} that is conceptually similar to FETI \cite{Farhat_94_IJNME} and the localized Lagrange multiplier schemes \cite{Park_01_CMAME,Ross_09_CMAME,Ross_08_CMAME}.
\end{remark}
}

\REV{We will use two versions of the IVR scheme to benchmark the data-driven loosely coupled scheme utilizing the dynamic flux surrogate estimate \eqref{eq:DMD-prelim}.  These versions, termed IVR(C) and IVR(L), respectively, form the Schur complement \eqref{eq:schur-matrix} using consistent and lumped mass matrices, respectively. 
One can show \cite{Bochev_19_CAMWA} that IVR(C) is second order accurate and that on grids with matching interface nodes the hybridized IVR(C) solution coincides with the solution of a non-hybrid monolithic discretization of the model TP when the diffusion coefficient is continuous across the interface.  In contrast, IVR(L) is only first order accurate but its complexity is similar to that of a loosely coupled scheme and it is cheaper than IVR(C) roughly by a factor of $\max_{i=1,2}n_{i,\gamma}/n_{i,D}$.
Thus, we shall use IVR(C) as a ``monolithic'' accuracy benchmark while IVR(L) will serve as a proxy for a loosely coupled scheme that provides the performance target for our data-driven partitioned method.}

%% file: DMD.tex
\REV{In this section we adapt Dynamic Mode Decomposition (DMD) \cite{Mezic_05_ND} to construct the dynamic flux surrogate  \eqref{eq:DMD-prelim}. Our goal is to learn an operator  $\mathbf{A}_\lambda$  whose computational cost is comparable to that of IVR(L), but whose accuracy approaches that of IVR(C). To achieve the first objective we will define the state of $\mathbf{A}_\lambda$ by taking into account the local nature of the interface flux. The second objective requires accurate snapshots of the interface flux. 
To that end we shall use IVR(C), which produces second-order accurate approximations of the flux and the states, to solve the hybrid monolithic discretization of \eqref{eq:TP}--\eqref{interface_conditions}.
We first discuss construction of $\mathbf{A}_\lambda$ for \eqref{eq:TP}--\eqref{interface_conditions}  with \emph{fixed} diffusion coefficients and then formulate a  parametric version of $\mathbf{A}_\lambda$ in Section \ref{sec:param-DMD} for the  \emph{parameterized} transmission problem.}

\subsection{\REV{Dynamic flux surrogate for transmission problems with fixed diffusion parameters (DMD-FS)}}\label{sec:single-DMD}
\REV{Assume that the diffusion parameter $\bm{\mu} = \{\kappa_1,\kappa_2\}$ in \eqref{eq:TP}--\eqref{interface_conditions}  is fixed and} let $\mathbf{u}_{1,k}$, $\mathbf{u}_{2,k}$, and $\bm{\lambda}_k$; $k=0,1,\ldots$ be the IVR(C) solution of the \REV{hybrid monolithic discretization of the} TP at time $t_k$. There are several possible ways to specialize the generic DMD system \eqref{eq:DMD-single} to this type of data. 
Since our goal is to develop a surrogate model for the dynamics of the interface flux, the most straightforward approach 
\REV{would be to construct the snapshot matrices $\mathbf{Y}$ and $\mathbf{Y}'$ in \eqref{operator_equation} using DMD states $\mathbf{y}_k = \bm{\lambda}_k$.}
However, this choice does not allow for any information exchange between the subdomains, which runs counter to the role of the \REV{interface flux} as the ``glue'' that keeps the subdomain states continuous across the interface. 
Indeed, \REV{an operator $\mathbf{A}_\lambda$ acting on a DMD state} $\mathbf{y}_k = \bm{\lambda}_k$ would generate  a dynamic \REV{Neumann} boundary condition for the subdomain equations without any input from their states. As a result, the flux \REV{predicted by $\mathbf{A}_\lambda$} can drift apart from the actual flux needed to satisfy the first coupling condition in \eqref{interface_conditions}.

This observation suggests that the surrogate needs to be made aware of the subdomain states so that it can generate fluxes that will keep these states continuous across $\gamma$. To that end, we consider a state $\mathbf{y}_k$ comprising the Lagrange multiplier concatenated with the subdomain solutions on either side of the interface. Depending on the time instances at which one samples these fields, such a state can be constructed in two different ways. 

Let $t_k$ be the current time step at which we seek an estimate of the interface flux. 
The first way to define a state for our flux surrogate is to view the DMD operator as an explicit time integrator and sample all fields at the previous time instance $t_{k-1}$. This construction agrees with the traditional DMD utilization in which an initial system state $\mathbf{y}_0$  is specified and propagated forward in time to all future states by a repeated application of the DMD operator to $\mathbf{y}_0$. 
A flux surrogate that adheres to this viewpoint would have to act on a \REV{DMD} state defined as $\mathbf{y}_{k-1} = \left(\bm{\lambda}_{k-1}, \mathbf{u}_{1,k-1},\mathbf{u}_{2,k-1}\right)^T$ in order to produce $\bm{\lambda}_{k-1}$. However, such a state does not account for the fact that more current solution information is already available at $t_k$.
This information can be incorporated into the dynamics of the DMD operator by defining its state as
\begin{equation}
\mathbf{y}_{k-1} := \begin{bmatrix}\bm{\lambda}_{k-1}\\ \mathbf{u}_{1,k}\\\mathbf{u}_{2,k}\end{bmatrix}\,,\quad
k=1,2,\ldots \,.
\label{flux_dmd_state}
\end{equation}
In so doing, the surrogate model is combining the most recent solution information computed by the subdomain equations on both sides of the interface with its previous prediction of the \REV{interface flux}. 
Application of the DMD approach to such ``staggered'' states departs from its traditional use and can be interpreted as advancing the \REV{interface flux} in time in a semi-implicit manner. 

\begin{figure}[t!]
  \begin{center}
    \includegraphics[width=0.65\textwidth]{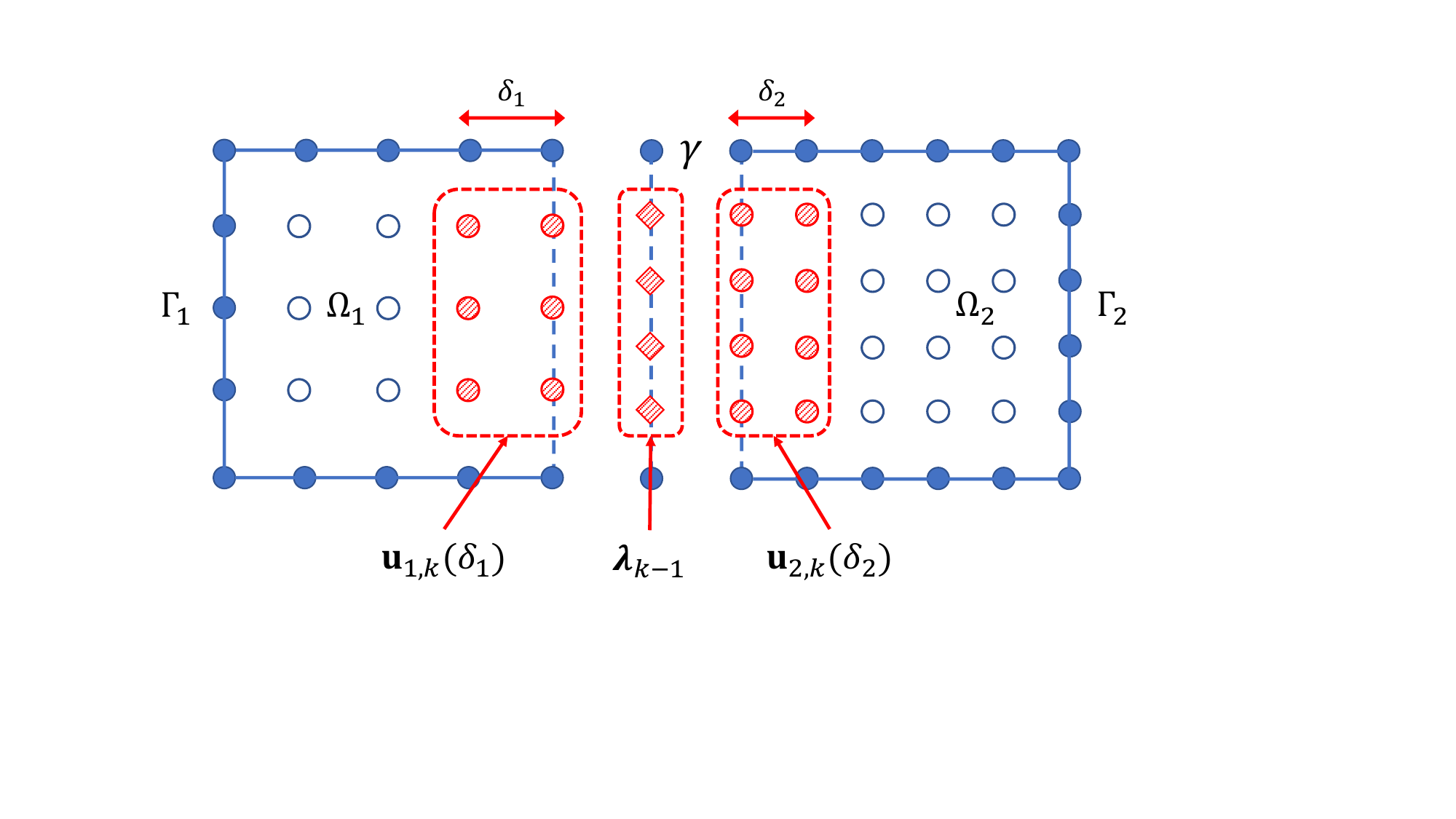}
  \end{center}
\vspace{-3ex}
\caption{The  staggered DMD state $\mathbf{y}_{k-1}$ comprises interface patches $\mathbf{u}_{i,k}(\delta_i)\subset \mathbf{u}_{i,k}$ of subdomain solution coefficients at the current time $t_k$ stacked together with the coefficient vector $\bm{\lambda}_{k-1}$ of the interface flux at the previous time $t_{k-1}$.} \label{fig:DMD-state}
\end{figure}

However, because the state \eqref{flux_dmd_state}  includes the complete subdomain coefficient vectors, its length is 
$N = n_{l,\gamma} + n_{1,D}+ n_{2,D}$. As a result, the computational cost of a surrogate acting on \eqref{flux_dmd_state} is comparable to that of IVR(C), i.e., such a surrogate will not meet our efficiency goal. At the same time it stands to reason that since the role of the interface flux is to maintain solution continuity across $\gamma$, its dynamics should not be strongly influenced by solution values away from the interface. 
Thus, a more economical DMD state, meeting the efficiency goal, can be designed by including only solution coefficients that are close to the interface. To formalize this idea, consider a distance threshold $0< \delta_i <\mbox{diam}(\Omega_i)$, $i=1,2$ and let
\begin{equation}\label{eq:patch-def}
\mathbf{u}_{i,k}(\delta_i) = 
\left\{
(\mathbf{u}_{i,k})_j \, |\, \exists \bm{x}_j\in\Omega^h_i\ \mbox{s.t.}\ d(\bm{x}_j,\gamma) < \delta_i
\right\}
\in \mathbb{R}^{n_{\delta_i,D}}
\end{equation}
be the subset of $\mathbf{u}_{i,k}$ containing all solution coefficients located on nodes that are within distance $\delta_i$ from $\gamma$; see Fig. \ref{fig:DMD-state}. We shall refer to $\mathbf{u}_{i,k}(\delta_i)$ as the interface \emph{patch} of $\mathbf{u}_{i,k}$. We now redefine the DMD state \eqref{flux_dmd_state} by replacing the complete coefficient vectors by their interface patches as follows:
\begin{equation}
\mathbf{y}_{k-1} := 
\begin{bmatrix}
\bm{\lambda}_{k-1}\\ 
\mathbf{u}_{1,k}(\delta_1)\\
\mathbf{u}_{2,k}(\delta_2)
\end{bmatrix}\,,\quad
k=1,2,\ldots \,.
\label{flux_dmd_state_short}
\end{equation}
The length of the redefined state is $N= n_{l,\gamma} + n_{\delta_1,D}+ n_{\delta_2,D}$. Thus, to achieve our efficiency goal we shall keep the interface patches as small as possible and roughly of the same order as the dimension of the interface space  $S^h_{i,\gamma}$. In other words, we shall require $n_{\delta_i,D} = O(n_{i,\gamma})$. Section \ref{sec:patch-sizes} provides further information about the selection of the interface patches.

Assume that a DMD operator $\mathbf{A}$ acting on the states \eqref{flux_dmd_state_short} has been identified. This operator has a 3-by-3 block structure given by 
$$
\mathbf{A} = 
\begin{bmatrix}
A_{\lambda,\lambda} & A_{\lambda,u_1} & A_{\lambda,u_2} \\
A_{u_1,\lambda}        & A_{u_1,u_1} & A_{u_1,u_2} \\
A_{u_2,\lambda}        & A_{u_2,u_1} & A_{u_2,u_2} \\
\end{bmatrix}
$$ 
where the subscripts indicate the range and the domain of each block. For example, $A_{\lambda,u_1}$ is a $n_{l,\gamma}\times n_{{\delta_1},D}$ matrix multiplying the second sub-vector of the input state $\mathbf{y}_{k-1}=(\bm{\lambda}_{k-1}, \mathbf{u}_{1,k}(\delta_1),\mathbf{u}_{1,k}(\delta_2))^T$ and contributing to the first sub-vector of the output state
$\mathbf{y}_{k}=(\bm{\lambda}_k, \mathbf{u}_{1,k+1}(\delta_1),\mathbf{u}_{1,k+1}(\delta_2))^T$. 
Since we only need predictions of the interface flux, the computational efficiency of the flux surrogate can be further improved  by discarding the second and the third row in $\mathbf{A}$ \REV{and setting}
$$
\mathbf{A}_{\lambda} =
\begin{bmatrix}
A_{\lambda,\lambda} & A_{\lambda,u_1} & A_{\lambda,u_2}
\end{bmatrix}\,.
$$
In so doing, we save approximately $(N - n_{l,\gamma})(2N - 1)$ flops compared to the cost of the full size operator $A$. The truncation of \REV{$\mathbf{A}$ to obtain $\mathbf{A}_\lambda$} is another key distinction between the conventional use of DMD and its application to \REV{obtain dynamic interface flux surrogates.} 

\REV{Implementation of the data-driven partitioned scheme in Section \ref{sec:closure} involves an offline training stage where one learns the operator $\mathbf{A}_\lambda$ from data and an online simulation stage where this operator is used to solve the model TP for a given diffusion parameter. Algorithm \ref{fsdmd_algorithm} summarizes these stages.}
%
\begin{algorithm}[t!]
\caption{\REV{DMD-FS partitioned method for TP with fixed diffusion parameters}}
\smallskip
\textbf{Offline}
\smallskip

For $i=1,2$:

\begin{itemize}
\item[1.] Given diffusion coefficients  $\kappa_1$ and $\kappa_2$, collect $S+1$ equally spaced in time solution snapshots $\mathbf{u}_{1,k}$, $\mathbf{u}_{2,k}$, and $\bm{\lambda}_k$; $k=0,\ldots,S$ by using IVR(C) to solve \REV{a hybrid monolithic discretization of} \eqref{eq:TP}--\eqref{interface_conditions}. 

\item[2.] Choose a distance threshold $0<\delta_i<\mbox{diam}(\Omega_i)$ and assemble the interface patches $\mathbf{u}_{i,k}(\delta_i)$ as in \eqref{eq:patch-def}.

\item[3.] Form the staggered states $\mathbf{y}_{k-1} = \left(\bm{\lambda}_{k-1}, \mathbf{u}_{1,k}(\delta_1),\mathbf{u}_{2,k}(\delta_2)\right)^T$ and the snapshot matrices $\mathbf{Y}$ and $\mathbf{Y}^\prime$. 

\item[4.] \REV{Choose a tolerance $0<\epsilon<1$ and perform  Algorithm \ref{dmd_algorithm} to identify the \emph{full size} DMD operator $\mathbf{A}$.}

\item[5.] Truncate $\mathbf{A}$ to obtain the DMD flux operator $\mathbf{A}_\lambda$.
\end{itemize}

\textbf{Online}
\smallskip

Assume a partition  $0=t_0<t_1<\ldots <t_K = T$ of the simulation time interval and compute the initial condition vectors 
$\mathbf{u}_{i,0}$, $i=1,2$. For $i=1,2$ and $k=0,1,\ldots,K-1$:

\begin{itemize}
\item[1.] \textbf{Synchronize:}

\begin{itemize}
\item[1.1]  \textbf{Construct state:} Assemble ${\mathbf{y}}_{k-1}$ from DMD prediction $\bm{\lambda}_{k-1}$, and subdomain patches $\mathbf{u}_{i,k}(\delta_i)$.
\item[1.2] \textbf{Predict  flux:} Apply $\mathbf{A}_\lambda$ to ${\mathbf{y}}_{k-1}$ to predict $\bm{\lambda}_k$:
$$
\bm{\lambda}_k =  \mathbf{A}_\lambda \mathbf{y}_{k-1} \,.
$$
 \end{itemize}
 
\item[2.]  \textbf{Step in time:} Use forward Euler to solve \eqref{eq:ODE-subd} with Neumann data $\bm{\lambda}_{k}$:

$$
\mathbf{u}_{i,k+1} = \mathbf{u}_{i,k} +
\Delta t \mathbf{M}_i^{-1} \left( \mathbf{f}_{i,k} - \mathbf{K}_{i}\mathbf{u}_{i,k} +(-1)^i \mathbf{G}^T_i\bm{\lambda}_{k}\right)\,.
$$
\label{fsdmd_algorithm}
\end{itemize}
\end{algorithm}

\subsection{\REV{Dynamic flux surrogate for parameterized transmission problems ($\mu$DMD-FS)}}\label{sec:param-DMD}
In this section we extend Algorithm \ref{fsdmd_algorithm} to handle the parameterized model transmission problem \eqref{eq:TP}. To that end we shall replace the \REV{dynamic flux surrogate operator $\mathbf{A}_\lambda$} by a parameterized \REV{version $\mathbf{A}^{I}_{\lambda}$} based on the reduced Koopman operator inference (rKOI) algorithm described in Section \ref{sec:DMD-param}. 
Application of rKOI requires modification of the offline phase in Algorithm \ref{fsdmd_algorithm}, where now one has to sample the parameter space of the coupled $\mu$PDE problem \eqref{eq:TP} . 
Recall that this problem is parameterized by the subdomain diffusion coefficients, i.e., $\bm{\mu}=\{\kappa_1,\kappa_2\}\in\mathcal{M}$. For simplicity we consider a rectangular parameter domain defined by lower and upper bounds for each diffusion coefficient:
\begin{equation}\label{eq:parspace}
\mathcal{M} := [\kappa_{1,\min},\kappa_{1,\max}]\times[\kappa_{2,\min},\kappa_{2,\max}]\subset \mathbb{R}^2\,.
\end{equation} 
Let $\mathcal{M}_m:=\{\bm{\mu}_j\}_{j=1}^m$ denote a representative set of parameter samples $\bm{\mu}_j = \{\kappa_{1,j},\kappa_{2,j}\}$. 
For every parameter $\bm{\mu}_j\in \mathcal{M}_m$ we generate a set of $S+1$ equally spaced in time solution snapshots of the states and the flux
$\{\mathbf{u}_{1,k}(\bm{\mu}_j),\mathbf{u}_{2,k}(\bm{\mu}_j),\bm{\lambda}_{1,k}(\bm{\mu}_j)\}$; $k=0,\ldots,S$ by using IVR(C) to solve \REV{a hybrid monolithic discretization of} \eqref{eq:TP}. 
Next, we select distance thresholds $\delta_i$, $i=1,2$ and construct the corresponding interface patches. Then, we assemble the patches and the interface flux into staggered states
$$
\mathbf{y}_{k-1}(\bm{\mu}_j) 
= \left(\bm{\lambda}_{1,k-1}(\bm{\mu}_j),\mathbf{u}_{1,k}(\delta_1;\bm{\mu}_j),\mathbf{u}_{2,k}(\delta_2;\bm{\mu}_j)\right)^T
$$
and use these states to precompute the full size DMD operators $\{\mathbf{A}(\bm{\mu}_j)\}_{j=1}^m$. Truncation of these operators yields the DMD flux surrogate operators $\{\mathbf{A}_{\lambda}(\bm{\mu}_j)\}_{j=1}^m$. This completes the offline phase of the algorithm.

During the online phase one uses the operators $\{\mathbf{A}_{\lambda}(\bm{\mu}_j)\}_{j=1}^m$ to compute the rKOI operator $\REV{\mathbf{A}^I_{\lambda}}(\bm{\mu})$ for any parameter value $\bm{\mu}\in \mathcal{M}$. \REV{This completes the extension of the data-driven partitioned approach in Section \ref{sec:closure} to the parameterized coupled problem. Implementation of the corresponding loosely coupled scheme is summarized in Algorithm \ref{parametric_fsdmd_algorithm}.}

\REV{
\begin{remark}\label{rem:basis}
The basis functions $\mathcal{N}_{\bm{\mu}_j}$ in \eqref{eq:rKOI} are typically chosen to satisfy the property $\mathcal{N}_{\bm{\mu}_j}(\bm{\mu})_i = \delta_{ij}$, where $\delta_{ij}$ is the Kroneker delta function. In this work we assume that the rectangular parameter domain $\mathcal{M}$ in \eqref{eq:parspace} is sampled on a uniform Cartesian grid and that the ball $B(\bm{\mu}, r)$ is such that the set $\mathcal{M}_m(\bm{\mu},r)$ comprises the four vertices of the mesh cell $\mathcal{S}$ containing the parameter value of interest $\bm{\mu}$; see Fig.~\ref{fig:param-DMD-1}. In such a case, the bilinear interpolant of the vertex values of $\mathcal{S}$; see \cite[\S3.2]{Hughes_00_BOOK} provides a convenient, second-order accurate choice for the rKOI operator \eqref{eq:rKOI}, which we shall use in this paper. We refer to Section \ref{sec:multimat} for further details.
\end{remark}
}

\begin{algorithm}[t!]
\caption{\REV{$\mu$DMD-FS partitioned method for parameterized TP}}
\medskip
\textbf{Offline}
\smallskip

\begin{itemize}
\item[1.] Sample $\mathcal{M}$ in a region of interest to define a parameter set $\mathcal{M}_m$.

\item[2.] For every $\bm{\mu}_j\in\mathcal{M}_m$ learn $\mathbf{A}_{\lambda}(\bm{\mu}_j)$ by performing steps 1--5 
 in the offline stage of Algorithm \ref{fsdmd_algorithm}. 
\end{itemize}

\textbf{Online}
\smallskip

\smallskip
Assume a partition  $0=t_0<t_1<\ldots <t_K = T$ of the simulation time interval and compute the initial condition vectors 
$\mathbf{u}_{i,0}$, $i=1,2$. 
\medskip

\textbf{Compute rKOI DMD surrogate:}  Given $\bm{\mu}\in \mathcal{M}$ choose a ball $B(\bm{\mu}, \REV{r})$, form the set $\mathcal{M}_m(\bm{\mu},r)$ as in \eqref{eq:intersect}, and set
\begin{equation}\label{eq:rKOI}
\REV{\mathbf{A}^{I}_{\lambda}}(\bm{\mu}):=\sum_{\bm{\mu}_j\in \mathcal{M}_m(\bm{\mu},r)}\REV{\mathcal{N}}_{\bm{\mu}_j}(\bm{\mu})\mathbf{A}_{\lambda}(\bm{\mu}_j)
\end{equation}
where $\REV{\mathcal{N}}_{\bm{\mu}_j}$ is \REV{a nodal interpolation basis} for the parameter set  $\mathcal{M}_m(\bm{\mu},r)$.
\medskip

For $i=1,2$ and $k=0,1,\ldots,K-1$

\begin{itemize}
\item[1.] \textbf{Synchronize:}

\begin{itemize}
\item[1.1]  \textbf{Construct state:} Assemble ${\mathbf{y}}_{k-1}$ from DMD prediction $\bm{\lambda}_{k-1}$, and subdomain patches $\mathbf{u}_{i,k}(\delta_i)$.
\item[1.2] \textbf{Predict  flux:} Apply the rKOI  operator $\REV{\mathbf{A}^{I}_{\lambda}}$ to ${\mathbf{y}}_{k-1}$ to predict $\bm{\lambda}_k$:
$$
\bm{\lambda}_k =  \REV{\mathbf{A}^{I}_{\lambda}} \mathbf{y}_{k-1} \,.
 $$
 \end{itemize}
 
\item[2.]  \textbf{Step in time:} Use forward Euler to solve \eqref{eq:ODE-subd} with Neumann data $\bm{\lambda}_{k}$:
$$
\mathbf{u}_{i,k+1} = \mathbf{u}_{i,k} +
\Delta t \mathbf{M}_i^{-1} \left( \mathbf{f}_{i,k} - \mathbf{K}_{i}\mathbf{u}_{i,k} +(-1)^i  \mathbf{G}^T_i  \bm{\lambda}_{k}\right)\,.
$$
\end{itemize}
\label{parametric_fsdmd_algorithm}
\end{algorithm}
\subsection{\REV{Extensions}}\label{sec:ext}
\REV{This section outlines extensions of the dynamic flux surrogate scheme to more general time integration settings and non-standard coupling conditions. A thorough development and demonstration of these extensions is beyond the scope of this work and will be addressed in a forthcoming paper.}

\paragraph{\REV{Extension to the $\theta$-method}}
\REV{We use the $\theta$-method to illustrate how the dynamic flux surrogate approach can be combined with more general time integration schemes. As in Section \ref{sec:closure}, our starting point is the ODE systems \eqref{eq:ODE-subd} representing the spatially discretized governing equations in each subdomain. We then 
apply the $\theta$-method with parameter $0 < \theta_i \le 1$,  $i=1,2$; to these systems  to obtain  the fully discrete equations\footnote{\REV{We have excluded the values $\theta_1=\theta_2=0$ from consideration because they reduce \eqref{eq:FE-theta} to the explicit formulation \eqref{eq:FE}.}}
\begin{equation}\label{eq:FE-theta}
\left\{
\begin{aligned}
\mathbf{u}_{1,k+1} = \mathbf{u}_{1,k} +
\Delta t \textbf{M}_1^{-1} \left( \mathbf{f}^{\theta_1}_{1,k} - K_{1}\mathbf{u}^{\theta_1}_{1,k} -	\textbf{G}_1^T\bm{\lambda}^{\theta_1}_{k}\right) \\[1ex]
\mathbf{u}_{2,k+1} = \mathbf{u}_{2,k} +
\Delta t \textbf{M}_2^{-1} \left( \mathbf{f}^{\theta_2}_{2,k} - K_{i}\mathbf{u}^{\theta_2}_{2,k} + \textbf{G}_2^T\bm{\lambda}^{\theta_2}_{k}\right) 
\end{aligned}
\right.\,,
\quad k=0,\ldots,K-1\,,
\end{equation}
where 
$\mathbf{u}^{\theta_i}_{i,k} = \theta_i \mathbf{u}_{i,k+1}+(1-\theta_i)\mathbf{u}_{i,k}$,
$\mathbf{f}^{\theta_i}_{i,k} = \theta_i \mathbf{f}_{i,k+1}+(1-\theta_i)\mathbf{f}_{i,k}$, and
$\bm{\lambda}^{\theta_i}_{k} = \theta_i \bm{\lambda}_{k+1} + (1-\theta_i)\bm{\lambda}_k$.
To advance the solution from the current time step $t_k$ to the new time step $t_{k+1}$, we need an estimate of the Neumann data $\bm{\lambda}^{\theta_i}_{k}$ in \eqref{eq:FE-theta}. This data is a linear combination of the flux $\bm{\lambda}_{k}$ at the current time, which is known and the flux $\bm{\lambda}_{k+1}$ at the future time, which is unknown. Thus, in contrast to the explicit problem \eqref{eq:FE} we now seek a flux surrogate $\mathbf{A}_\lambda$ that predicts  $\bm{\lambda}_{k+1}$ from the available state and flux data at the current time step. To construct such a surrogate we can consider a hybrid DMD state defined as
\begin{equation}
\mathbf{y}_{k} := 
\begin{bmatrix}
\bm{\lambda}_{k}\\ 
\mathbf{u}_{1,k}(\delta_1)\\
\mathbf{u}_{2,k}(\delta_2)
\end{bmatrix}\,,\quad
k=1,2,\ldots \,.
\label{flux_dmd_state_short_theta}
\end{equation}
where, as before, $\mathbf{u}_{i,k}(\delta_i)$ are interface solution patches. Because all components of  \eqref{flux_dmd_state_short_theta} are taken at the current time step, the resulting flux surrogate $\mathbf{A}_\lambda$ is similar to a traditional DMD operator.}

\REV{
\begin{remark}\label{rem-var-time}
The standard DMD approach requires solution snapshots at uniformly spaced time intervals. A recently developed DMD extension to variable time steps (VDMD) \cite{Smith_23_NSE} can be used to develop dynamic flux surrogates for partitioned methods that can predict the interface flux at irregular time intervals. 
\end{remark}
}

\paragraph{\REV{Extensions to imperfect interfaces}}
\REV{An important case of an imperfect interface arises in the coupled ocean-atmosphere system in earth system models.  To discuss this example assume that $\Omega_1$ and $\Omega_2$ represent the ocean and the atmosphere domains, respectively, and that $u_i$, $i=1,2$ are the associated ocean and atmosphere temperatures. Let  $\rho_i$, $c^p_i$, and $K^t_i$ denote the density, the specific heat coefficient, and the eddy diffusivity, respectively in $\Omega_i$. Then, the heat transfer between the ocean and the atmosphere can be modeled  by the coupled system \eqref{eq:TP}  in which the second condition in \eqref{interface_conditions} is replaced by the ``bulk'' condition
\begin{equation}\label{eq:bulk}
F_i(u_i):=\kappa_i \nabla u_i \cdot \bm{n}_\gamma  = \rho_2 c^p_2 \| [\bm{v}]\| (u_1-u_2)\,,
\end{equation}
obtained by parametrization of the ocean and atmosphere surface layers  \cite{Lemarie_15_PCS,Lemarie_13_ETNA}. In \eqref{eq:bulk} the ``diffusion'' parameter $\kappa_i = \rho_i c^p_i K^t_i$ and $[\bm{v}]$ is the velocity jump on the interface.}

\REV{Suppose we want to formulate a data-driven partitioned scheme for the coupled ocean-atmosphere system \eqref{eq:TP} and \eqref{eq:bulk} that uses the $\theta$-method described earlier.  Discretization in space and time of the components of this system will yield the exact same subdomain equations \eqref{eq:FE-theta} as in the case of the standard coupling conditions. The only difference is that now the interface flux specifying the Neumann condition for each subdomain equation is defined by the bulk condition \eqref{eq:bulk}. Thus, instead of the DMD state \eqref{flux_dmd_state_short_theta} we  can consider a state defined as
\begin{equation}
\mathbf{y}_{k} := 
\begin{bmatrix}
\rho_2 c^p_2 \| [\bm{v}]\| (\mathbf{u}_{1,k}-\mathbf{u}_{2,k})\big |_{\gamma}\\ 
\mathbf{u}_{1,k}(\delta_1)\\
\mathbf{u}_{2,k}(\delta_2)
\end{bmatrix}\,,\quad
k=1,2,\ldots \,.
\label{flux_dmd_state_bulk_theta}
\end{equation}
}

\begin{remark}\label{rem:imperfect}
\REV{
A general imperfect interface $\gamma$ in heat transfer applications  is modeled by a separate temperature equation for the interface \cite{Javili_14_CMAME}
\begin{equation}\label{eq:imperfect}
c_\gamma \partial_t u_\gamma = -\nabla_\gamma\cdot F_\gamma -[F]\cdot\bm{n}_\gamma + q_\gamma
\end{equation}
where $c_\gamma$ is the heat capacity coefficient on the interface, $\nabla_\gamma\cdot$ is the surface divergence on $\gamma$,  $u_\gamma$ is the interface temperature, $F_\gamma$ is the interface heat flux, $[F]$ is the jump of the bulk heat flux, and $q_\gamma$ is interface heat source. The bulk condition \eqref{eq:bulk} is an instance of \eqref{eq:imperfect} corresponding to a lowly-conducting imperfect interface with Kapitza thermal resistance.
A dynamic flux surrogate approach is particularly attractive for imperfect interfaces because the relationships in \eqref{eq:imperfect} cannot be modeled by a static surrogate approximating the Poincare-Steklov operator. In contrast,  a DMD operator can be trained to learn the flow map of this equation. }
\end{remark}

%% file: Stability.tex
In this section we examine the stability of the surrogate-based partitioned schemes presented in Section \ref{sec:DMD}. We start with analysis of the IVR(C) scheme, which is a hybridized solution method for the model transmission problem based on a Schur complement calculation of the interface flux. In Section \ref{sec:IVR-stab} we show that this scheme maintains the correct balance of energy transfers between the subdomains. As such IVR(C) can be viewed as a stability benchmark for any scheme that employs an approximation of the interface flux, such as the surrogate-based scheme in the focus of this paper. Stability of DMD-FS is investigated in Section \ref{sec:DMD-stab}. In both cases we follow the classical analysis of forward Euler applied to diffusive systems. We recall that this analysis reveals the Courant-Friedrichs-Lewy (CFL) condition $\kappa \Delta t \leq C h^2$ as the critical scaling for stability when $h$ is small, where 
\[ 
\kappa := \max \{ \kappa_1 , \kappa_2\}\quad \text{and}\quad h:=\min \{ h_1 , h_2\}.
\] 
Stability of IVR(C) and DMD-FS partitioned schemes will be analyzed in the context of this CFL condition. 
We first review a few additional technical details that are necessary for the rigorous solution bounds. 

We recall that in this paper the Dirichlet data is treated as a system forcing through the decomposition of the semi-discrete finite element solution into a sum of the boundary interpolant  $g^h_i(x,t)\in S^h_{i,\Gamma}$ of this data and an unknown function $u^h_{i,D}(x,t)\in S^h_{i,D}$; see eq. \eqref{eq:dirichlet-data}.  We remind that for notational clarity the coefficients of $u^h_{i,D}(x,t)$ are denoted as $\mathbf{u}_i(t)$, and that the coefficient vector $\mathbf{u}_{i,k}\approx \mathbf{u}_i(t_k)$ corresponds to the fully discrete finite element solution $u^h_{i,k}\approx u^h_{i,D}(x,t_k)$. Finally, recall that the forcing term $\mathbf{f}_i(t)$ in \eqref{eq:FE} combines the contributions from the source term $f_i(x,t)$ and the Dirichlet data $g_i(x,t)$ in \eqref{eq:strong_subd}. Thus, this term is given by 
\begin{equation}\label{eqn:fHat} 
\mathbf{f}_{i,k} := \hat{\mathbf{f}}_{i,k} - \mathbf{K}_i \mathbf{g}_{i,k} -\frac{1}{\Delta t} \mathbf{M}_i\left( \mathbf{g}_{i,k+1}-\mathbf{g}_{i,k}\right) \,,
\end{equation} 
where $\hat{\mathbf{f}}_{i,k}$ are the coefficients of the projection of  $f_{i,k}(x):=f_i(x,t_k)$ onto $S^h_{i,D}$, and $\mathbf{g}_{i,k}$ are the nodal coefficients of $g^h_{i,k}(x) := g^h_i(x,t_k)$.
%
Similarly, the vectors $\mathbf{b}_{i,k}$ in the Schur complement system \eqref{eq:schur} are given by 
\begin{equation}\label{eqn:bHat}
\mathbf{b}_{i,k} := \mathbf{f}_{i,k} - \mathbf{K}_i \mathbf{u}_{i,k} 
= \hat{\mathbf{f}}_{i,k} - 
\mathbf{K}_i \mathbf{g}_{i,k} -
\frac{1}{\Delta t} \mathbf{M}_i\left( \mathbf{g}_{i,k+1}-
\mathbf{g}_{i,k}\right) 
- \mathbf{K}_i \mathbf{u}_{i,k} \,.
\end{equation} 
The following bound is standard.  
\begin{lem}\label{lem:fBound} 
Given any $v^h_i \in S_{i}^h$, 
\[
\left|\left(\mathbf{v}_i \right)^T  \mathbf{f}_{i,k} \right| \leq 
C_{i,k}\| v^h_i \|_{1,\Omega_i} , 
\] 
where 
\begin{equation*}
C_{i,k} := \left\| f_{i,k}\right\|_{0,\Omega_i} +\frac{1}{\Delta t}\left\| g^h_{i,k+1}-g^h_{i,k}\right\|_{0,\Omega_i} 
+ C \left\| g^h_{i,k}\right\|_{1,\Omega_i}  , 
\end{equation*} 
and $C>0$ depends on $\kappa_i$ and the advective velocity $\bm{v}$.  
\end{lem}  
Thus,  $C_{i,k}$ can be bound in terms of the Dirichlet data ${g}_i(x,t)$.

\begin{defn}[Elliptic bilinear form]
Given any $v^h_i \in S_{i}^h$ and $w^h_i \in S_{i}^h$, we define 
\[
a(v^h_i , w^h_i) := \kappa_i \left( \nabla v^h_i , \nabla w^h_i \right)_{0,\Omega_i} -\left( \bm{v} v^h_i , \nabla w^h_i \right)_{0,\Omega_i} \,.
\]
\end{defn}
One can show \cite{Johnson_92_BOOK} that under some common assumptions about the velocity field $\bm{v}$ there exist constants $c_0>0$ and $c_1>0$ independent of $\Delta t$, $i$ and $h_i$ such that 
\begin{align*}
\quad c_0 \| v^h_i \|_{1,\Omega_i}^2 & \leq a(v^h_i , v^h_i) , \hspace{12ex}   \forall v^h_i \in S_{i,D}^h&\text{(coercivity)} \\ 
\quad  a(v^h_i , w^h_i) & \leq  c_1 \| v^h_i \|_{1,\Omega_i} \|  w^h_i \|_{1,\Omega_i} ,\ \ \forall v^h_i \in S_{i}^h &\text{(continuity)} 
\end{align*}

\subsection{Stability of the IVR(C) scheme} \label{sec:IVR-stab}
Since IVR(C) calculates the flux via \eqref{eq:schur-estimate}, the following analysis shows that the energy transfer is perfectly balanced between subdomains and we retain the standard monolithic energy bounds.  
The proof uses the \textit{polarization identity}, in the form 
\[
\left( a_{k+1}-a_k \right) a_{k+1} = \frac{1}{2} \left( \left| a_{k+1} \right|^2 +\left| a_{k+1}-a_{k} \right|^2 - \left| a_k \right|^2 \right) .  
\]
\noindent 
We use the following inverse estimate; see, \textit{e.g.} \cite{Brenner_02_BOOK}.  
\begin{lem}\label{lem:inverseEst} 
There exists $C_{-1}>0$ independent of $h_i$ (and $i$) such that 
\[
\left\| v^h_i\right\|_{1,\Omega_i} \leq C_{-1} h_i^{-1} \left\| v^h_i \right\|_{0,\Omega_i}, \ \forall v^h_i\in S_i^h.  
\] 
\end{lem}
\noindent 
Also, we recall Young's inequality: given any $a>0$, $b>0$ and $\epsilon >0$, 
\begin{equation}
ab \leq \frac{1}{2\epsilon} a^2 + \frac{\epsilon}{2} b^2 . \label{eqn:Youngs} 
\end{equation} 

\begin{thm}[Stability of IVR(C)]\label{thm:stab_IVRC} 
For $i=1,2$, $k=0,1,\ldots, K-1$ let $\mathbf{u}_{i,k}$ and $u^h_{i,k}\in S^h_{i,D}$ denote the solution  of the IVR(C) scheme and the associated fully discrete finite element solution, respectively.  Assume the following interface compatibility of the initial coefficient vectors: 
\begin{equation}\label{eq:assume}
\mathbf{G}_1 \mathbf{u}_{1,0}  - \mathbf{G}_2 \mathbf{u}_{2,0} = 0.  
\end{equation}
Given a CFL restriction of the form $\Delta t \leq C h^2$, the IVR(C) solutions  satisfy the bound 
\begin{equation}
\sum_{i=1,2}  \left\| u^h_{i,k+1} \right\|_{0,\Omega_i}^2
+ c_0\sum_{i=1,2} \Delta t \sum_{n=0}^k  \left\| u^h_{i,n+1} \right\|_{1,\Omega_i}^2 
\leq \sum_{i=1,2} \left\| u^h_{i,0} \right\|_{0,\Omega_i}^2
+ \frac{2}{c_0}\sum_{i=1,2} \Delta t \sum_{n=0}^k C_{i,n}^2
\end{equation} 
for $0\leq k \leq K-1$. 
\end{thm} 
\begin{proof}
Left-multiply the subdomain ODE systems in  \eqref{eq:FE} by $(\mathbf{u}_{i,k+1})^T\mathbf{M}_i$ and sum over $i=1,2$. 
The result can be rewritten using the polarization identity and bilinear form notation as 
\begin{align}
&\frac{1}{2}\sum_{i=1,2}  \left\| u^h_{i,k+1} \right\|_{0,\Omega_i}^2
+\frac{1}{2}\sum_{i=1,2} \left\| u^h_{i,k+1}-u^h_{i,k} \right\|_{0,\Omega_i}^2 \label{eqn:sp1} \\ 
&=\frac{1}{2}\sum_{i=1,2} \left\| u^h_{i,k} \right\|_{0,\Omega_i}^2
+\Delta t  \sum_{i=1,2}\left( \mathbf{u}_{i,k+1}\right)^T \mathbf{f}_{i,k} 
- \Delta t \sum_{i=1,2}a\left( u^h_{i,k}, u^h_{i,k+1} \right) 
- \Delta t E_k , \nonumber 
\end{align} 
where 
\begin{equation} \label{eq:balance}
E_k :=  \left( \boldsymbol{\lambda}_k \right)^T \left( \mathbf{G}_1 \mathbf{u}_{1,k+1}  - \mathbf{G}_2 \mathbf{u}_{2,k+1}\right) . 
\end{equation}
Owing to assumption \eqref{eq:assume} we have that $E_k=0$, indicating the correct balance of energy transfer between subdomains.  Indeed, it follows from \eqref{eq:FE} and \eqref{eqn:bHat} that 
\begin{equation} \label{eqn:u_Jump}  
\mathbf{G}_1 \mathbf{u}_{1,k+1}  - \mathbf{G}_2 \mathbf{u}_{2,k+1} = 
\mathbf{G}_1 \mathbf{u}_{1,k}  - \mathbf{G}_2 \mathbf{u}_{2,k} 
+ \Delta t \left( \mathbf{G}_1 \mathbf{M}_1^{-1} \mathbf{b}_{1,k} -\mathbf{G}_2 \mathbf{M}_2^{-1} \mathbf{b}_{2,k} - \mathbf{S}\boldsymbol{\lambda}_k \right)
\end{equation}
where  $\mathbf{b}_{i,k}$ are the vectors defined in \eqref{eqn:bHat}.
Then \eqref{eq:schur-estimate} may be applied to conclude 
\begin{equation*}
\mathbf{G}_1 \mathbf{u}_{1,k+1}  - \mathbf{G}_2 \mathbf{u}_{2,k+1} = 
\mathbf{G}_1 \mathbf{u}_{1,k}  - \mathbf{G}_2 \mathbf{u}_{2,k} = 
\ldots = 
\mathbf{G}_1 \mathbf{u}_{1,0}  - \mathbf{G}_2 \mathbf{u}_{2,0} = 0.
\end{equation*}

We return to \eqref{eqn:sp1} and add the terms 
\[
\Delta t \sum_{i=1,2}a\left( u^h_{i,k+1}, u^h_{i,k+1} \right)
\]
to both sides.  Via coercivity, this implies the inequality 
\begin{align}
&\frac{1}{2}\sum_{i=1,2}  \left\| u^h_{i,k+1} \right\|_{0,\Omega_i}^2
+\frac{1}{2}\sum_{i=1,2} \left\| u^h_{i,k+1}-u^h_{i,k} \right\|_{0,\Omega_i}^2 
+ c_0\Delta t \sum_{i=1,2} \left\| u^h_{i,k+1} \right\|_{1,\Omega_i}^2 \label{eqn:sp2} \\ 
&\leq \frac{1}{2}\sum_{i=1,2} \left\| u^h_{i,k} \right\|_{0,\Omega_i}^2
+\Delta t  \sum_{i=1,2}\left( \mathbf{u}_{i,k+1}\right)^T \mathbf{f}_{i,k} 
+ \Delta t \sum_{i=1,2}a\left( u^h_{i,k+1} -u^h_{i,k}, u^h_{i,k+1} \right) \,. \nonumber 
\end{align} 
Appropriate bounds are needed for the right side.  
We apply Lemma \ref{lem:fBound} first, followed by Young's inequality: 
\begin{equation*}
\left| \left( \mathbf{u}_{i,k+1}\right)^T \mathbf{f}_{i,k} \right| 
\leq 
C_{i,k}\| u^h_{i,k+1} \|_{1,\Omega_i} 
\leq \frac{1}{2\epsilon_1}C_{i,k}^2 + \frac{\epsilon_1}{2} \| u^h_{i,k+1} \|_{1,\Omega_i}^2 . 
\end{equation*}
The bilinear form terms are bounded by continuity, then Lemma \ref{lem:inverseEst} and Young's inequality: 
\begin{align*}
\left| a\left( u^h_{i,k+1}   -u^h_{i,k}, u^h_{i,k+1} \right) \right| 
&\leq c_1 \left\| u^h_{i,k+1} -u^h_{i,k}\right\|_{1,\Omega_i} \left\| u^h_{i,k+1}\right\|_{1,\Omega_i} \\[1ex] 
&\leq c_1 C_{-1} h_i^{-1} \left\| u^h_{i,k+1} -u^h_{i,k}\right\|_{0,\Omega_i} \left\| u^h_{i,k+1}\right\|_{1,\Omega_i} \\[1ex]
&\leq
\frac{(c_1 C_{-1})^2}{2\epsilon_2 h_i^2} \left\| u^h_{i,k+1} -u^h_{i,k}\right\|_{0,\Omega_i}^2 
+ \frac{\epsilon_2}{2} \left\| u^h_{i,k+1}\right\|_{1,\Omega_i}^2 .
\end{align*} 
After inserting these bounds into \eqref{eqn:sp2}, choosing $\epsilon_1 = \epsilon_2 = c_0/2$, 
and assuming the time step restriction 
\[
\Delta t \frac{(c_1 C_{-1})^2}{2\epsilon_2 h_i^2} 
\leq \Delta t \frac{(c_1 C_{-1})^2}{2\epsilon_2 h^2} 
\leq \frac{1}{2} 
\]
we find that 
\begin{equation*}
\frac{1}{2}\sum_{i=1,2}  \left\| u^h_{i,k+1} \right\|_{0,\Omega_i}^2
+ \frac{c_0}{2}\Delta t \sum_{i=1,2} \left\| u^h_{i,k+1} \right\|_{1,\Omega_i}^2 
\leq \frac{1}{2}\sum_{i=1,2} \left\| u^h_{i,k} \right\|_{0,\Omega_i}^2
+ \frac{1}{c_0}\Delta t \sum_{i=1,2} C_{i,k}^2 \,.
\end{equation*} 
This concludes the proof.
\end{proof}

\subsection{Stability of the dynamic surrogate-based scheme}\label{sec:DMD-stab}
The term $E_k$ defined in \eqref{eq:balance} provides a measure of the imbalance in the energy transfer between the subdomains at the current time step. Assuming interface compatibility \eqref{eq:assume} of the initial data, equation \eqref{eqn:u_Jump} shows that this term vanishes for the IVR(C) scheme at every time step because the Neumann data  is  computed by solving the Schur complement equation  \eqref{eq:schur}.

This is not the case for the surrogate-based scheme where the Neumann data is computed by \eqref{eq:DMD-prelim}. Even if \eqref{eq:assume} is satisfied at the initial time, the resulting flux $\bm{\lambda}_k$ does not solve equation \eqref{eq:schur} and, as a result, $E_k$ is not guaranteed to be zero at subsequent time steps. This creates a numerical imbalance of energy transfer between subdomains that could potentially destabilize the solution of the surrogate-based partitioned scheme. Consequently, stability of the latter requires a bound on the extra interface terms $E_k$.
We will show that if the dynamic flux surrogate model is well-constructed, this imbalance has a minimal effect on the scaling of the time step, which will remain governed by the stabilization of diffusion.  The following analysis formalizes these considerations.  

We begin with tools to bound interface quantities.  
Let $( \lambda , \mu )_\gamma$ denote the inner-product of $L^2 (\gamma)$, with norm $\| \lambda \|_\gamma = \sqrt{(\lambda , \lambda)_\gamma}$.  
Given $v \in H^1 (\Omega_i)$, recall the lifting bound \cite{Brenner_02_BOOK}
\[
\| v \|_\gamma \leq C \| v \|_{0,\Omega_i}^{1/2} \| v \|_{1,\Omega_i}^{1/2} , 
\]
for some constant $C>0$ that we may take to be independent of $i$ here.  
\begin{lem}\label{lem:lift} 
There exists $C_{L}>0$ independent of $h_i$ (and $i$) such that 
\[
\left\| v^h_i \right\|_\gamma \leq C_{L} h_i^{-1/2} \left\| v^h_i \right\|_{0,\Omega_i}, \ \forall v^h_i \in S_i^h.  
\] 
\end{lem}
\begin{proof} 
The proof follows by applying the lifting bound and then the inverse estimate from Lemma \ref{lem:inverseEst}.
\end{proof}
For the stability analysis we will also need the following technical lemma.
\begin{lem}\label{lem:S_norm} 
Let $\bm{\lambda}$ be the coefficient vector of an arbitrary $\lambda^h\in\Lambda^h$ and let 
$w^h_i\in S_{i,D}^h$ be the finite element function with coefficient vector $\mathbf{w}_i =\mathbf{M}_i^{-1} \mathbf{G}_i^T \boldsymbol{\lambda}$.  Then 
\[
\left\| w^h_i \right\|_{0,\Omega_i} 
\leq C_{L} h_i^{-1/2} \left\| \lambda^h \right\|_\gamma . 
\]
\end{lem} 
\begin{proof} 
We apply Lemma \ref{lem:lift}: 
$$
\left\| w^h_i \right\|_{0,\Omega_i}^2 
= \left( \mathbf{w}_i\right)^T \mathbf{M}_i \mathbf{w}_i
= \left( \mathbf{w}_i\right)^T \mathbf{G}_i^T \boldsymbol{\lambda} 
= (\lambda^h, w^h_i)_\gamma \
\leq \| \lambda^h \|_\gamma \| w^h_i \|_\gamma  
\leq C_{L} h_i^{-1/2} \| \lambda^h \|_\gamma \left\| w^h_i \right\|_{0,\Omega_i}. 
$$
The lemma now easily follows from the last inequality.  
\end{proof}

We may now assert the stability of the surrogate-based partitioned scheme that computes fluxes via \eqref{eq:DMD-prelim} under an assumption that the DMD operator $A_{\lambda}$ is constructed with a sufficient accuracy.  
Define the \textit{target} fluxes $\bm{\lambda}^{\sf S}_k$ at the current time $t_k$ as the solution of the Schur complement system \eqref{eq:schur}, i.e., 
\begin{equation} 
\bm{\lambda}^{\sf S}_k := \mathbf{S}^{-1} \left( \mathbf{G}_1 \mathbf{M}_1^{-1} \mathbf{b}_{1,k} -\mathbf{G}_2 \mathbf{M}_2^{-1} \mathbf{b}_{2,k} \right) , \label{eqn:targetFlux} 
\end{equation} 
and let   
\[
\mathbf{e}_k :=\bm{\lambda}^{\sf S}_k -\boldsymbol{\lambda}_k , 
\] 
be error in the flux computed by the dynamic flux surrogate $A_{\lambda}$ based on its current state. 
In principle, such errors may be controlled by using enough snapshots and retaining enough modes in the construction of the DMD operator; see \cite{Lu_19_SISC}.
Let $e^h_k\in \Lambda^h$ be the finite element function with coefficient vector $\mathbf{e}_k$.
We shall assume that up to time $t_k$ the DMD operator satisfies the accuracy requirement 
\begin{equation}\label{eqn:DMD_acc}
\sqrt{\Delta t \sum_{n=0}^k \| {e}^h_n \|_\gamma^2} \leq \sqrt{C_F \Delta t \, h} 
\end{equation}
for some $C_F>0$.  
Then we prove a stability bound for the next step.  
Under the CFL restriction $\Delta t \leq C\, h^2$,  condition \eqref{eqn:DMD_acc} would be satisfied, for example, by requiring 
\begin{equation}\label{eq:DMD-scale}
\sqrt{\Delta t \sum_{n=0}^k \| {e}^h_n \|_\gamma^2} \leq \sqrt{C_F} \left(  \Delta t^{3/4} \right) , 
\end{equation}
which does not even require optimal time accuracy, if the DMD operator accuracy is interpreted in terms of the accuracy of forward Euler.  
We also note that the proof of the next result shows that the time step restriction to bound any instabilities introduced by an energy imbalance across the interface due to the DMD operator is actually \emph{relaxed} to $\Delta t\leq Ch^{3/2}$.  
This would also relax the CFL-based accuracy condition \eqref{eq:DMD-scale} where the scaling on the right hand side would now become $\Delta t^{7/8}$.  
However, since we must stabilize the diffusion operator, the surrogate-based scheme is still subject to the same time step restriction as IVR(C).  

\begin{thm}[Stability of DMD-FS]\label{thm:stab_DMD}
For $i=1,2$, $k=0,1,\ldots, K-1$ let $\mathbf{u}_{i,k}$ and $u^h_{i,k}\in S^h_{i,D}$ denote the solution  of the surrogate-based scheme and the associated fully discrete finite element solution, respectively. 
Assume that $\boldsymbol{\lambda}_k$ is computed as in \eqref{eq:DMD-prelim} by a DMD operator 
that satisfies the accuracy requirement \eqref{eqn:DMD_acc} at time $t_k$.  
Then, under the assumptions of Theorem \ref{thm:stab_IVRC}, there exists $C>0$ independent of $\Delta t$ and $h_i$ such that at time $t_{k+1}$ 
\begin{multline*}
\sum_{i=1,2}  \left\| u^h_{i,k+1} \right\|_{0,\Omega_i}^2
+ \frac{c_0}{2} \sum_{i=1,2} \Delta t \sum_{n=0}^k \left\| u^h_{i,n+1} \right\|_{1,\Omega_i}^2 \\ 
\leq \sum_{i=1,2} \left\| u^h_{i,0} \right\|_{0,\Omega_i}^2 
+ \frac{c_0}{2} \Delta t \sum_{i=1,2} \left\| u^h_{i,0} \right\|_{1,\Omega_i}^2 
+ C\left( \sum_{i=1,2} \Delta t \sum_{n=0}^k C_{i,n}^2 
+ C_F\, T \right).\qquad
\end{multline*} 
\end{thm} 
\begin{proof} 
It follows from \eqref{eqn:u_Jump} that 
$$
\mathbf{G}_1 \mathbf{u}_{1,k+1}  - \mathbf{G}_2 \mathbf{u}_{2,k+1} = 
\mathbf{G}_1 \mathbf{u}_{1,k}  - \mathbf{G}_2 \mathbf{u}_{2,k} 
+\Delta t  \mathbf{S}\left( \bm{\lambda}^{\sf S}_k- \boldsymbol{\lambda}_k \right)    
$$
and so, 
$$
\mathbf{G}_1 \mathbf{u}_{1,k+1}  - \mathbf{G}_2 \mathbf{u}_{2,k+1} = 
\Delta t \sum_{n=0}^k \mathbf{S}\left( \boldsymbol{\lambda}^{\sf S}_n- \boldsymbol{\lambda}_n \right) . 
$$
We can then manipulate the expression $E_k$ defined in \eqref{eq:balance} (proof of Theorem \ref{thm:stab_IVRC}) into the form: 
\begin{equation}\label{eqn:Ek1} 
\Delta t\, E_k = -\Delta t^2 \sum_{n=0}^k \left( \mathbf{e}_k\right)^T \mathbf{S}  \mathbf{e}_n
+ \Delta t^2 \sum_{n=0}^k \left( \bm{\lambda}^{\sf S}_k\right)^T \mathbf{S} \mathbf{e}_n . 
\end{equation}
Let $\mathbf{w}_{i,n} = \mathbf{M}_i^{-1} \mathbf{G}_i^T \mathbf{e}_n$ and let $w^h_{i,n}\in S^h_{i,D}$ be the associated finite element function.  
Then we may expand out 
\begin{equation*}
\left( \mathbf{e}_k\right)^T \mathbf{S}  \mathbf{e}_n 
= \sum_{i=1,2} \left( \mathbf{w}_{i,k}\right)^T \mathbf{M}_i  \mathbf{w}_{i,n} 
= \sum_{i=1,2} \left( w^h_{i,n} ,w^h_{i,k} \right)_{0,\Omega_i} 
\end{equation*}
and bound this term using Lemma \ref{lem:S_norm}: 
\begin{equation*}
\left| \left( \mathbf{e}_k\right)^T \mathbf{S}  \mathbf{e}_n \right| 
\leq \sum_{i=1,2} \left\|  w^h_{i,n}\right\|_{0,\Omega_i} \left\| w^h_{i,k} \right\|_{0,\Omega_i} 
\leq (C_L)^2 \sum_{i=1,2} h_i^{-1} \| e^h_k \|_\gamma \|e^h_n \|_\gamma . 
\end{equation*}
We use $h\leq h_i$, the Cauchy-Schwarz inequality and $(k+1)\Delta t \leq T$ to bound 
\begin{equation}
\Delta t^2 
\left| \sum_{n=0}^k\left( \mathbf{e}_k\right)^T \mathbf{S}  \mathbf{e}_n \right| 
\leq C \Delta t^2 h^{-1} \| e^h_k \|_\gamma \sum_{n=0}^k \|e^h_n \|_\gamma 
\leq C \Delta t^2 h^{-1} \left( \sum_{n=0}^k \|e^h_n \|_\gamma \right)^2 
\leq C \frac{\Delta t}{h} \sum_{n=0}^k \|e^h_n \|_\gamma^2  .
\label{eqn:Ek2} 
\end{equation}

We turn our attention to bounding the terms $\left( \bm{\lambda}^{\sf S}_k\right)^T \mathbf{S} \mathbf{e}_n$ in \eqref{eqn:Ek1}.  
It follows from the definition of $\bm{\lambda}^{\sf S}_k$ that 
\begin{equation*} 
\left( \bm{\lambda}^{\sf S}_k\right)^T \mathbf{S} \mathbf{e}_n 
=\left( \mathbf{b}_{1,k} \right)^T  \mathbf{M}_1^{-1} \mathbf{G}_1^T \mathbf{e}_n - \left( \mathbf{b}_{2,k} \right)^T \mathbf{M}_2^{-1}  \mathbf{G}_2^T \mathbf{e}_n . 
\end{equation*}
Define $\bm{v}_{i,k} =\mathbf{M}_i^{-1} \mathbf{b}_{i,k}$ and $\mathbf{w}_{i,n} =\mathbf{M}_i^{-1} \mathbf{G}_i^T \mathbf{e}_n$ and let $v^h_{i,k},w^h_{i,n}\in S^h_{i,D}$ be the associated finite element functions.
We have 
\begin{equation} 
\left| \left( \bm{\lambda}^{\sf S}_k\right)^T \mathbf{S} \mathbf{e}_n \right| 
= \left| \left( \bm{v}_{1,k} \right)^T \mathbf{M}_1 \mathbf{w}_{1,n} - \left( \bm{v}_{2,k} \right)^T \mathbf{M}_2 \mathbf{w}_{2,n} \right|  
\leq \sum_{i=1,2} \| v^h_{i,k} \|_{0,\Omega_i} \| w^h_{i,n} \|_{0,\Omega_i}  . 
\label{eqn:Ek3} 
\end{equation}

In order to bound $\| v^h_{i,k} \|_{0,\Omega_i}$, we use \eqref{eqn:bHat} and note first that 
\[
\| v^h_{i,k} \|_{0,\Omega_i}^2 
= \left( \bm{v}_{i,k} \right)^T \mathbf{M}_i \bm{v}_{i,k} 
= \left( \bm{v}_{i,k} \right)^T \mathbf{b}_{i,k} 
= \left( \bm{v}_{i,k} \right)^T \left( \mathbf{f}_{i,k} - \mathbf{K}_i \mathbf{u}_{i,k} \right) .
\]
Next, apply Lemma \ref{lem:fBound}, continuity of the bilinear form $a(\cdot , \cdot )$ and Lemma \ref{lem:inverseEst}: 
\begin{multline*}
\| v^h_{i,k} \|_{0,\Omega_i}^2 
= \left( \bm{v}_{i,k} \right)^T \mathbf{f}_{i,k} - a(u^h_{i,k} , v^h_{i,k}) \\
\leq C_{i,k} \left\| v^h_{i,k}\right\|_{1,\Omega_i} 
+ c_1  \left\| u^h_{i,k} \right\|_{1,\Omega_i} \left\| v^h_{i,k}\right\|_{1,\Omega_i}  
\leq C_{-1} h_i^{-1}\left( C_{i,k} + c_1  \left\| u^h_{i,k} \right\|_{1,\Omega_i} \right)   \left\| v^h_{i,k}\right\|_{0,\Omega_i} .
\end{multline*}
Thus we have the bound
$$
\| v^h_{i,k} \|_{0,\Omega_i} 
\leq  C_{-1} h_i^{-1}\left( C_{i,k} + c_1  \left\| u^h_{i,k} \right\|_{1,\Omega_i} \right) . 
$$
We insert this bound in \eqref{eqn:Ek3} and estimate $\| w^h_{i,n} \|_{0,\Omega_i}$ via Lemma \ref{lem:S_norm}.  
The result is 
\begin{equation*} 
\left| \left( \bm{\lambda}^{\sf S}_k\right)^T \mathbf{S} \mathbf{e}_n \right| 
\leq C_L C_{-1} \| e^h_n \|_\gamma \sum_{i=1,2} 
 h_i^{-3/2}\left( C_{i,k} + c_1  \left\| u^h_{i,k} \right\|_{1,\Omega_i} \right)  .
\end{equation*}
Next, using $h\leq h_i$ we find that 
\begin{equation} 
\Delta t^2 \left| \sum_{n=0}^k \left( \bm{\lambda}^{\sf S}_k\right)^T \mathbf{S} \mathbf{e}_n \right| 
\leq C\Delta t^2 h^{-3/2}\sum_{i=1,2}\left( C_{i,k} + c_1  \left\| u^h_{i,k} \right\|_{1,\Omega_i} \right) \sum_{n=0}^k \| e^h_n \|_\gamma . 
\label{eqn:Ek4} 
\end{equation}
Application of Cauchy-Schwarz and Young's inequalities then yields
\begin{align*}
C\Delta t^2 h^{-3/2} C_{i,k} \sum_{n=0}^k \| e^h_n \|_\gamma &\leq 
C\Delta t^2 h^{-3/2} \left( \frac{1}{2\epsilon_3}C_{i,k}^2 + \frac{\epsilon_3}{2} \left( \sum_{n=0}^k \| e^h_n \|_\gamma \right)^2 \right) \\ 
&\leq 
C\Delta t^2 h^{-3/2} \left( \frac{1}{2\epsilon_3}C_{i,k}^2 + \frac{\epsilon_3}{2}(k+1)  \sum_{n=0}^k \| e^h_n \|_\gamma^2 \right) \\ 
&\leq 
C\frac{\Delta t^2}{\epsilon_3 h^{3/2}}  C_{i,k}^2 + C\frac{\Delta t \epsilon_3}{h^{3/2}}  \sum_{n=0}^k \| e^h_n \|_\gamma^2 .
\end{align*} 
We proceed by analogy and bound 
\begin{equation*}
C\Delta t^2 h^{-3/2} c_1 \left\| u^h_{i,k} \right\|_{1,\Omega_i} \sum_{n=0}^k \| e^h_n \|_\gamma \leq 
C\frac{\Delta t^2}{\epsilon_4 h^{3/2}}  \left\| u^h_{i,k} \right\|_{1,\Omega_i}^2 + C\frac{\Delta t \epsilon_4}{h^{3/2}}  \sum_{n=0}^k \| e^h_n \|_\gamma^2 .
\end{equation*}
We use the last two results to bound \eqref{eqn:Ek4}.  
Together with \eqref{eqn:Ek2}, we have bounded $\Delta t\, E_k$ in \eqref{eqn:Ek1}.
In turn, we may use this bound in \eqref{eqn:sp1} and then add the extra terms to the last inequality in the proof of Theorem \ref{thm:stab_IVRC}.  
The result is 
\begin{multline*}
\frac{1}{2}\sum_{i=1,2}  \left\| u^h_{i,k+1} \right\|_{0,\Omega_i}^2
+ \frac{c_0}{2}\Delta t \sum_{i=1,2} \left\| u^h_{i,k+1} \right\|_{1,\Omega_i}^2 \\ 
\leq \frac{1}{2}\sum_{i=1,2} \left\| u^h_{i,k} \right\|_{0,\Omega_i}^2
+ C\left( \frac{1}{c_0}+\frac{\Delta t}{\epsilon_3 h^{3/2}}\right) \Delta t \sum_{i=1,2} C_{i,k}^2 \\ 
+ C\frac{\Delta t^2}{\epsilon_4 h^{3/2}}  \sum_{i=1,2} \left\| u^h_{i,k} \right\|_{1,\Omega_i}^2 
+ C\left( \frac{1}{h} +\frac{(\epsilon_3 +\epsilon_4 )}{h^{3/2}}\right)  \Delta t \sum_{n=0}^k \| e^h_n \|_\gamma^2  \,.
\end{multline*} 
There is already a time step restriction of the form $\Delta t\leq C^* h^2$ to bound the diffusion terms.  
By comparison, the restriction to bound the additional terms here due to the DMD flux approximation turns out to be less stringent. 
Given $0<h<1$, it holds that $\Delta t \leq C^* h^{3/2+p}$, where $0\leq p \leq 1/2$.  
Then choose $\epsilon_3 = h^p$ and $\epsilon_4 = 4C^* Ch^p / c_0$, and the stability bound now reduces via these scalings to the form 
\begin{multline*}
\frac{1}{2}\sum_{i=1,2}  \left\| u^h_{i,k+1} \right\|_{0,\Omega_i}^2
+ \frac{c_0}{2}\Delta t \sum_{i=1,2} \left\| u^h_{i,k+1} \right\|_{1,\Omega_i}^2 \\ 
\leq \frac{1}{2}\sum_{i=1,2} \left\| u^h_{i,k} \right\|_{0,\Omega_i}^2
+ C \Delta t \sum_{i=1,2} C_{i,k}^2 \\ 
+ \frac{c_0}{4} \Delta t \sum_{i=1,2} \left\| u^h_{i,k} \right\|_{1,\Omega_i}^2 
+ C\frac{\Delta t}{h^{3/2-p}} \sum_{n=0}^k \| e^h_n \|_\gamma^2  
\end{multline*} 
The proof may be completed by summing over $k$, using $p=1/2$ and applying the scaling assumption \eqref{eqn:DMD_acc}.  
However, we see that the time step restriction to bound the extra terms due to the DMD approximation only requires $p=0$, so for small $h$ the stability restriction is still dominated by the stiffness of the diffusion.  
\end{proof}

%% file: Training.tex
%
\REV{Success of the data-driven partitioned approach is contingent upon the ability to train \emph{predictive} dynamic flux surrogates, i.e., operators $\mathbf{A}_\lambda$ that can provide accurate approximations of the interface flux for general initial conditions not seen during the training process. 
Construction of such surrogates requires selection of training data that allows one to characterize the dynamics of the interface flux as accurately and completely as possible. This section describes the procedure used in the paper to generate such training data for the model transmission problem (TP).} 

Section \ref{sec:config} states the discretization setting for this problem and Section \ref{sec:train-nosource} considers the case when this problem is augmented with homogenous Dirichlet boundary conditions and has no source term. Then, in Section \ref{sec:train-source}, we consider generation of training data for general Dirichlet conditions and source terms. The construction of the interface patches is discussed in Section \ref{sec:patch-sizes}. We recall that in all cases the training data is obtained by using IVR(C) to solve a \REV{hybrid monolithic discretization of a} properly configured \REV{model TP. This approach ensures that the snapshots of the interface flux are second-order accurate.}

\subsection{\REV{Discretization setting}}\label{sec:config}
\begin{figure}[htbp] 
   \centering
   \includegraphics[width=0.45\linewidth]{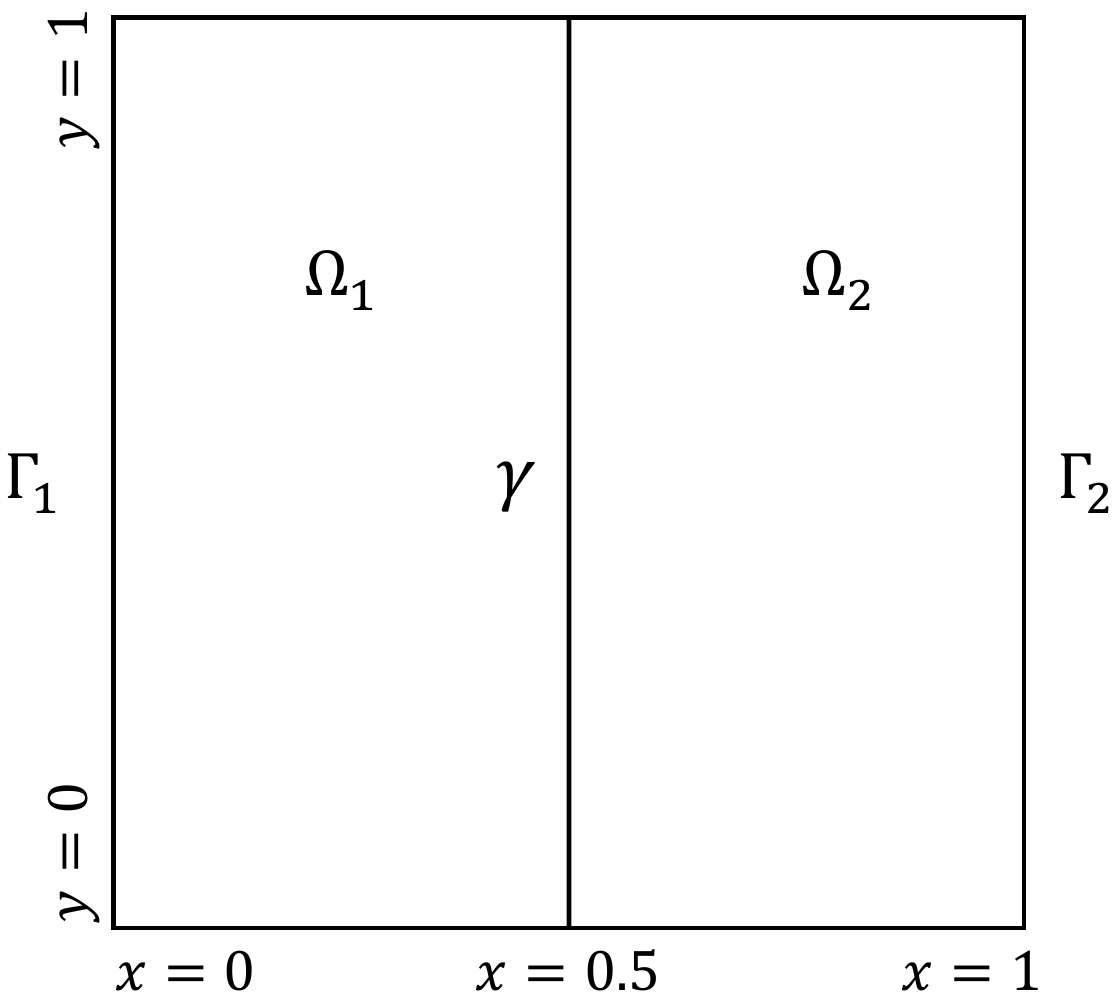} \hfill
   \includegraphics[width=0.45\linewidth]{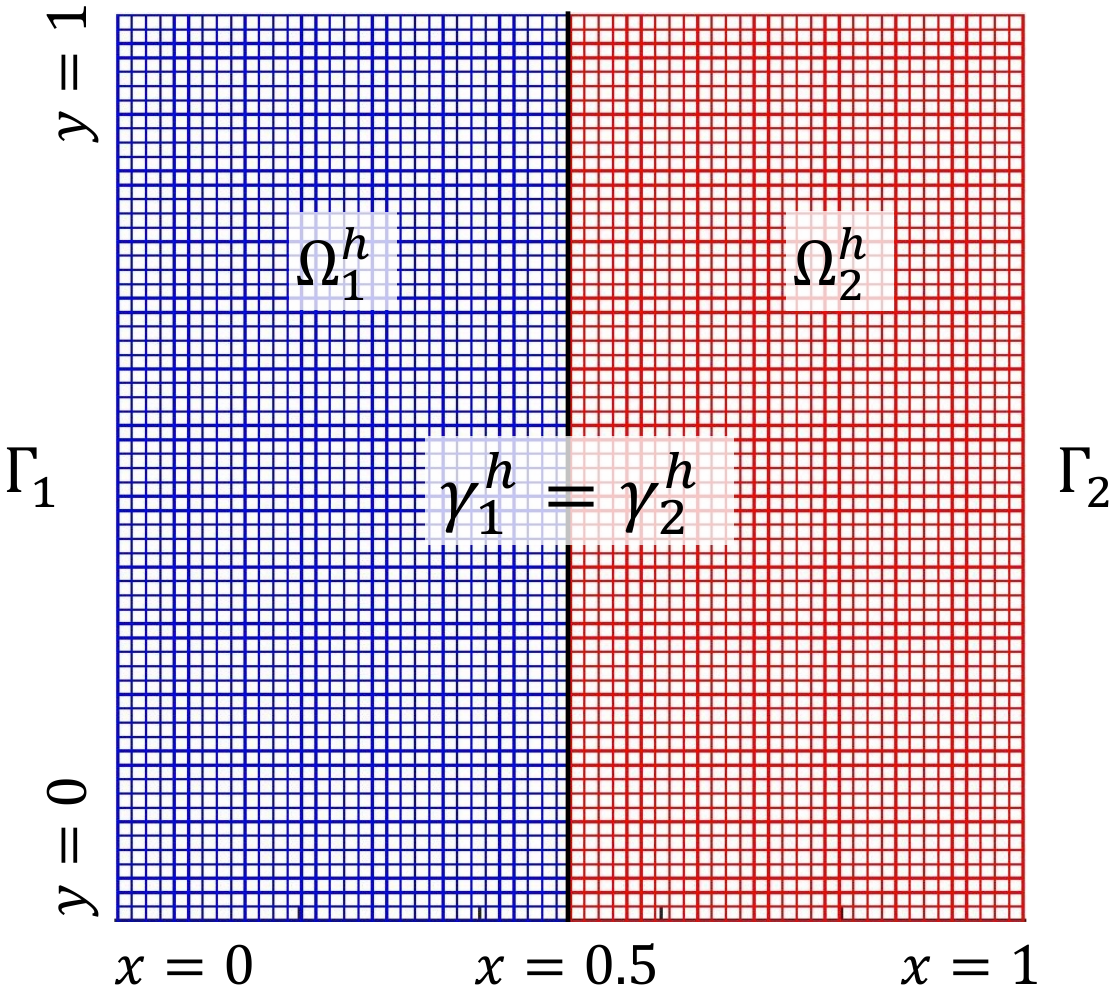} \vspace{2ex}
   \caption{\REV{Computational domain and its discretization} for the generation of the training data.
   Left: Computational domain $\Omega$ and its partition into subdomains $\Omega_1$ and $\Omega_2$. Right: finite element partition $\Omega^h$ and the induced subdomain and interface grids $\Omega^h_i$ and $\gamma^h_i$; $i=1,2$, respectively, for $n=64$.}
   \label{fig:domain}
\end{figure}
For simplicity we consider a computational domain $\Omega$ defined as the unit square $[0,1]\times[0,1]$ with $\Omega_1 = [0,0.5]\times[0,1]$ and $\Omega_2 = [0.5,1]\times[0,1]$. Thus, $\gamma$ is the line $x=0.5$; see Figure \ref{fig:domain} (left). However, the  procedure described here is applicable to general domain and interface configurations. 
Recall that the model transmission problem is parameterized by the diffusion coefficients, i.e.,  $\bm{\mu}=\{\kappa_1,\kappa_2\}$, while  the source term, the boundary conditions and the advective term are assumed fixed. Here we shall define the latter using the rotating velocity field $\bm{v} = \left(0.5-y, x-0.5\right)$. We set the simulation time interval to $[0,2\pi]$, i.e., $T=2\pi$. Thus, given an initial condition $u_{i,0}(\bm{x})$, $i=1,2$, the solution of   \eqref{eq:TP} represents one full rotation with diffusion of this initial condition. 
To discretize \eqref{eq:TP} in space we consider a finite element partition $\Omega^h$ comprising 
$n\times n$ uniform quadrilateral elements with mesh size $h=1/n$. Restriction of $\Omega^h$ to $\Omega_i$ and $\gamma$ induces subdomain meshes $\Omega^h_i$ and interface partitions $\gamma^h_1$ and $\gamma^h_2$ with matching grid nodes; see Figure \ref{fig:domain} (right). 

To generate the training data we solve \eqref{eq:TP}, augmented with appropriate initial and boundary conditions, on this mesh using the IVR(C) scheme with a uniform time step $\Delta t$. This time step is selected to satisfy the  Courant–Friedrichs–Lewy condition on $\Omega^h$, required for the stability of the explicit Euler scheme \eqref{eq:FE} employed by IVR(C). Thus, the training data comprises time series for the subdomain solutions and the interface flux, each of length $S=2\pi/\Delta t$. Since $\Delta t$ varies with $h$, the length of these time series also varies with the mesh size.


\subsection{Training data generation without source terms}\label{sec:train-nosource}
We first consider the case when the model transmission problem is augmented with homogeneous boundary conditions and has no source terms. The ``combination'' test case in Section \ref{sec:num} is one example of this configuration. 
We seek to define a training set for this type of problems that will enable the inference of DMD flux surrogates \REV{that can provide accurate flux approximations for \emph{arbitrary} initial conditions in  \eqref{eq:TP}.}

To define \REV{such a} training set we shall invoke an analogy with the identification of linear time invariant systems (LTIs), specifically the fact that an LTI system is completely characterized by its impulse response. 
Applying this analogy to the interface flux suggests that we can characterize its dynamics by collecting data about the spatial ``impulse response'' of \eqref{eq:ODE-subd} along the interface. Thus, we shall generate the training data by computing the solution of \eqref{eq:ODE-subd} for a set of initial conditions $\{{u}^j_{i,0}\}_{j=1}^{P}$, comprising Gaussian hills with standard deviation $\sigma$ and centers $(x_0,y_0)$, i.e., functions having the following general form:
\begin{equation}
    \psi(x, y;x_0, y_0) = e^{-\frac{(x-x_0)^2+(y-y_0)^2}{2\sigma^2}}.
\end{equation}
\begin{figure}[!t]
\begin{center}
	\subfigure[A family of Gausian hills along the segment $\mathcal{C}$.]{\includegraphics[width=0.5\linewidth]{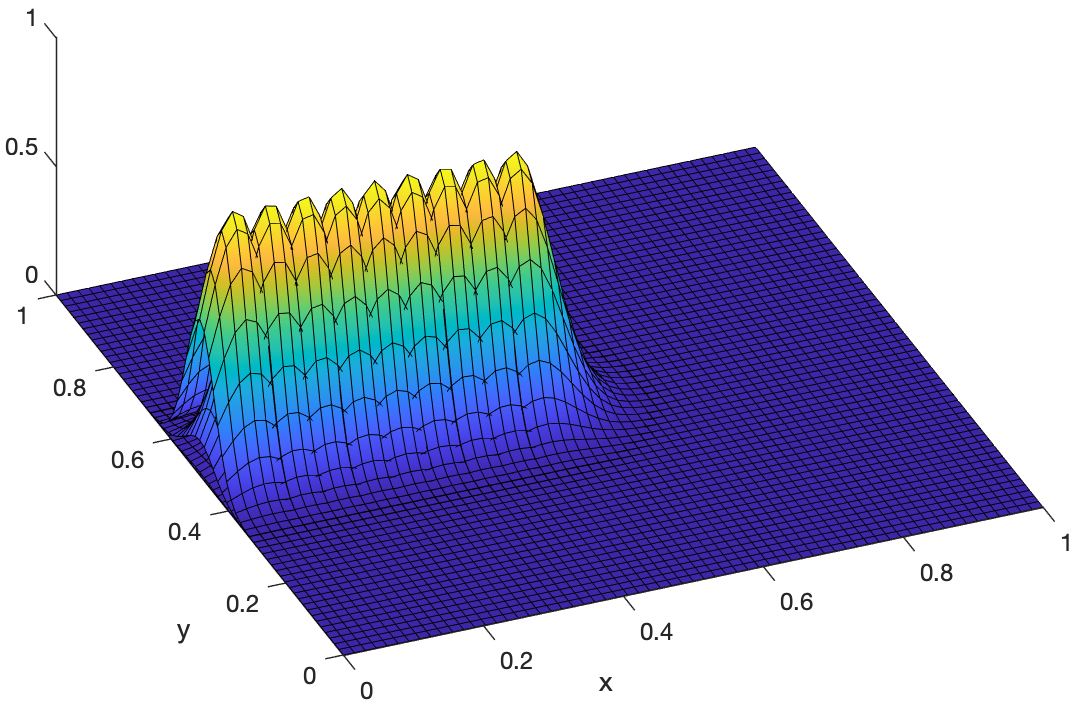}\label{fig:gauss-hill}} \hspace{-3ex}
	\subfigure[A single Gausian hill along the segment $\mathcal{C}$.]{\includegraphics[width=0.5\linewidth]{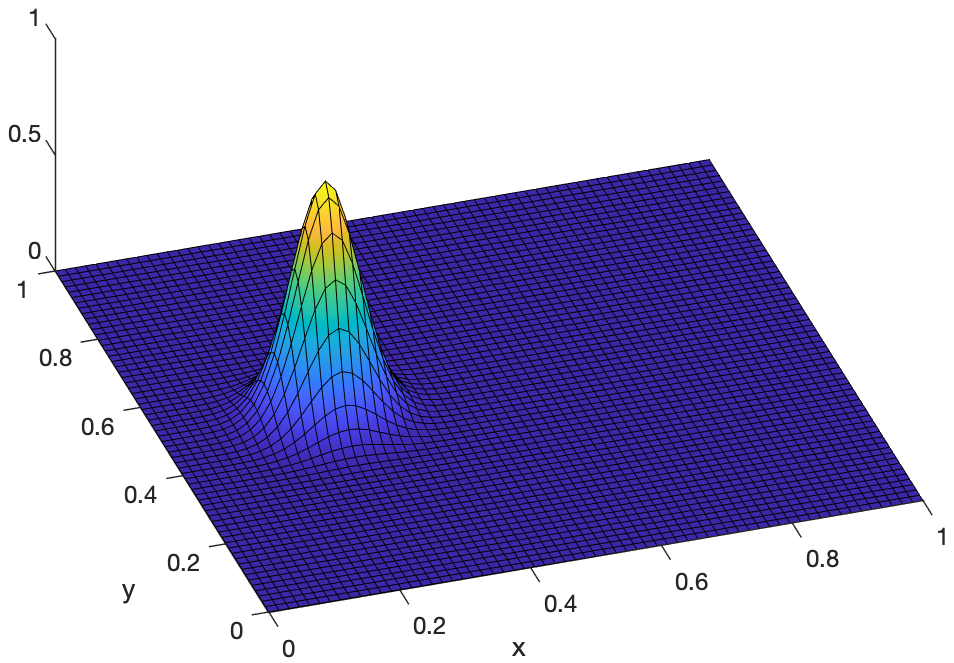}\label{fig:single-gauss-hill}}
	\caption{Typical examples of Gaussian hill initial conditions used to generate the training data for the DMD flux surrogate.}
\end{center}
\end{figure}
To define the \REV{set of initial conditions} we place the centers of the Gaussian hills at uniform distances from each other along the segment
$\mathcal{C}=\{(x,y)\in\Omega_1\,|\, 0\le x\le 0.5\ \mbox{and}\ y=0.5\}$; \REV{see Figures \ref{fig:gauss-hill}-\ref{fig:single-gauss-hill}.} The spacing between these centers is chosen to be of the same order as the mesh size $h$ of the finite element partition $\Omega^h$. Likewise, we set the standard deviation for each Gaussian to be $O(h)$. 
As a result, these Gaussians can be thought of  as smooth approximations of the Dirac delta function on the finite element mesh, further highlighting the parallels with an impulse response of a system.

The training data is then generated by using IVR(C) to solve a \REV{hybrid monolithic discretization of} \eqref{eq:TP} for each initial condition in the set $\{{u}^j_{i,0}\}_{j=1}^{P}$. The exact solutions corresponding to these initial conditions are  circular paths of diffusing Gaussians about the domain center $(x,y)=(0.5, 0.5)$. With this choice of Gaussian sizes and initial spacings, the interface $\gamma$ defined by $x=0.5$ experiences crossings everywhere when the entire set of initial conditions is used. In so doing, we obtain a training set that contains detailed information about \REV{the dynamical response of} the interface flux along the entire interface.

\subsection{Training data generation with source terms}\label{sec:train-source}
In Section \ref{sec:train-nosource} we described generation of training data for homogeneous Dirichlet conditions and no source terms. 
Here we briefly discuss the general case when the model problem is augmented with inhomogeneous boundary conditions and/or source terms. Besides being of a practical importance, such a configuration arises when testing a numerical scheme for  \eqref{eq:TP} using the method of manufactured solutions. In  such cases it is common for a manufactured source term and boundary conditions to appear. We will need such training data to perform the ``patch'' test in Section \ref{sec:num}. 

It is important to keep in mind that in this paper we restrict attention to a $\mu$PDE version of \eqref{eq:TP} parameterized solely by the diffusion coefficients. Thus, we do not consider a case where the boundary conditions and/or the source terms are also a part of the problem parameterization. Such settings are beyond the main scope of this paper, which is to demonstrate a proof-of-principle for a dynamic flux surrogate-based partitioned scheme.

The introduction of a source term changes the dynamics of the original system and thus destroys the applicability of the DMD operators created using homogeneous Dirichlet boundary conditions and no source terms in Section \ref{sec:train-nosource}. In order to incorporate the effects of source term and boundary conditions into the DMD surrogate model, we must generate the Gaussian initial condition data as in Section \ref{sec:train-nosource}, but with the appropriate boundary and source terms.  Specifically, given a source term $f_i$ and Dirichlet data $g_i$, $i=1,2$ we use IVR(C) to solve  \eqref{eq:TP} for every initial condition in the set $\{{u}^j_{i,0}\}_{j=1}^{P}$.

\subsection{Construction of the interface patches}\label{sec:patch-sizes}
In order to enable information exchange between the subdomains, the DMD state $\mathbf{y}_k$ must include information from the subdomain solutions. Recall that if we define this state as in \eqref{sec:single-DMD}, i.e., by including the entire subdomain solution coefficient vectors, the computational cost of the DMD surrogate is comparable to that of the IVR(C) scheme and does not meet our efficiency goal. Essential to achieving this goal is the utilization of interface patches $\mathbf{u}_{i,k}(\delta_i)\subset \mathbf{u}_{i,k}$ instead of the entire coefficient vectors. 

Implementation of the interface patch definition \eqref{eq:patch-def} on non-uniform grids can be performed by using $k$-rings or a $k-d$ tree search, which is generally applicable to DoFs that have a correlation to physical space, such as those in the Lagrangian $C^0$ finite element spaces \REV{used here}. 
For the unit square domain and the uniform quadrilateral grids considered in this paper, implementation of \eqref{eq:patch-def} is fairly straightforward and amounts to selecting all DoFs located on vertical grid lines within the prescribed distance threshold from the interface; see Figure \ref{fig:DMD-state}. 

In this case it is also convenient to measure the patch sizes by the number of grid lines included in their definition. 
Thus, in what follows, we shall say that $\mathbf{u}_{i,k}(\delta_i)$ is an interface patch of size $R$ if it contains all solution coefficients located on the interface mesh $\gamma^h_i$ and the $R-1$ adjacent grid lines. For example, the patches shown in Figure \ref{fig:DMD-state} both have size 2.

%

%% file: Numerics.tex
In this section we demonstrate numerically the performance of the \REV{DMD-FS and $\mu$DMD-FS} partitioned schemes formulated  in Section \ref{sec:DMD}. To that end we compare and contrast the accuracy and efficiency of these schemes with that of the IVR(C) and IVR(L) methods using two different solutions and two distinct parameter settings for the diffusion coefficient in the model transmission problem \eqref{eq:TP}. \REV{As a reference, we use a non-hybrid monolithic discretization of this problem.}  In all cases the advective term in \eqref{eq:TP} is defined by the rotating velocity field $\bm{v} = \left(0.5-y, x-0.5\right)$.

\begin{figure}[!t]
\centering
\subfigure[Multi-material]{\includegraphics[width=0.45\linewidth]{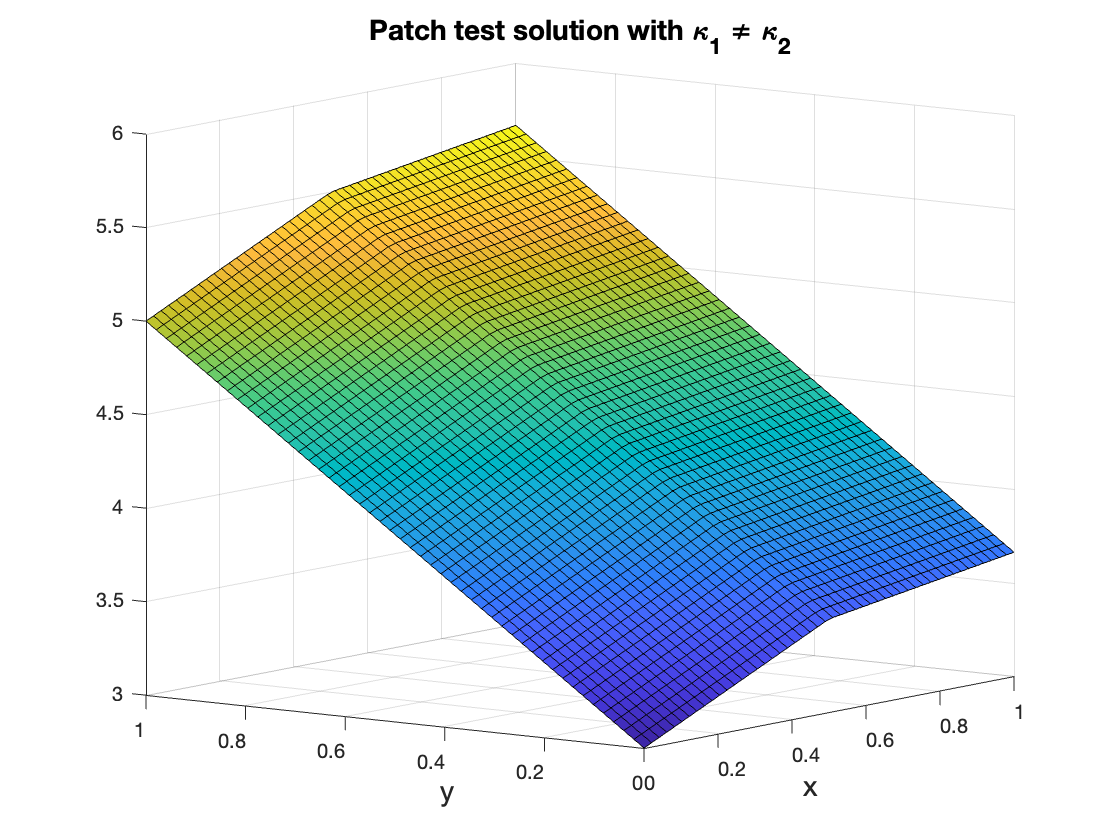}
\label{fig:patch_multimat}} 
\subfigure[Single material]{\includegraphics[width=0.45\linewidth]{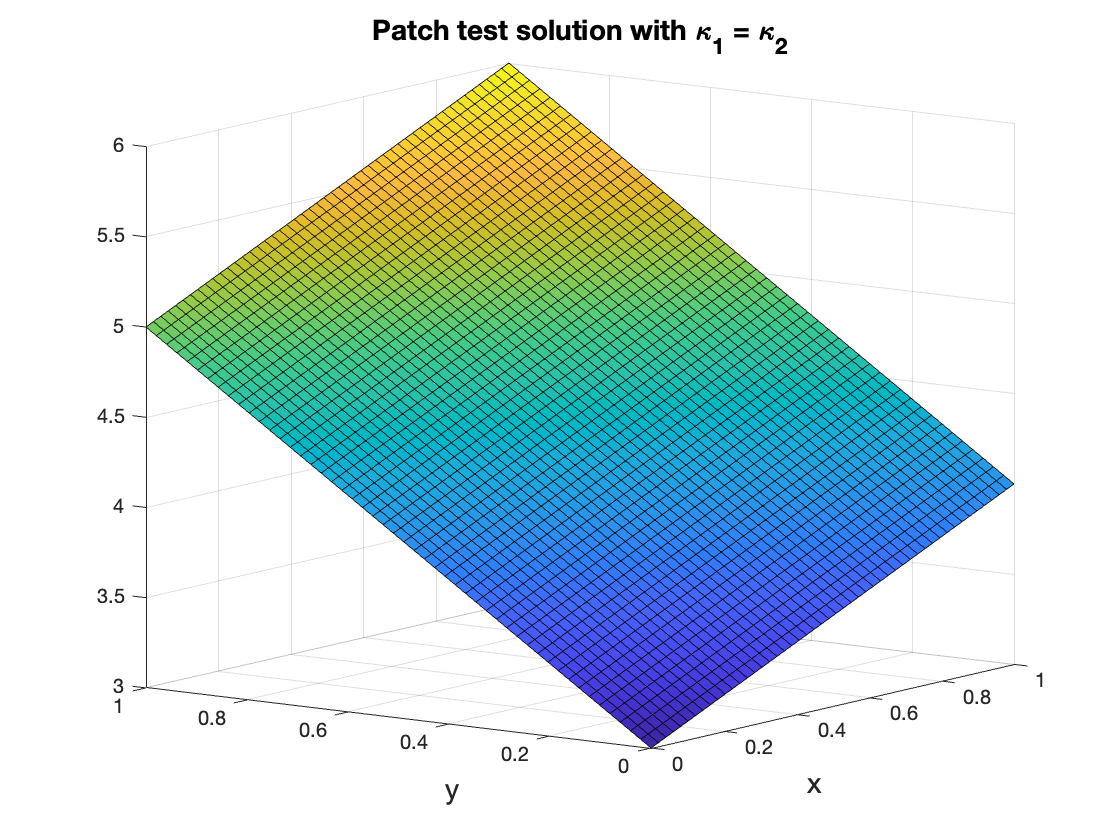}
\label{fig:patch_singlemat}} 
\caption{A multi-material patch test solution with $\kappa_1 = 1\times 10^{-3}$, $\kappa_2 = 3\times 10^{-3}$ and single material patch test solution with $\kappa_1 = \kappa_2 = 1\times 10^{-3}$. Both solutions shown at $t=1$.}
\end{figure}

\paragraph{\REV{Patch test}} 
The first solution is given by the linear in time and piecewise linear in space function
\begin{equation}
u(x,y) = \left\{ 
\begin{array}{ll}
\displaystyle
t(x + 2y + 3)  & \mbox{if $(x,y)\in\bar{\Omega}_1$} \\[2ex]
\displaystyle
t\left(\frac{\kappa_1}{\kappa_2}x + 2y + \frac{\kappa_2-\kappa_1}{2\kappa_2} +3\right)    & \mbox{if $(x,y)\in      {\Omega}_2$}
\end{array}
\right. \,.
\label{patch_equation}
\end{equation}
\REV{The coefficients of \eqref{patch_equation}} on $\Omega_2$ are defined so that the manufactured solution satisfies the coupling conditions in \eqref{interface_conditions} for any combination of positive diffusion \REV{parameters} $\kappa_i$, $i=1,2$. 
Note that for $\kappa_1\neq\kappa_2$ this solution has a ``kink'' along the interface that is necessary to match the fluxes on both sides of $\gamma$; see Figure \ref{fig:patch_multimat}.
We define source terms and boundary data matching the manufactured solution  by inserting \eqref{patch_equation} into the governing equations and the boundary condition of the model problem \eqref{eq:TP}, respectively. As mentioned in Section \ref{sec:train-source}, these manufactured source terms and boundary conditions are used to generate training data that represents the dynamics of the interface flux produced by  \eqref{patch_equation}. 

\REV{Because on meshes with matching interface nodes both hybrid and non-hybrid monolithic discretizations of \eqref{eq:TP} recover  \eqref{patch_equation} to machine precision, the patch test will serve as a measure for the ``consistency'' gap of the IVR(L), DMD-FS and $\mu$DMD-FS schemes. Since IVR(C) is a consistent solution method for the hybrid discretization of the model problem, we expect its solution to recover \eqref{patch_equation} within the roundoff error.}

%

\begin{figure}[t!]
\centering
\includegraphics[width=0.65\linewidth]{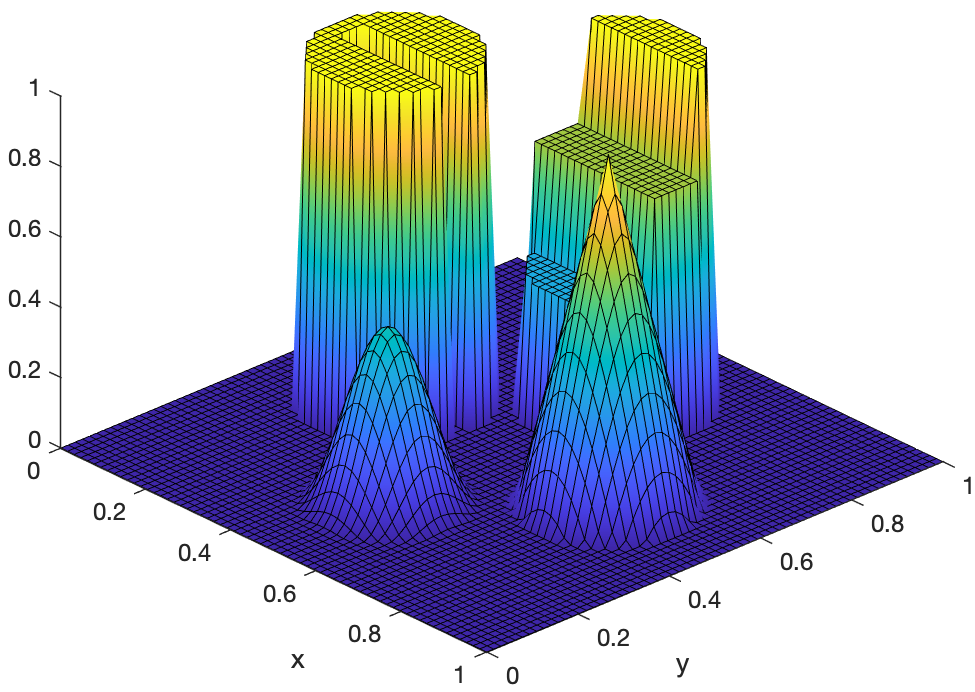}
\caption{Initial condition for combination test.}
\label{combo_initial}
\end{figure}

\paragraph{\REV{Combination test}} 
The second solution is defined by homogeneous Dirichlet boundary conditions, homogeneous source terms and an initial condition considered in \cite{Bochev_20_CMAME}. This initial condition comprises a Gaussian hill, cone, and slotted cylinder from \cite{Leveque_96_SINUM} augmented by a ``staircased" cylinder; see  Figure \ref{combo_initial}.
We refer to this test as the ``combination test.'' The combination test is designed to examine how well a scheme can handle initial condition sets with varying degrees of smoothness.

\paragraph{\REV{Material configurations}} 
We pair both of these solutions with two different combinations of the diffusion coefficients in \eqref{eq:TP}. The first one is characterized by diffusion coefficients that are discontinuous along the interface, i.e., $\kappa_1\neq \kappa_2$. We refer to this case as the ``multi-material'' configuration of the model problem. The second combination uses the same diffusion coefficient in both subdomains, i.e., $\kappa_1=\kappa_2$, and is referred to as the ``single material'' configuration of \eqref{eq:TP}.

We will use the multi-material configuration of \eqref{eq:TP} to exercise the parametric  $\mu$DMD-FS scheme defined in Algorithm \ref{parametric_fsdmd_algorithm}. \REV{In particular, we examine the accuracy, efficiency and robustness of $\mu$DMD-FS  by using different combinations of target parameters $\bm{\mu}$ and sampling sets $\mathcal{M}_m(\bm{\mu},r)$ in Algorithm \ref{parametric_fsdmd_algorithm}; see Figures \ref{fig:param-DMD-1}-\ref{fig:param-DMD-3}.}
The single material case will be used to compare and contrast the convergence of \REV{the data-driven partitioned} scheme with that of the IVR(C) and IVR(L) methods. To that end we shall use the fixed parameter scheme DMD-FS given in  Algorithm \ref{fsdmd_algorithm}. 

\paragraph{\REV{Solution errors}}
In all numerical studies we report the solution error of the IVR(C), IVR(L) and DMD-FS schemes relative to a reference  solution computed by a \REV{non-hybrid} monolithic discretization of \eqref{eq:TP} on the same mesh $\Omega^h$ that induces the subdomain and interface meshes for these schemes. The errors are measured at the final time $T=2\pi$ after the initial condition has completed one full revolution. 

Specifically, let $u^h_{M}$ denote the \REV{non-hybrid} monolithic solution of  \eqref{eq:TP} on $\Omega_h$ at $T=2\pi$ and let $u^h_{i,M}$, $i=1,2$ be the restrictions of this solution to $\Omega^h_i$. Let $u^h_{i,C}$, $u^h_{i,L}$, and $u^h_{i,D}$ denote the solutions of the IVR(C), IVR(L) and DMD-FS schemes on the same mesh at the same time. For $X\in \{C,L,D\}$ we define the relative $L^2$ and $H^1$ errors of the solution $u^h_{i,X}$, $i=1,2$ as
\begin{equation}
\mathcal{E}^{r}_{X} = \frac{1}{2} \sum_{i=1}^2 \frac{\| u^h_{i,X} - u^h_{i,M}\|_{r,\Omega_i}}{\|u^h_{i,M}\|_{r,\Omega_i}} \,,
r=0,1\,.
\end{equation}
Since the combination test problem does not have a closed form solution, this approach ensures consistency in reporting the errors.

\paragraph{\REV{Flux surrogate configuration}}
\REV{The flux surrogates for all examples in this section are constructed by following steps 1--5 in the offline phase of Algorithm \ref{fsdmd_algorithm}. This phase requires selection of a distance threshold $\delta>0$ for the interface patches and a snapshot energy tolerance $\epsilon$ that determines the rank of the DMD operator. 
In all cases we set $\delta_i =  5/2h$ in \eqref{eq:patch-def}, where $h$ is the size of the underlying finite element mesh. With this choice the interface patch size remains the same on all grids and equals two. 
We select $\epsilon$ to yield the ``best possible'' DMD operator by choosing the smallest possible value of $\epsilon$ for which the modes retained in Algorithm  \ref{dmd_algorithm} remain linearly independent in floating point arithmetic. This ``optimal'' value of $\epsilon$ may change from case to case and can be different on different meshes.}

\paragraph{\REV{Numerical stability}}
\REV{Analysis in Section \ref{sec:stability} establishes \eqref{eqn:DMD_acc} as a sufficient condition for the stability of the flux surrogate-based schemes. Moreover, the proof of Theorem \ref{thm:stab_DMD} reveals that this CFL-based condition can be further relaxed. 
Thus, assuming that \eqref{eqn:DMD_acc} holds, the use of the DMD flux surrogate should not lead to a more stringent restriction on the time step than already established by the CFL condition.
To test this conclusion, in all examples below we advance the solution of the DMD schemes in time using the same CFL-restricted time step that was used in the IVR(C) scheme to generate the training data. For all combinations of material configurations and test solutions we observed stable and accurate numerical solutions, thereby confirming that the surrogate based scheme does not impose additional time step restrictions as long as the DMD operator is sufficiently accurate.} 

\subsection{Multi-material configuration tests}\label{sec:multimat}
%
For both multi-material tests we use a uniform $n\times n$ grid with $n=64$ and mesh size $h=1/64$. The time step $\Delta t = 3.37\times 10^{-3}$ is selected to satisfy the  Courant–Friedrichs–Lewy (CFL) condition on this mesh, necessary for the stability of the explicit Euler scheme \eqref{eq:FE}. This time step results in training data comprising time series of length $S=1866$ for the subdomain solutions and the interface flux, respectively.  

%
\begin{figure}[t!]
  \begin{center}
    \subfigure[Single $\bm{\mu}$]{\includegraphics[width=0.3\textwidth]{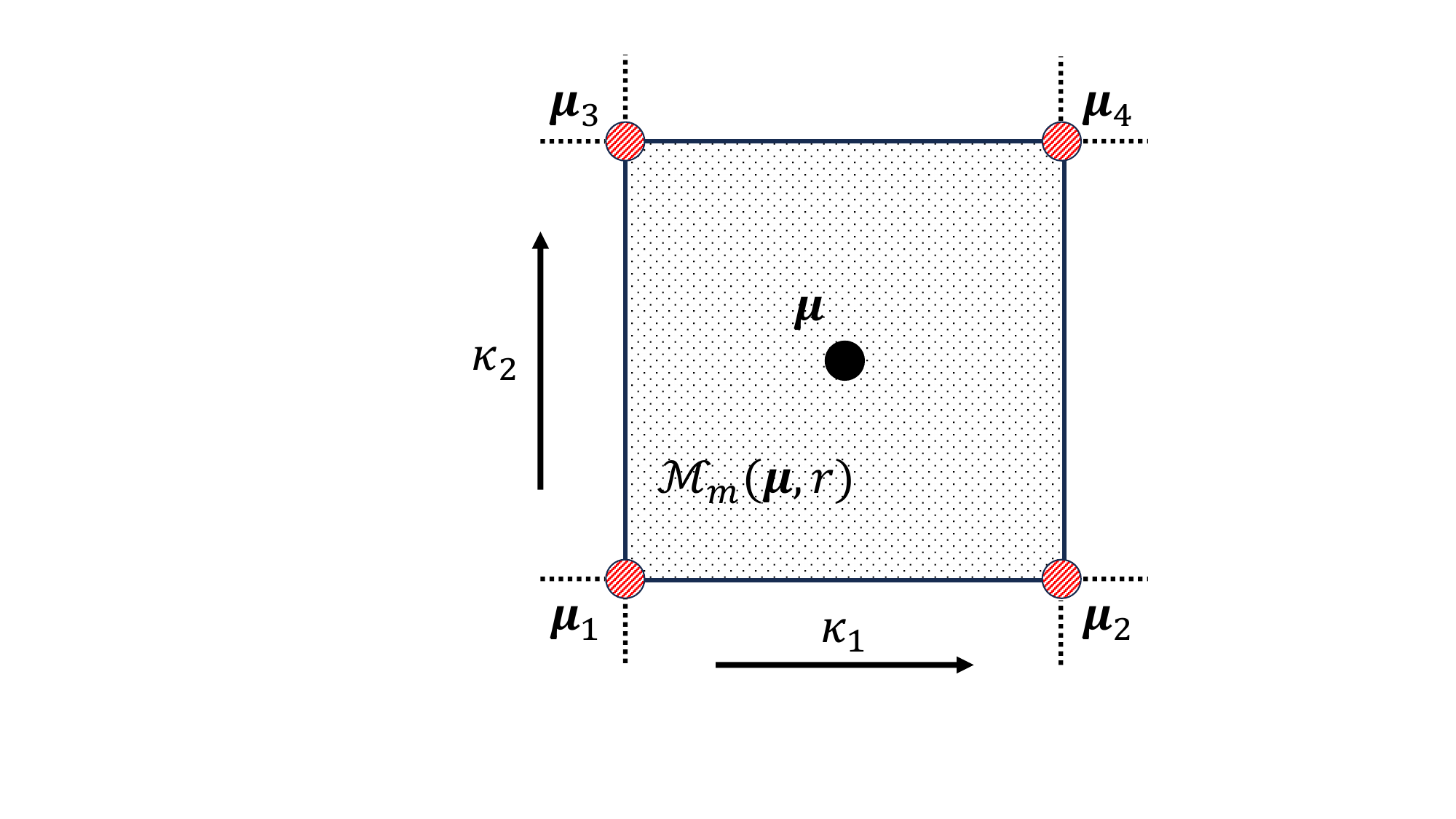}
    \label{fig:param-DMD-1}} 
    \subfigure[Random $\bm{\mu}$]{\includegraphics[width=0.3\textwidth]{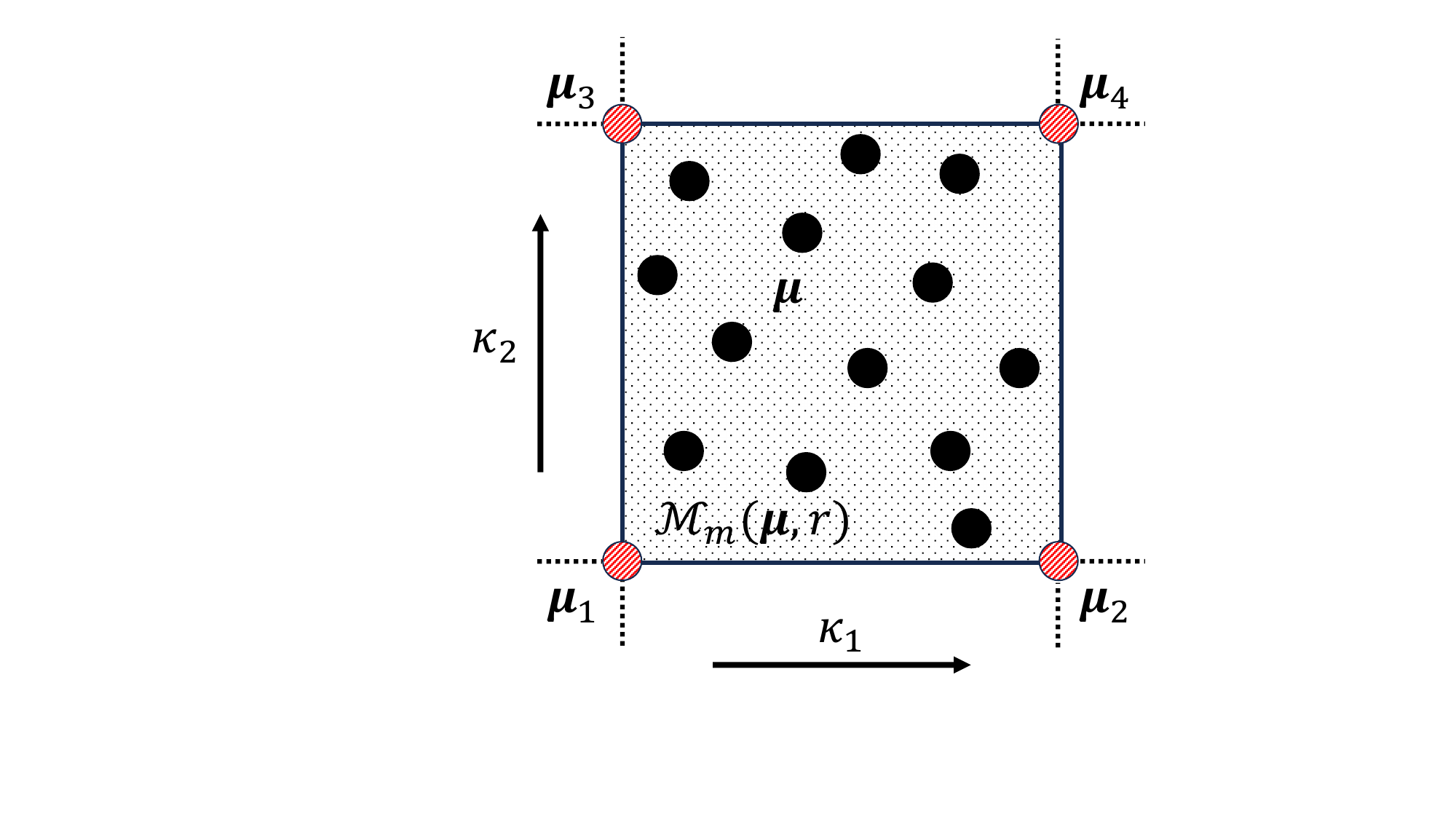}
    \label{fig:param-DMD-2}} 
    \subfigure[Nested $\mathcal{M}_m$]{\includegraphics[width=0.3\textwidth]{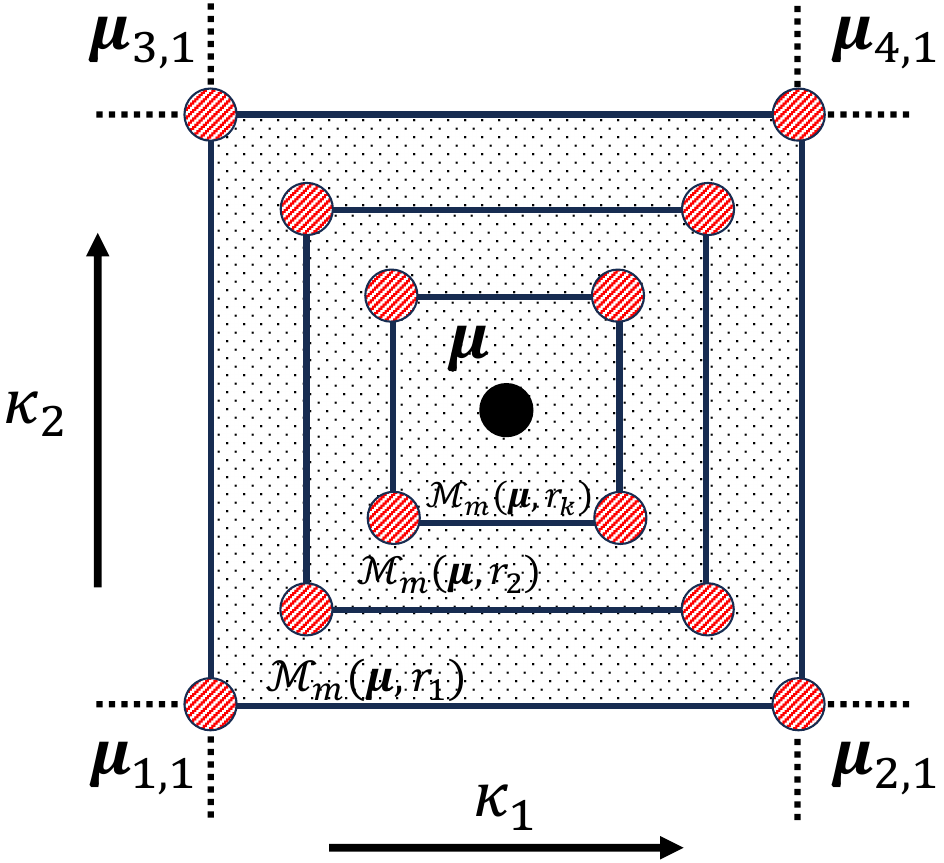}
    \label{fig:param-DMD-3}} 
  \end{center}
\vspace{-3ex}
\caption{\REV{Combinations of parameter values of interest (solid black circles) and sampling sets $\mathcal{M}_m(\bm{\mu},r)$ (hashed red circles) used to study performance of the $\mu$DMD-FS scheme in the multi-material configuration.}}
\end{figure}
\REV{To study the performance of the parametric $\mu$DMD-FS scheme, we consider three different combinations of parameter values $\bm{\mu}$ and sampling sets $\mathcal{M}_m(\bm{\mu},r)$; see Figures \ref{fig:param-DMD-1}-\ref{fig:param-DMD-3}. 
In all three cases we set $m=4$ and $\mathcal{M}_4(\bm{\mu},r) = \{\bm{\mu}_j\}_{j=1}^4$, where $\bm{\mu}_j\in\mathcal{M}$ are the vertices of a $\sigma\times\sigma$ square $\mathcal{S}\subset\mathcal{M}$ containing the parameter(s) of interest. Then we define  the rKOI operator  \eqref{eq:rKOI} as the bilinear interpolant of the vertex values of $\mathcal{S}$, i.e., the operators $\mathbf{A}_\lambda(\bm{\mu}_j)$, $\bm{\mu}_j\in\mathcal{M}_4(\bm{\mu},r)$; see Remark \ref{rem:basis}. This setting corresponds to a sampling of the parameter domain \eqref{eq:parspace} on a uniform Cartesian grid with grid size $\sigma$ and choosing $r=\sigma/\sqrt{2}$ in $B(\bm{\mu}, r)$.}

\REV{In the first parameter/sampling set combination we consider a single parameter of interest $\bm{\mu}$ and assume that $\mathcal{S}$ is centered at $\bm{\mu}$, see  Fig.~\ref{fig:param-DMD-1}. We will use this combination in both the patch test and the combination test to examine the accuracy and the speedup afforded by the $\mu$DMD-FS scheme. 
In the second combination, shown in Fig. \ref{fig:param-DMD-2}, we fix  $\mathcal{S}$ and consider a random sample of parameters of interest $\{\bm{\mu}\}\subset\mathcal{S}$ drawn from a uniform distribution on $\mathcal{S}$.  We use this combination to study the robustness of rKOI operator  \eqref{eq:rKOI} with respect to the position of $\bm{\mu}$ inside the box $\mathcal{S}$.
In the final combination, we fix the parameter of interest $\bm{\mu}$ and take a sequence of nested squares $\mathcal{S}_{i}$ with decreasing sizes $\sigma_i$, all centered at $\bm{\mu}$; see Fig. \ref{fig:param-DMD-3}. This combination will be used to examine the convergence \eqref{eq:rKOI} as a function of the parameter space sampling density. For brevity, we present results for the last two combinations only for the patch test.}

\paragraph{Patch test}
For this test we \REV{take $\bm{\mu} = (1.5,2.5)\times 10^{-3}$ as a parameter of interest and define $\mathcal{S} = ([1,2]\times [2,3])\times 10^{-3}$. Thus, } 
\begin{equation}\label{eq:sample-set-patch}
\mathcal{M}_4(\bm{\mu},r) =\left \{(1, 2), (1, 3), (2, 2), (2, 3)\right \}\times 10^{-3} \,.
\end{equation}
Training data is generated \REV{using the procedure described in Section \ref{sec:trainDMD} for every $\bm{\mu}_j = (\kappa_{1,j},\kappa_{2,j}) \in \mathcal{M}_4$.} The manufactured source term and boundary data for a given $\bm{\mu}_j$  are defined by inserting this parameter into the definition of the exact solution \eqref{patch_equation}. 
\REV{The ``optimal'' tolerance in the relative snapshot energy condition \eqref{eq:POD-energy} for this test is  $\epsilon=10^{-13}$. We use this value to construct the DMD flux surrogates $\mathbf{A}_\lambda(\bm{\mu}_j)$, $\bm{\mu}_j \in \mathcal{M}_4$.}
\begin{table}[!ht]
\renewcommand{\arraystretch}{1.15}
\centering
\textsf{
\begin{tabular}{p{0.15\linewidth}  p{0.15\textwidth}p{0.15\textwidth}p{0.15\textwidth}}
 \multicolumn{4}{c}{Multi-material Patch Test} \\
 \hline
  & $\mathcal{E}^{0}_{X}$ & $\mathcal{E}^{1}_{X}$ & Speedup \\
 \hline
 IVR(C) &  1.18E-14   & 2.03E-12  &  N/A  \\
 IVR(L) &  1.69E-4  & 1.94E-3 &  $\times 3.41$ \\ 
 $\mu$DMD-FS & \REV{1.04E-5} & \REV{8.71E-4} & $\times 16.66$ \\ 
 \hline
\end{tabular}
}
\caption{Relative $L^2$ and $H^1$ errors of the multi-material patch test solutions computed by the IVR(C), IVR(L) and $\mu$DMD-FS schemes at the final time $T=2\pi$, along with the speedup of  IVR(L) and $\mu$DMD-FS relative to IVR(C).}
\label{tab:multimat_patch}
\end{table}

\REV{To test $\mu$DMD-FS we solve \eqref{eq:TP} with the parameter of interest $\bm{\mu}$ using Algorithm \ref{parametric_fsdmd_algorithm}.} Table \ref{tab:multimat_patch} reports the relative solution errors and the speedups of the IVR(L) and $\mu$DMD-FS schemes relative to IVR(C).  

\begin{figure}[!h]
\begin{center}
\subfigure[\REV{Non-hybrid} monolithic]{\includegraphics[width=0.4\linewidth]{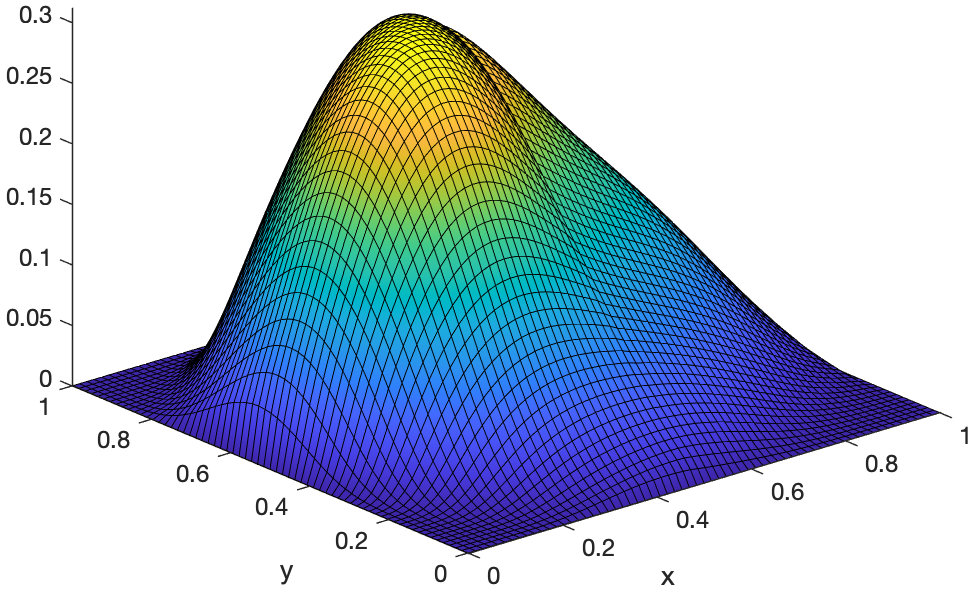}
\label{fig:monolithic_multimat}} 
\subfigure[IVR(C)]{\includegraphics[width=0.4\linewidth]{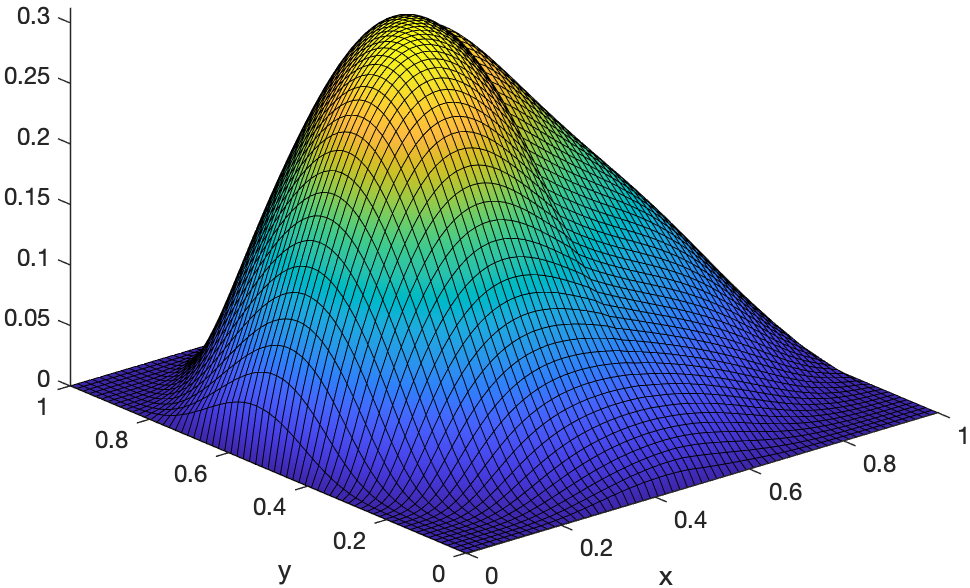}
\label{fig:consistent_multimat}} 
\end{center}
\vspace{-5ex}
\begin{center}
\subfigure[IVR(L)]{\includegraphics[width=0.4\linewidth]{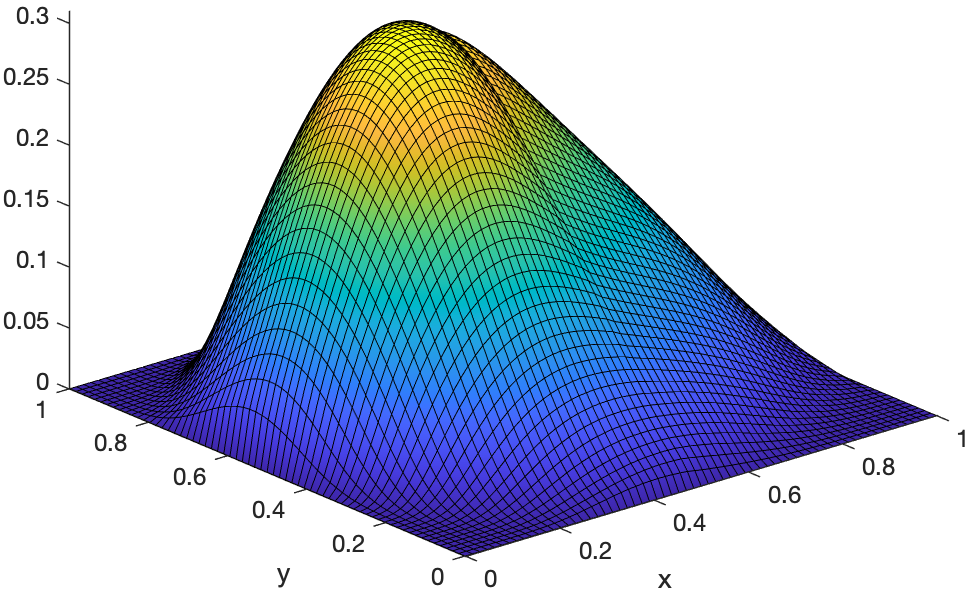}
\label{fig:lumped_multimat}}
\subfigure[$\mu$DMD-FS]{\includegraphics[width=0.4\linewidth]{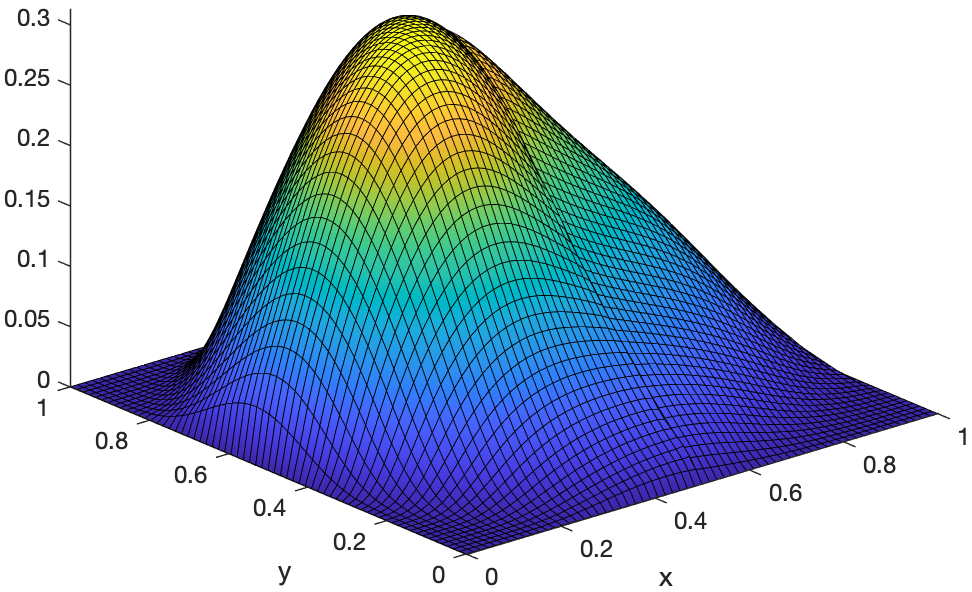}\label{fig:DMD_multimat}}
\end{center}
\vspace{-4ex}
\caption{Surface plots of the reference \REV{non-hybrid} monolithic solution and the solutions of the partitioned schemes for the multi-material combination test at the final simulation time $T=2\pi$.}
\end{figure}

\begin{figure}[!t]
\centering
	\subfigure[IVR(C)]{\includegraphics[width=0.75\linewidth]{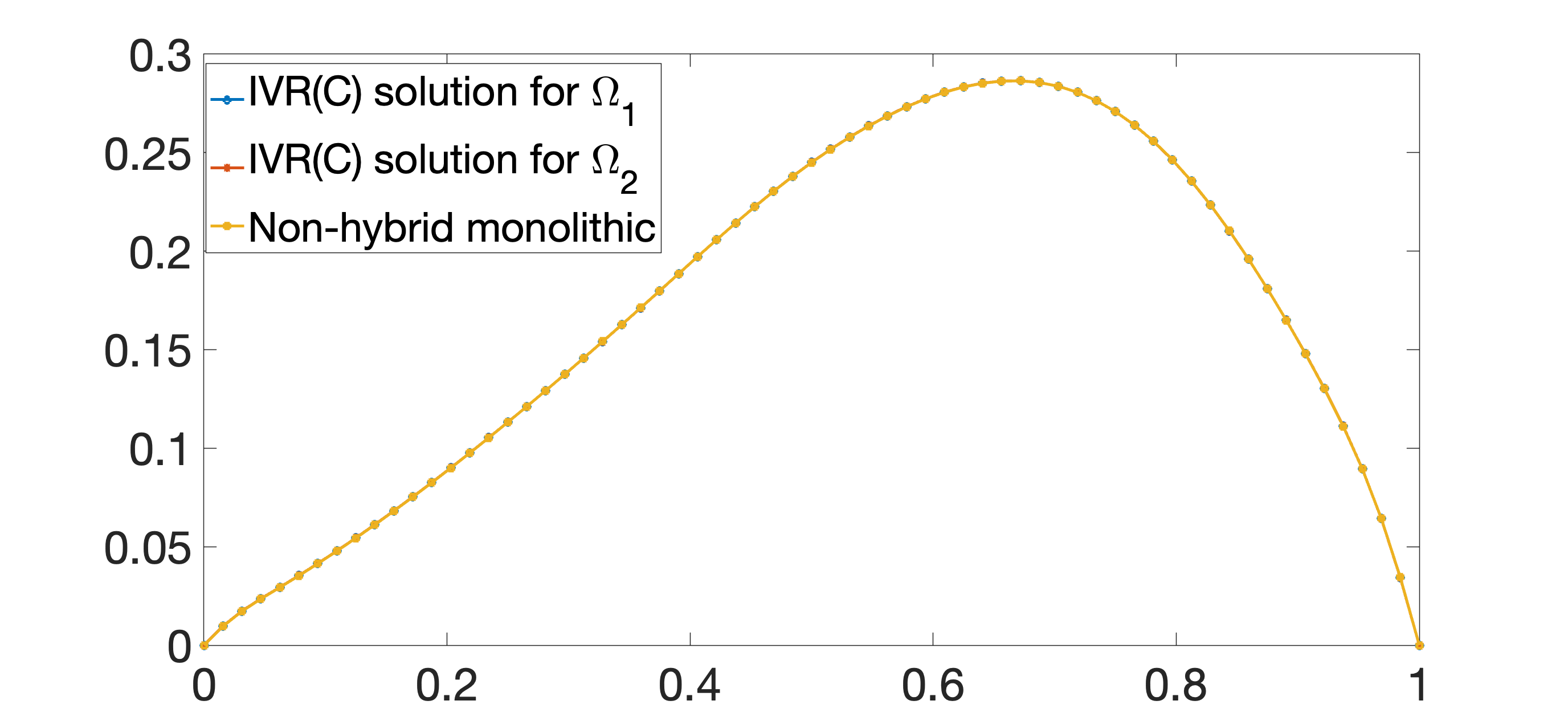}\label{fig:interface_consistent_multimat}} 
	\subfigure[IVR(L)]{\includegraphics[width=0.75\linewidth]{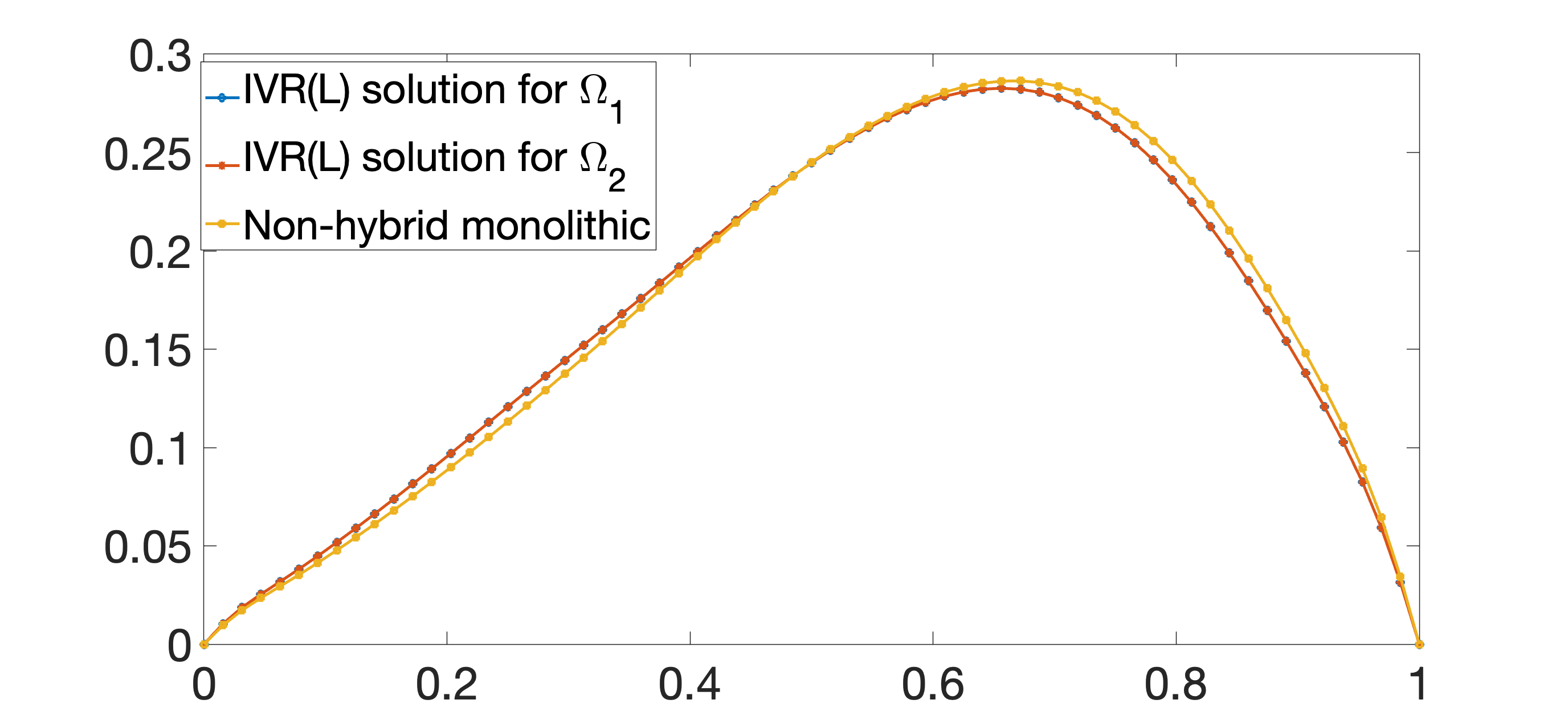}\label{fig:interface_lumped_multimat}}
	\subfigure[$\mu$DMD-FS]{\includegraphics[width=0.75\linewidth]{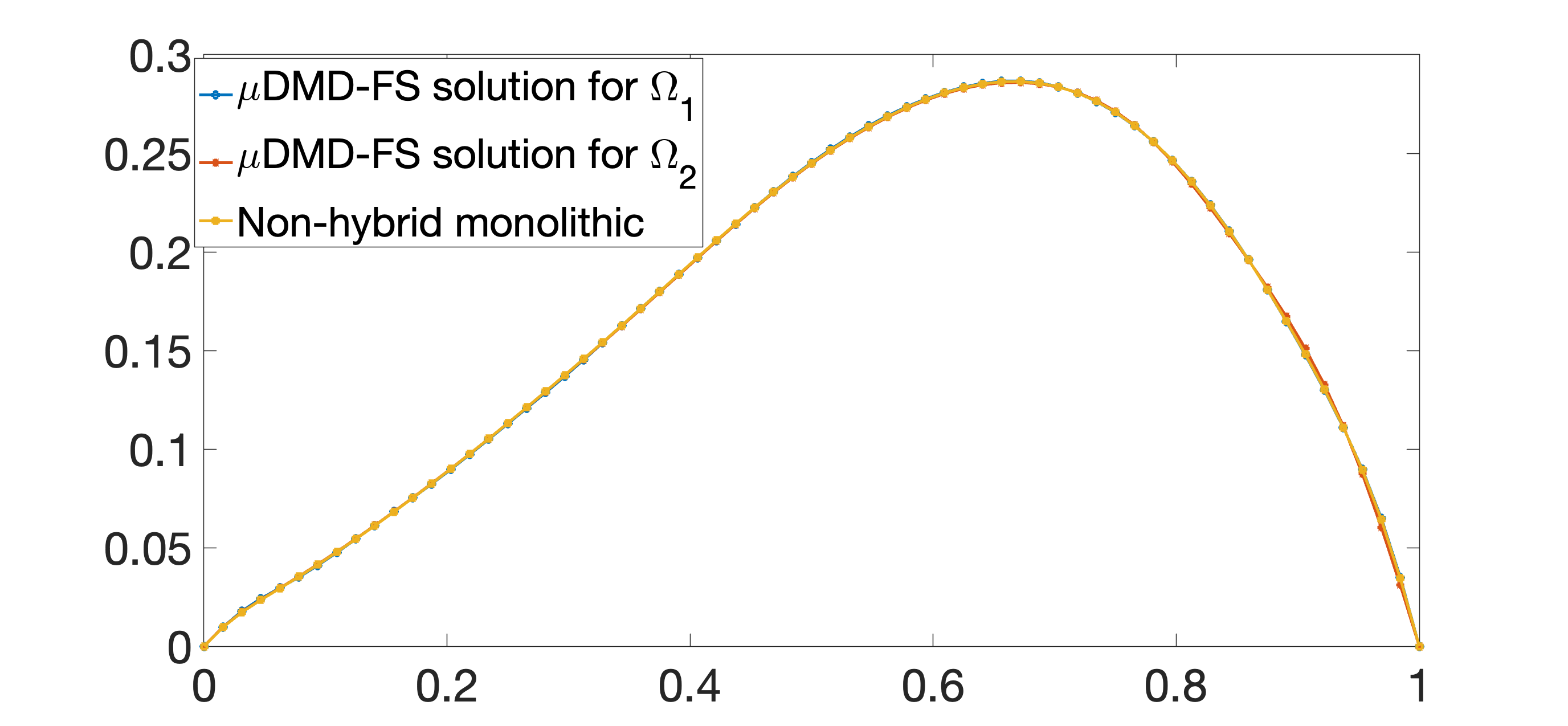}\label{fig:interface_DMD_multimat}}
	\caption{Interface restrictions of the reference \REV{non-hybrid} monolithic solution and the solutions of the partitioned schemes for the  multi-material combination test at the final simulation time $T = 2\pi$.}
\end{figure}

\paragraph{Combination test}
For this test we \REV{select  $\bm{\mu} = (1.5, 3.5)\times 10^{-3}$ as a parameter of interest and define  $\mathcal{S} = ([1,2]\times [3,4])\times 10^{-3}$, so that}
$$
\mathcal{M}_4(\bm{\mu},r) =\left \{(1, 3), (1, 4), (2, 3), (2, 4)\right \}\times 10^{-3} \,.
$$
 \REV{As in the patch test case}, training data is generated by using IVR(C) to solve  \eqref{eq:TP} for all $\bm{\mu}_j\in \mathcal{M}_m$, except that now the model problem has a homogeneous source term and homogeneous boundary data for all parameter values. 
\REV{The ``optimal'' tolerance for this test is $\epsilon=10^{-8}$ and we use this value to obtain the operators $\mathbf{A}_\lambda(\bm{\mu}_j)$, $\bm{\mu}_j \in \mathcal{M}_4$.}
\begin{table}[ht]
\renewcommand{\arraystretch}{1.15}
\centering
\textsf{
\begin{tabular}{p{0.15\linewidth}  p{0.15\textwidth}p{0.15\textwidth} p{0.15\textwidth}}
 \multicolumn{4}{c}{Multi-material Combination Test} \\
 \hline
  & $\mathcal{E}^{0}_{X}$ & $\mathcal{E}^{1}_{X}$ & Speedup \\
 \hline
 IVR(C) &   2.49E-03  &  7.70E-03  &  N/A \\
 IVR(L) & 3.61E-02     & 8.35E-02 & $\times$ 1.39 \\ 
 $\mu$DMD-FS & 2.92E-03 & 9.15E-03 & $\times$ 11.63 \\ 
 \hline
\end{tabular}
}
\caption{Relative $L^2$ and $H^1$ errors of the multi-material combination test solutions computed by the IVR(C), IVR(L) and $\mu$DMD-FS schemes at the final time $T=2\pi$, along with average speedup of the latter two relative to IVR(C).}
\label{tab:multimat_combo}
\end{table}
\REV{We then proceed to solve the model problem with the selected parameter of interest $\bm{\mu}$ using Algorithm \ref{parametric_fsdmd_algorithm}.} Results from the multi-material combination test are presented in Table \ref{tab:multimat_combo}, Figures  \ref{fig:monolithic_multimat}--\ref{fig:DMD_multimat}, and Figures  \ref{fig:interface_consistent_multimat}--\ref{fig:interface_DMD_multimat}.


\paragraph{\REV{Robustness and convergence}}
\REV{Below we use the multi-material patch test with the parameter and sampling set combinations shown in Figures \ref{fig:param-DMD-2}-\ref{fig:param-DMD-3} to examine the robustness of the $\mu$DMD-FS scheme and its accuracy as a function of the sampling grid size $\sigma$.}

\begin{figure}[!t]
\includegraphics[width=0.5\linewidth]{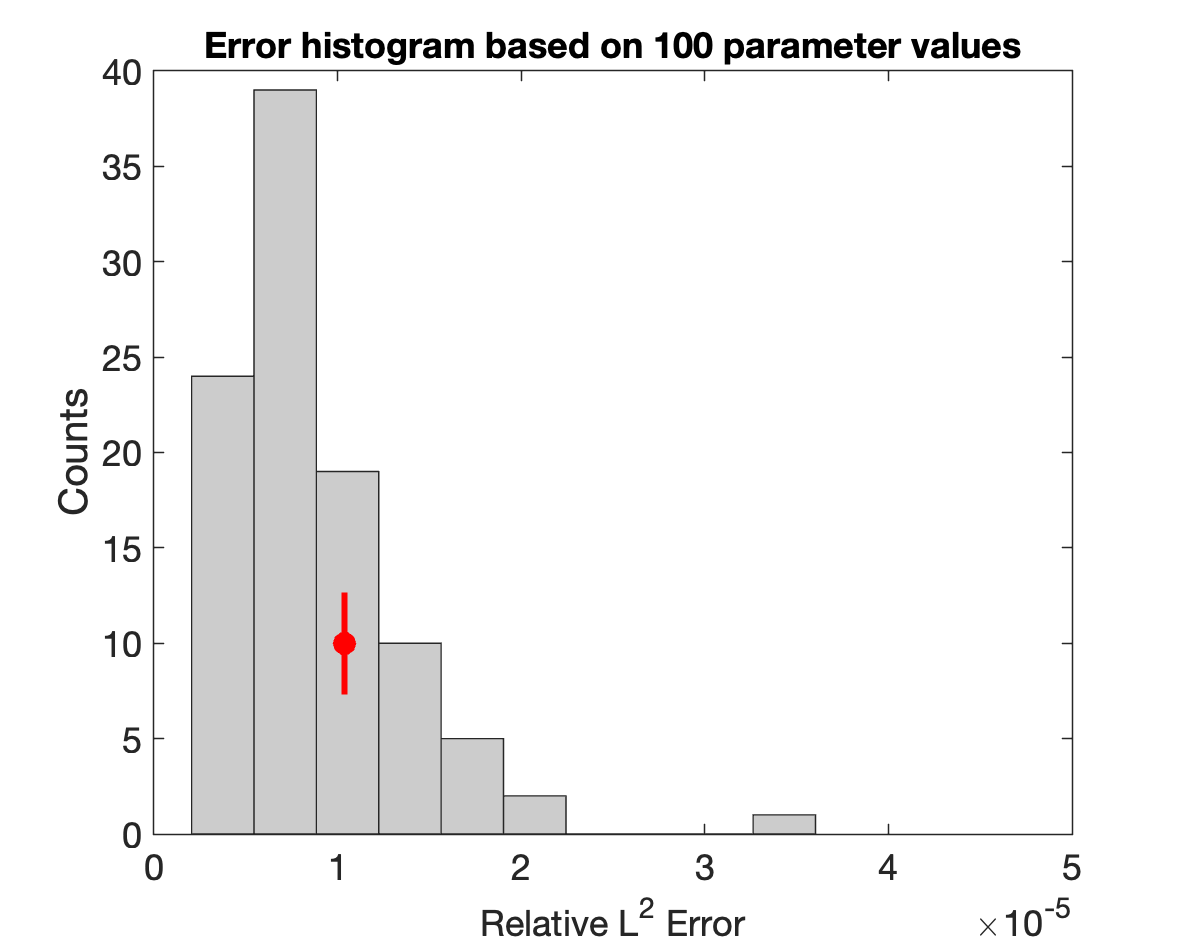} 
\hfill
\includegraphics[width=0.5\linewidth]{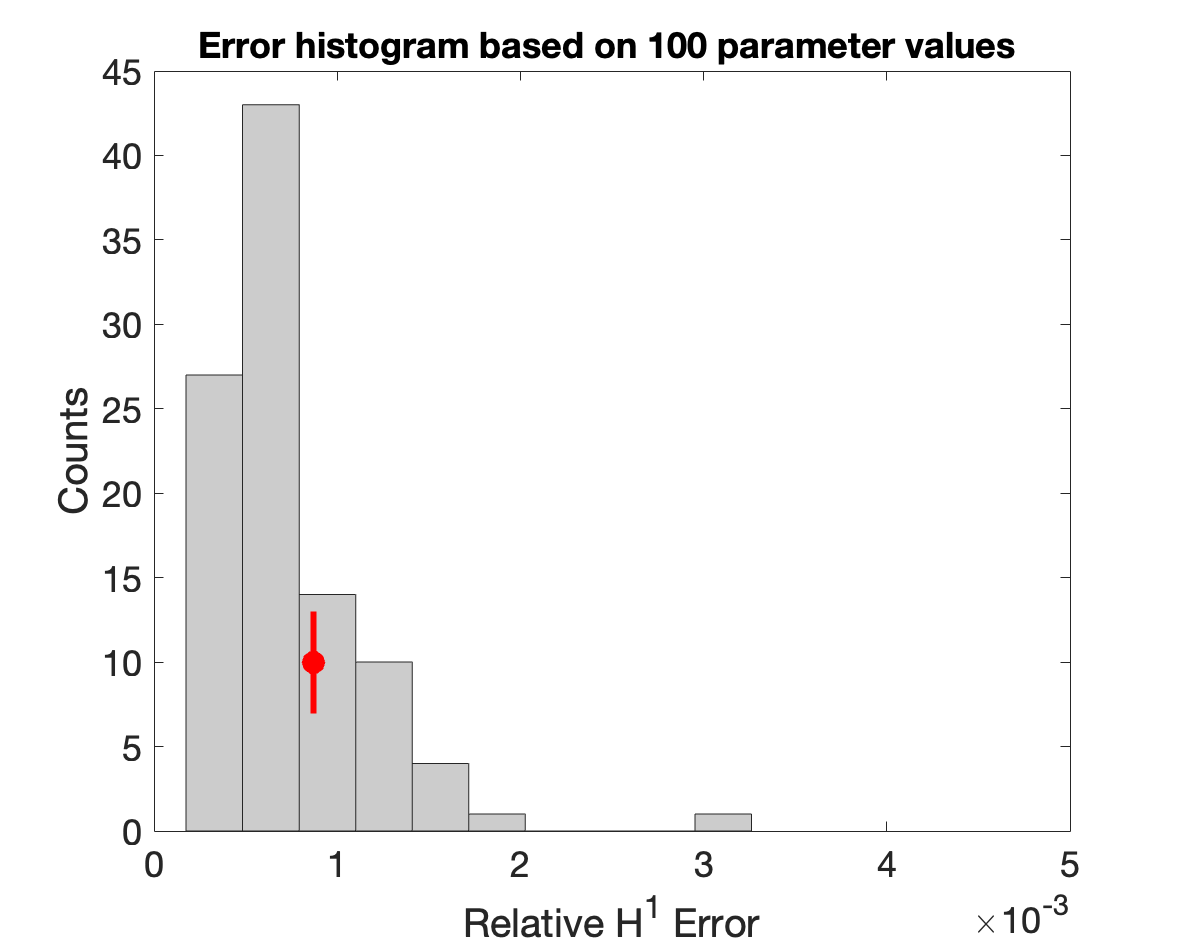}
\caption{\REV{Histograms of the relative $L^2$ and $H^1$ errors of the $\mu$DMD-FS solution obtained by solving the multi-material patch test problem for a random parameter sample of size 100 drawn from a uniform point distribution on the square $\mathcal{S}$ defining the set \eqref{eq:sample-set-patch}.
The red dots mark the errors from Table \ref{tab:multimat_patch}, which correspond to the parameter $\bm{\mu}$ at the center of $\mathcal{S}$.}}
\label{fig:histograms}
\end{figure}

\REV{For the first task we retain the sampling set \eqref{eq:sample-set-patch} and the associated square $\mathcal{S}$, and consider a random parameter sample $\{\bm{\mu}\}$ of size $100$ drawn from a uniform distribution on $\mathcal{S}$. For each one of these random parameter values we use Algorithm \ref{parametric_fsdmd_algorithm} with the \emph{fixed sampling set} \eqref{eq:sample-set-patch}  to solve the model problem and collect the error data. We then use this data to construct the histograms shown in Fig.\ref{fig:histograms}.}

\begin{figure}[!t]
\includegraphics[width=0.5\linewidth]{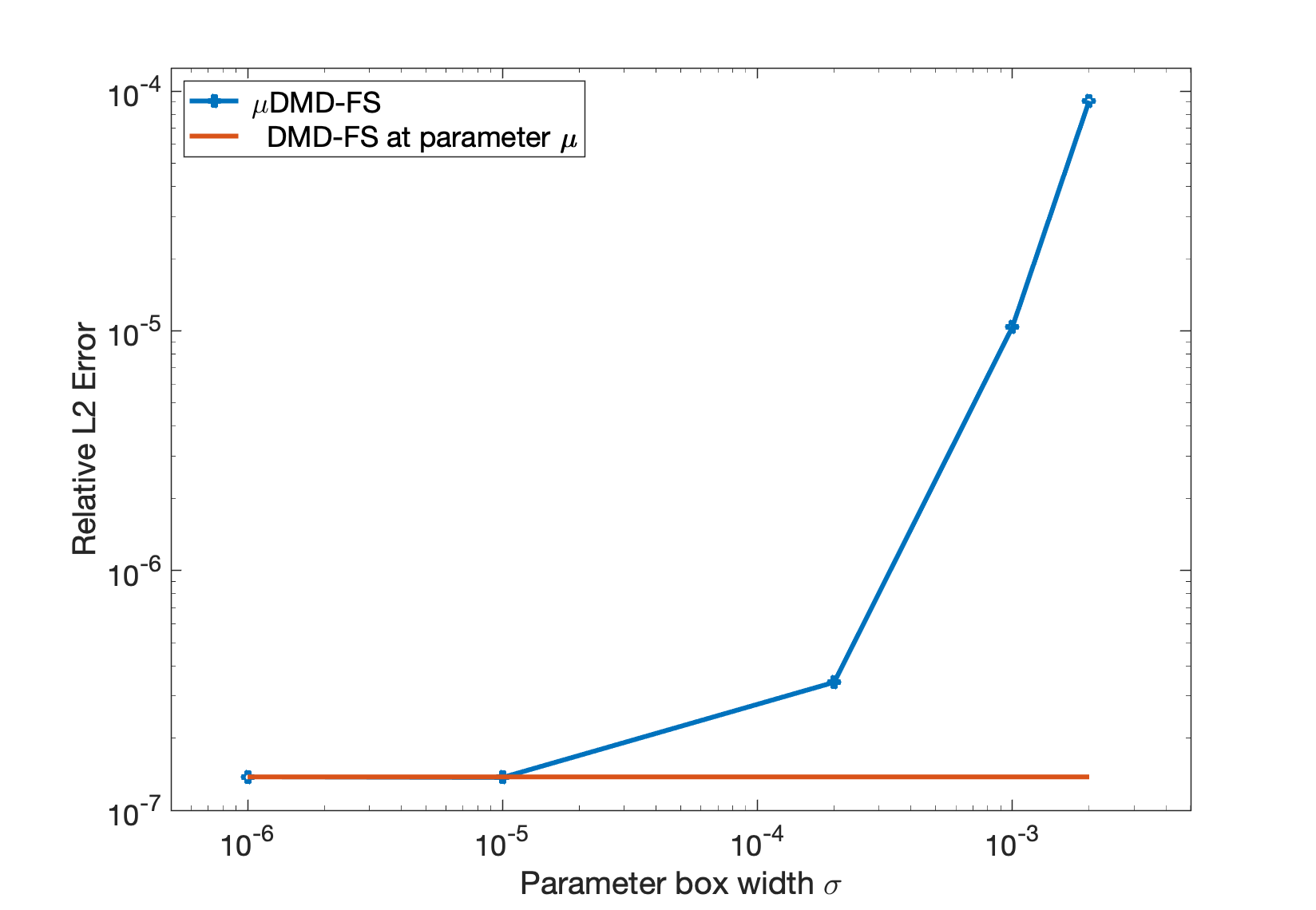} 
\includegraphics[width=0.5\linewidth]{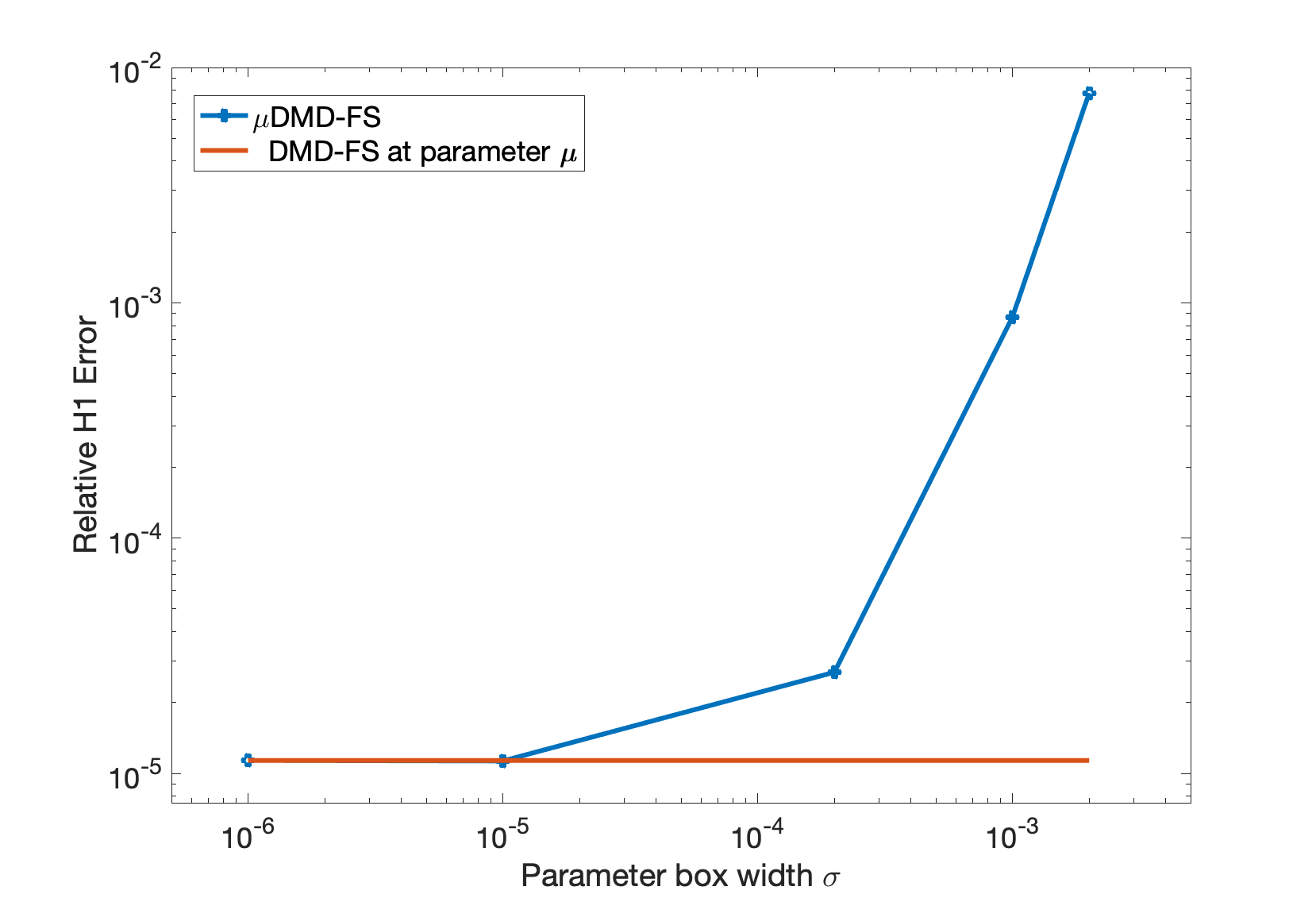}
\caption{
\REV{Relative $L^2$ and $H^1$ errors of the $\mu$DMD-FS solution of the multi-material patch test with $\bm{\mu} = (1.5,2.5)\times 10^{-3}$ as functions of the width of the parameter box $\mathcal{S}_i$ 
centered at $\bm{\mu}$.}
}
\label{fig:param-convergence}
\end{figure}

\REV{For the second task we fix the parameter of interest $\bm{\mu} = (1.5,2.5)\times 10^{-3}$ and consider five nested squares $\mathcal{S}_i$  with widths $\sigma_1= 10^{-6}$, $\sigma_2=  10^{-5}$, $\sigma_3= 2\times 10^{-4}$, $\sigma_4= 10^{-3}$, and  $\sigma_5= 2\times 10^{-3}$, centered at $\bm{\mu}$. These squares can be thought of as representing a refinement of the sampling grid in the parameter space \eqref{eq:parspace}. 
We then use Algorithm \ref{parametric_fsdmd_algorithm}  to solve the model problem with the fixed parameter $\bm{\mu}$ and \emph{varying sampling sets} $\mathcal{M}_4(\bm{\mu},r_i)$ defined by the vertices of the nested squares $\mathcal{S}_i$. Figure \ref{fig:param-convergence} shows the relative $L^2$ and $H^1$ solution errors of the parametric $\mu$DMD-FS scheme as functions of the box width $\sigma$. As a reference we plot the relative $L^2$ and $H^1$ solution errors of the non-parametric DMD-FS scheme defined in  Algorithm \ref{fsdmd_algorithm}.}

\paragraph{Discussion of results}
As expected, the patch test data in Table \ref{tab:multimat_patch} confirms that the IVR(C) solution matches the \REV{non-hybrid} monolithic solution of \eqref{eq:TP} and  the manufactured solution \eqref{patch_equation}, to machine precision. 
Furthermore, the relative errors reported in Tables  \ref{tab:multimat_patch}--\ref{tab:multimat_combo} suggest that the $\mu$DMD-FS scheme meets our accuracy goal. \REV{Indeed, in both cases the relative $L^2$ and $H^1$  errors of the $\mu$DMD-FS solution are approximately one order of magnitude better than those of the IVR(L) scheme and for the multi-material combination test these errors are  within $18\%$ of the respective IVR(C) errors}. 
While these distinctions are hardly noticeable in the surface plots of the solutions in Figures  \ref{fig:monolithic_multimat}--\ref{fig:DMD_multimat}, they become visible when examining the profiles of the IVR(C), IVR(L) and $\mu$DMD-FS solutions along the interface. From the plots in Figures \ref{fig:interface_consistent_multimat}--\ref{fig:interface_DMD_multimat}, one can clearly see that IVR(L) solution deviates the most from the  \REV{reference non-hybrid} monolithic solution, while the profile of the $\mu$DMD-FS solution is \REV{almost indistinguishable from that of the reference one}. 

Insofar as the computational efficiency of $\mu$DMD-FS is concerned, recall that our goal was a scheme whose cost is comparable to that of the IVR(L) scheme. The data in Tables \ref{tab:multimat_patch}--\ref{tab:multimat_combo} shows that DMD-FS clearly exceeds this goal for both examples.  It is almost five times faster that IVR(L) for the patch test and more than eight times faster for the combination test. The speedups over IVR(C) are even more pronounced, confirming the potential of the surrogate-based partitioned approach formulated in this paper.

\REV{Let us now examine the results presented in Figures \ref{fig:histograms}--\ref{fig:param-convergence}. The histograms of the relative $L^2$ and $H^1$ errors in the first figure reveal that bulk of these errors is clustered to the left of the errors reported in Table \ref{tab:multimat_patch}, which are marked by red dots on the plots. Thus, one can conclude that the accuracy of the $\mu$DMD-FS scheme is roughly the same for all parameter values in $\mathcal{S}$, i.e., the scheme is not sensitive with respect to the relative position of $\bm{\mu}$ inside the box.}

\REV{We now turn attention to Figure \ref{fig:param-convergence}, which shows plots of $\mathcal{E}_{D}^0(\sigma)$ and $\mathcal{E}_{D}^1(\sigma)$, i.e., the relative $L^2$ and $H^1$ solution errors of the $\mu$DMD-FS solution as functions of the  sampling grid size $\sigma$. We recall that the data for these plots was obtained by constructing the rKOI operator  \eqref{eq:rKOI} on a sequence of five nested squares $\mathcal{S}$ centered at $\bm{\mu} = (1.5,2.5)\times 10^{-3}$.
Alongside these errors, for reference, we plot the relative $L^2$ and $H^1$ errors of the \emph{non-parametric} DMD-FS scheme utilizing a DMD flux operator $\mathbf{A}_\lambda(\bm{\mu})$ constructed at this parameter. The horizontal lines corresponding to the DMD-FS errors  represent the best possible accuracy attainable by the $\mu$DMD-FS scheme.}
\REV{The first conclusion that can be drawn from the plots in Figure \ref{fig:param-convergence} is that the errors of the  \emph{parametric} $\mu$DMD-FS scheme do converge to the errors of the \emph{non-parametric} DMD-FS scheme. Thus, the accuracy of the former can be made arbitrarily close to that of the latter by sampling the parameter domain on a sufficiently fine grid.}

\REV{The second conclusion is that the rate of this convergence matches that of the bilinear interpolant $\mathbf{A}^I_\lambda(\bm{\mu})$, which is second-order accurate in  $\sigma$. We confirm this conclusion by estimating the slopes of $\mathcal{E}_{D}^0(\sigma)$ and $\mathcal{E}_{D}^1(\sigma)$. To that end we drop the error data corresponding to the smallest two values of $\sigma$ and perform linear regression of the logarithms of the remaining three data points. The slopes of the resulting linear fits are $2.4$, thereby confirming that  $\mathcal{E}_{D}^k(\sigma)=O(\sigma^2)$, $k=0,1$.}

\REV{
\begin{remark}
It is important to keep in mind that the second order of convergence of the relative errors only applies to the case when they are treated as functions of the sampling grid size $\sigma$ and the underlying finite element mesh size $h$ is fixed.  When treated as functions of $h$, $\mathcal{E}^1_{D}(h)$ is one order less accurate than $\mathcal{E}^0_{D}(h)$.
\end{remark}
}

\subsection{Single material configuration tests}
For the single material configuration we use  $\kappa_1 = \kappa_2 = 1\times 10^{-3}$ for both the patch test and the combination test problems.  We solve these problems using IVR(C), IVR(L) and the \REV{\emph{non-parametric}} DMD-FS scheme on a sequence of uniform $n\times n$ quadrilateral grids with $n=16$, $n=32$, $n=64$, and $n=128$.  
The time steps for each grid size are selected to satisfy the CFL condition and are given by  $\Delta_{16} t = 1.42\times 10^{-2}$, $\Delta_{32} t = 6.84\times 10^{-3}$, $\Delta_{64} t = 3.37\times 10^{-3}$, and $\Delta_{128} t = 1.67\times 10^{-3}$, respectively. As a result, the number of time steps required to reach the final time $T=2\pi$ is different on every grid and is given by  $S_{16}=444$, $S_{32}=918$, $S_{64}=1866$, and $S_{128}=3761$, respectively. 

\REV{Thus,} for every grid size $n$ we train a separate DMD flux surrogate operator using training data specific to the grid size and the test problem. This data is generated by prescribing the initial conditions given by the Gaussian hills in Section \ref{sec:train-nosource} and then augmenting \eqref{eq:TP} with source terms and boundary conditions corresponding to the patch test and the combination test problems. \REV{Once the model problem is properly configured, its hybrid monolithic formulation is solved using the IVR(C) scheme.} 

To account for variations in the grid size, we scale the Gaussian hills in a manner proportional to the mesh size $h=1/n$, i.e., we double the number of Gaussian hills used in training and half their width when we reduce the mesh size by half. For example, a $128\times 128$ grid will use twice as many Gaussians for training the DMD operator as a $64\times 64$ grid. Because the number of time steps required to reach the final time $T=2\pi$ is different for each grid, the lengths of the time series for the subdomain solutions and the interface flux comprising the training data are also different and are given by the numbers $S_n$, $n\in\{16,32,64,128\}$ defined earlier. 

\begin{table}[!h]
\renewcommand{\arraystretch}{1.15}
\centering
\textsf{
\begin{tabular}{p{0.1\linewidth} | p{0.15\textwidth}p{0.15\textwidth} | p{0.15\textwidth}p{0.15\textwidth}}
 \multicolumn{5}{c}{Optimal energy tolerances and DMD operator ranks} \\
 \hline
      & \multicolumn{2}{c|}{Combination test} & \multicolumn{2}{c}{Patch test} \\
 \hline
  Grid &\quad  $\epsilon$ &\quad  Rank  &  $\quad \epsilon$ &\quad  Rank  \\
 \hline
$16\times 16$ & \quad  1E-8 & \quad 29  & \quad 1E-8  & \quad 14 \\
 $32\times 32$ & \quad   1E-8 & \quad 30 & \quad 1E-11 & \quad 29 \\
$64\times 64$ & \quad  1E-8 & \quad  42    &\quad  1E-13 &\quad  45\\
$128\times 128$ & \quad  \REV{1E-8} &\quad  \REV{49}  &\quad  1E-15 &\quad  59 \\
 \hline
\end{tabular}
}
\caption{``Optimal'' tolerances $\epsilon$ for the relative snapshot energy condition \eqref{eq:POD-energy} for each mesh size and the resulting ranks of the DMD-FS operators for the combination and the patch test problems.}
\label{tab:DMD_eps_ranks}
\end{table}

\begin{figure}[t] 
   \centering
   \subfigure[Patch test]{\includegraphics[width=0.475\linewidth]{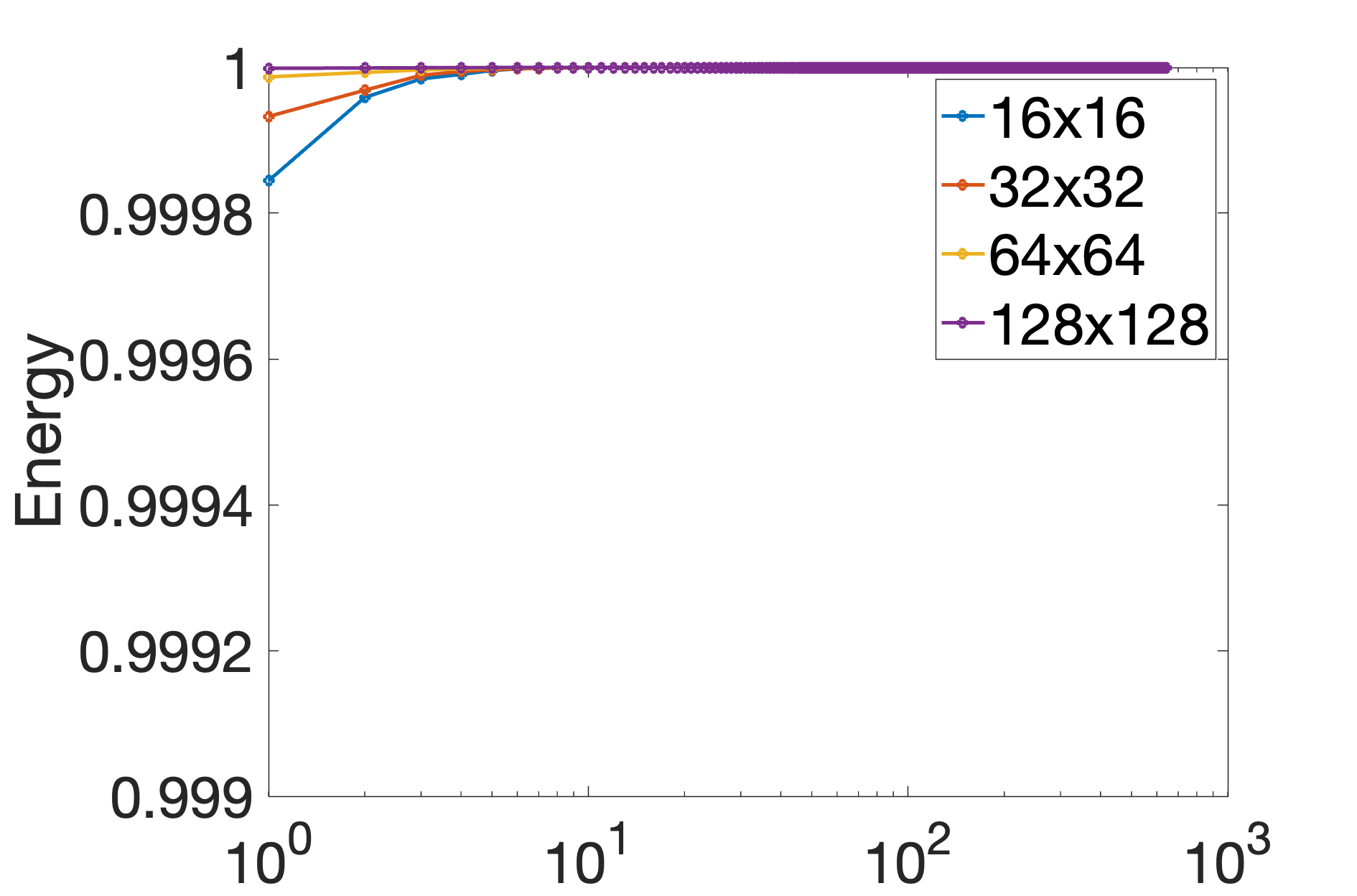}
   \label{fig:dmd-energy-patch}}
\subfigure[Combination test]{\includegraphics[width=0.475\linewidth]{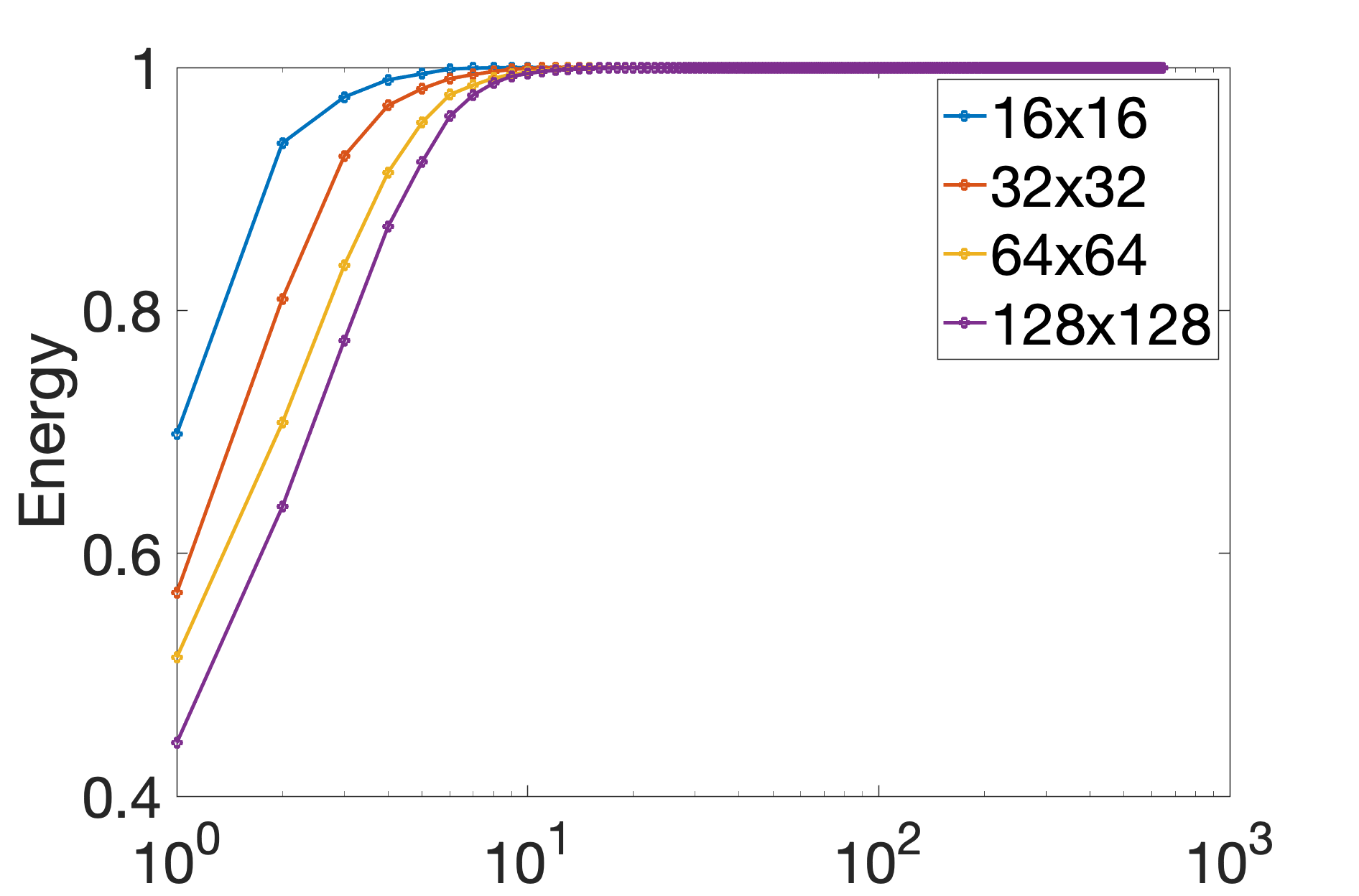} 
 \label{fig:dmd-energy-combo}}
   \caption{Snapshot energy as a function of the DMD rank for the grids used in the single material configuration.}
\end{figure}

\REV{To train the DMD flux surrogate operators we use the same definitions of the distance threshold $\delta$ and the snapshot energy tolerance $\epsilon$ as in Section \ref{sec:multimat}. Table  \ref{tab:DMD_eps_ranks} shows the values of the ``optimal'' energy tolerance and the ranks of the resulting DMD flux operators for every grid size and test problem considered in this section.
The data in this table reveals that for the combination test the ``optimal'' tolerance value remains the same on all meshes and that the rank of the DMD operator grows by less than a factor of 2 from the coarsest to the finest mesh. 
For the patch test the ``optimal'' value of $\epsilon$ becomes smaller as the mesh size increases. However, the  growth of the corresponding DMD rank remains moderate and does not exceed a factor of approximately 4 between the coarsest and the finest meshes.}
This observation is further confirmed by the plots of the snapshot energies in Figures \ref{fig:dmd-energy-patch}--\ref{fig:dmd-energy-combo} which show that the bulk of the snapshot energy is captured by a relatively small number of modes on every grid size used in this section.
\REV{These results indicate that the flux surrogates remain computationally efficient with grid refinement.}

\begin{figure}[!t]
\centering
\subfigure[\REV{Non-hybrid} monolithic]{\includegraphics[width=0.45\linewidth]{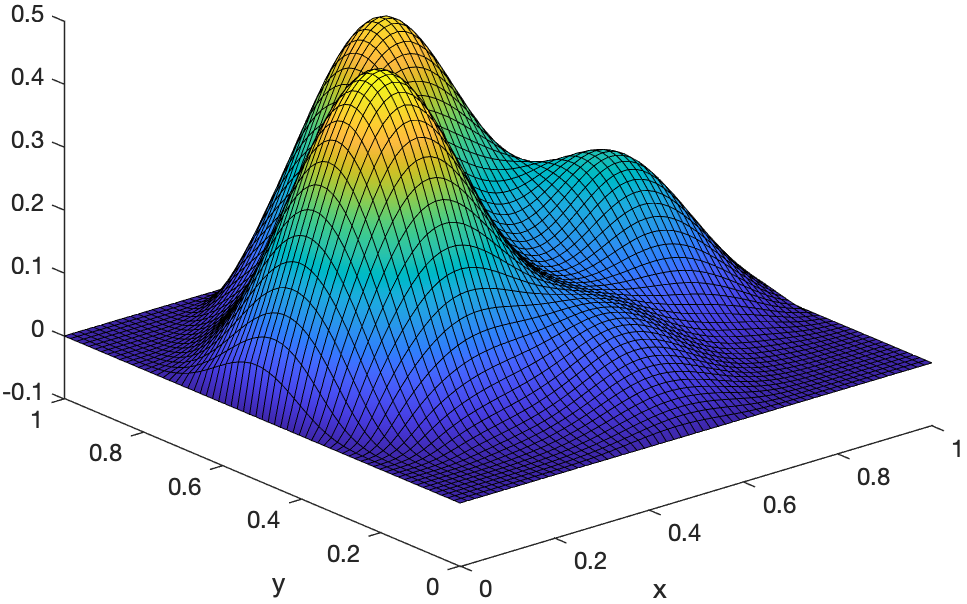}\label{fig:monolithic_singlemat}} 
	\subfigure[IVR(C)]{\includegraphics[width=0.45\linewidth]{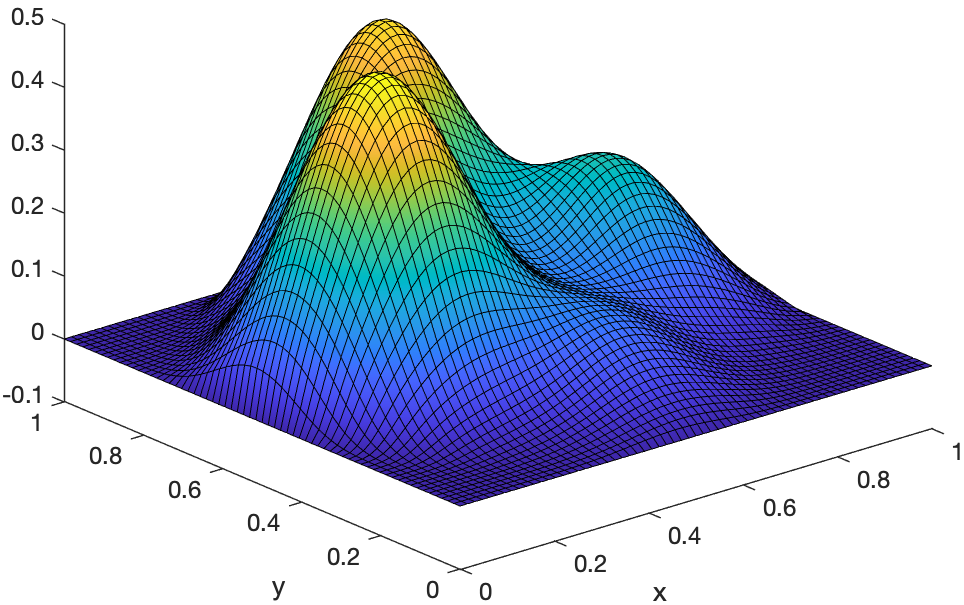}\label{fig:consistent_singlemat}} 
\centering
	\subfigure[IVR(L)]{\includegraphics[width=0.45\linewidth]{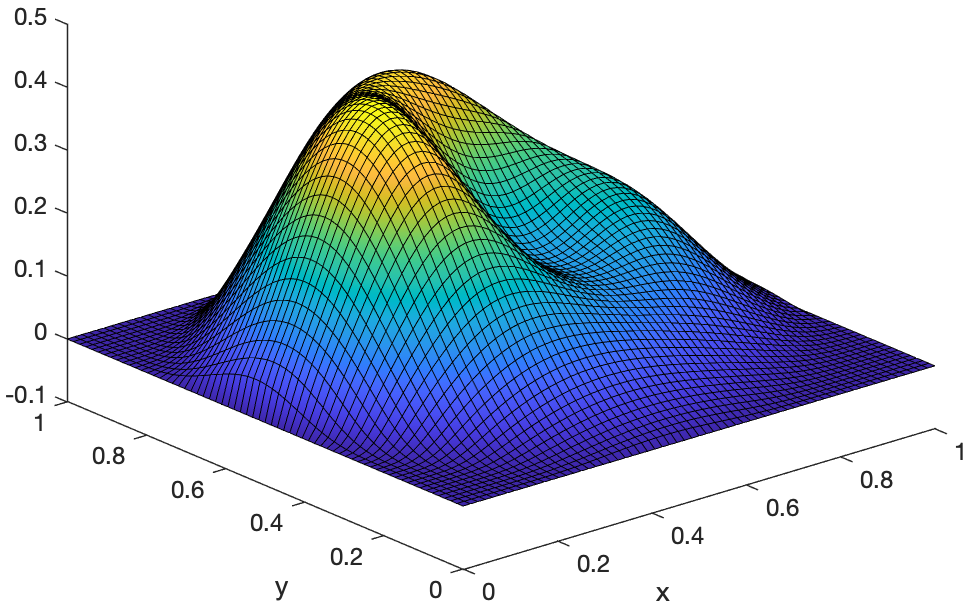}\label{fig:lumped_singlemat}}
	\subfigure[DMD-FS]{\includegraphics[width=0.45\linewidth]{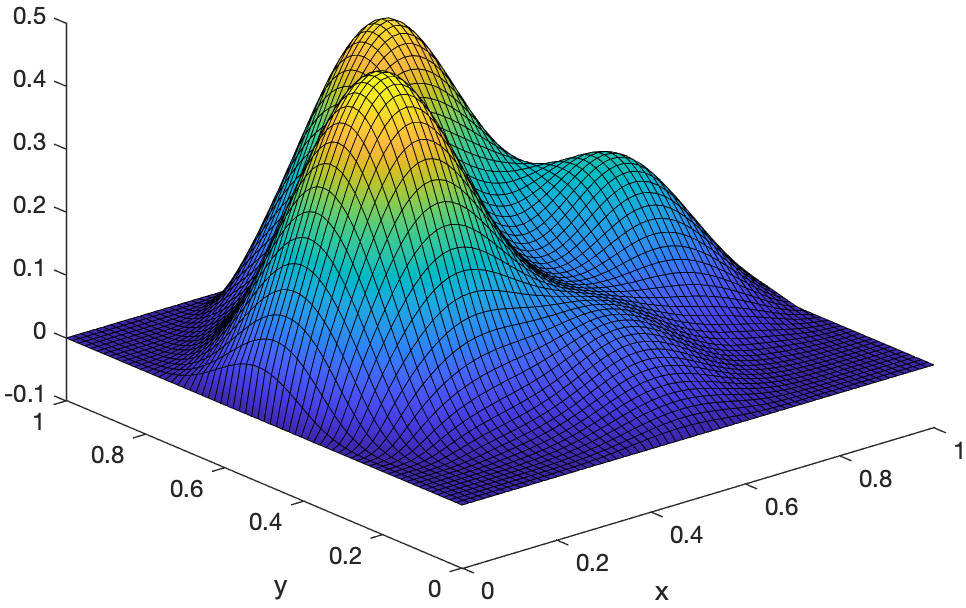}\label{fig:DMD_singlemat}}
	\caption{Surface plots of the reference \REV{non-hybrid} monolithic solution and the solutions of the partitioned schemes for the  single material combination test at the final simulation time $T = 2\pi$.}
\end{figure}

\begin{figure}[!ht]
\centering
	\subfigure[IVR(C)]{\includegraphics[width=0.75\linewidth]{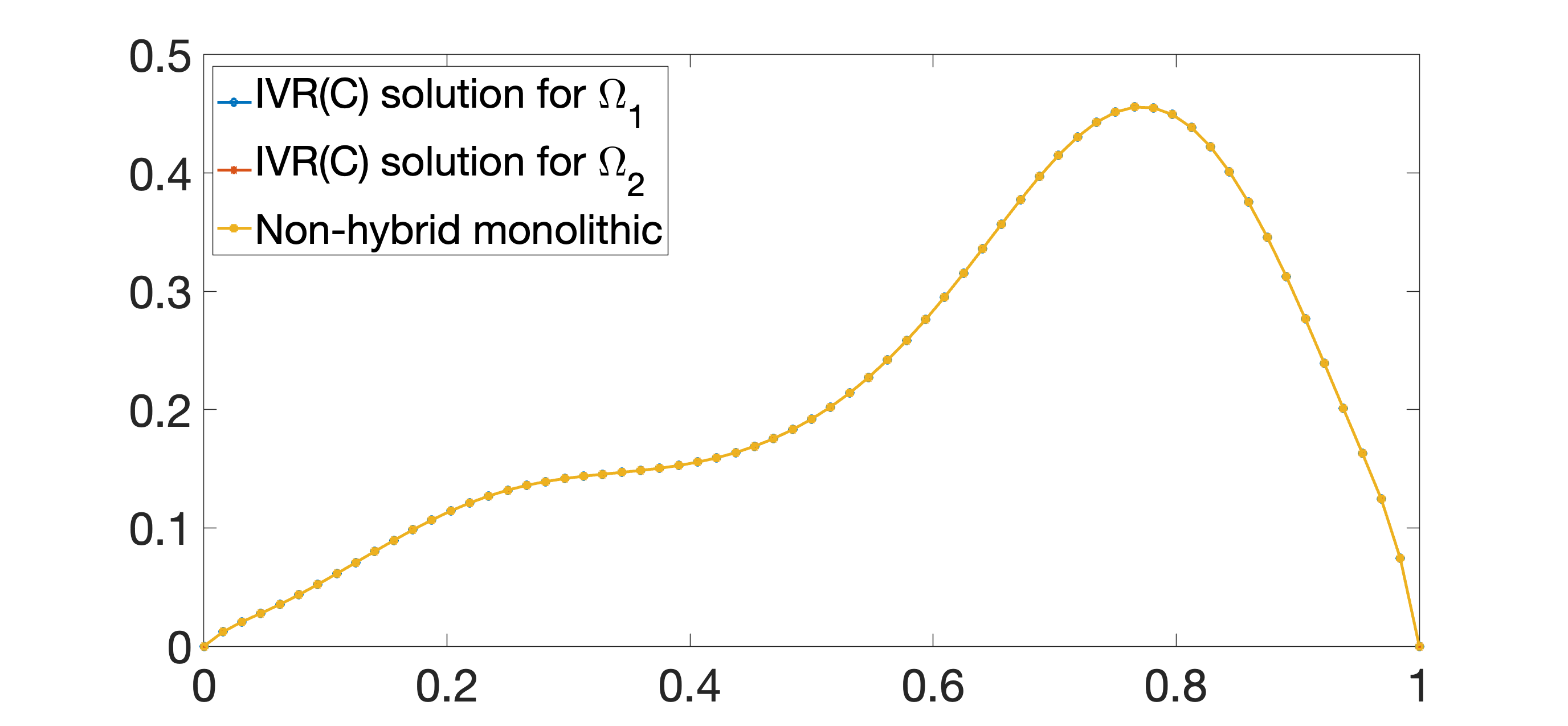}\label{fig:interface_consistent_singlemat}} 
	\subfigure[IVR(L)]{\includegraphics[width=0.75\linewidth]{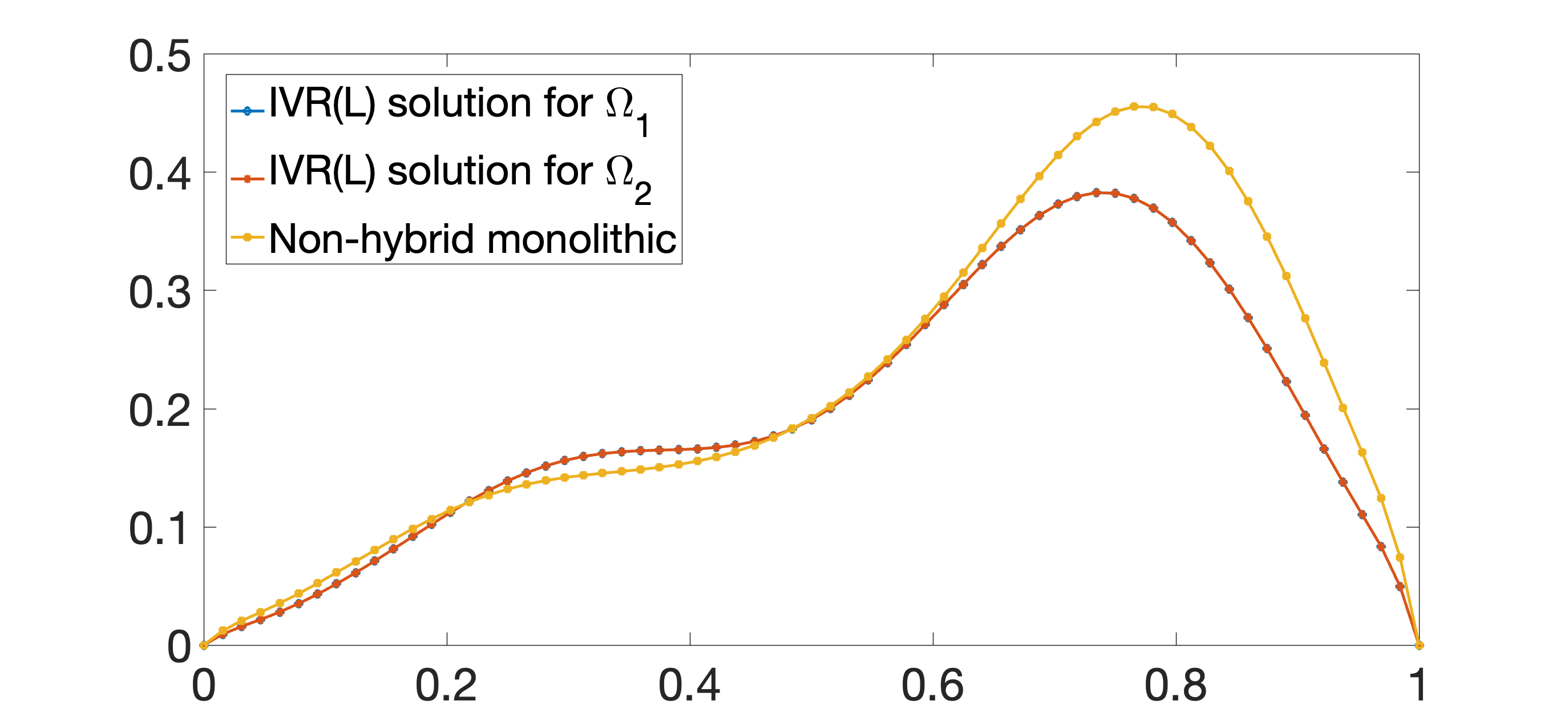}\label{fig:interface_lumped_singlemat}}
	\subfigure[DMD-FS]{\includegraphics[width=0.75\linewidth]{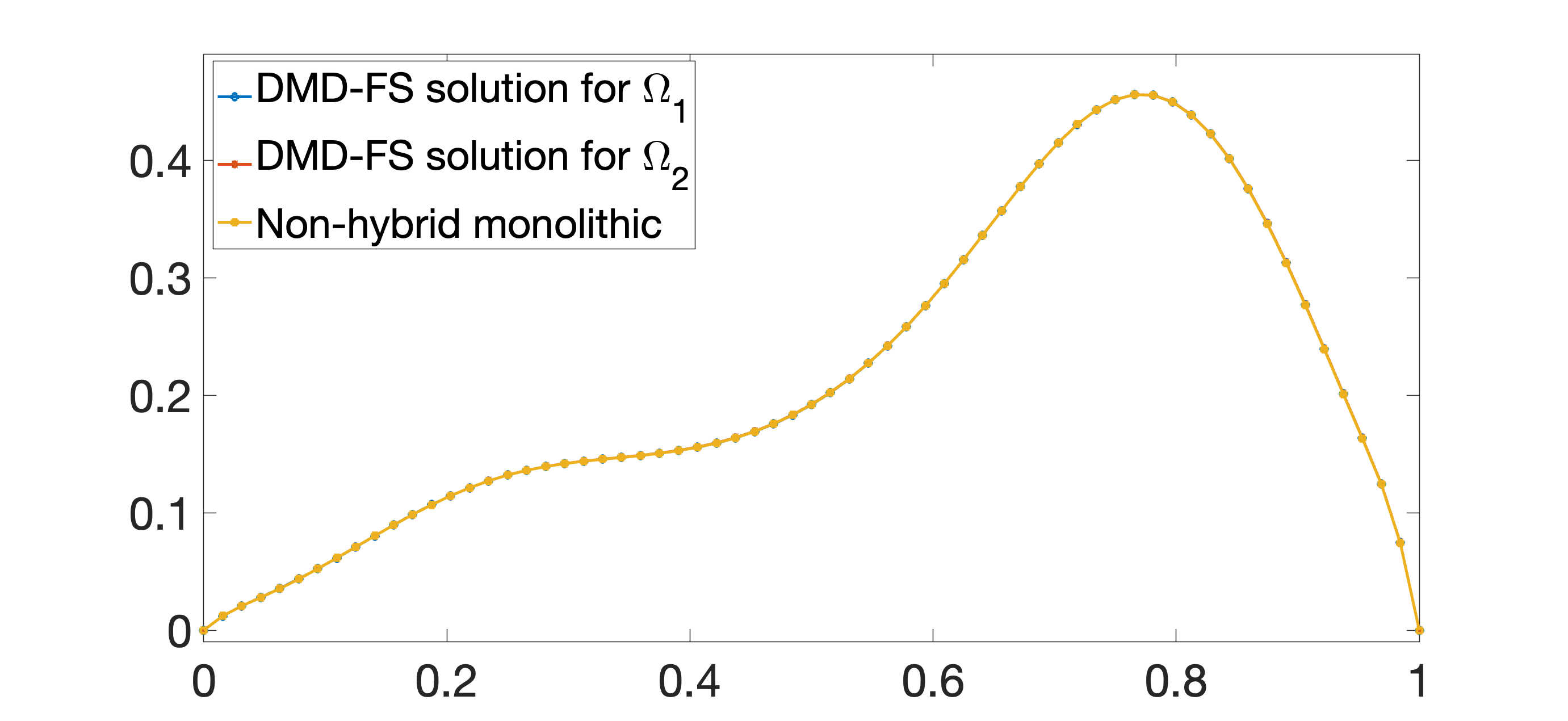}\label{fig:interface_DMD_singlemat}}
	\caption{Interface restrictions of the reference \REV{non-hybrid} monolithic solution  and the solutions of the partitioned schemes for the  single material combination test at the final simulation time $T = 2\pi$.}
\end{figure}

Results for the single material configuration are collected in two tables and two figures. Table \ref{tab:singlemat_patch} shows relative solution errors and speedups for the  patch test, whereas Table \ref{tab:singlemat_combo} shows this data for the combination test. For the latter we also provide surface plots of the \REV{reference non-hybrid} monolithic solution, and the IVR(C), IVR(L), and DMD-FS solutions in Figures \ref{fig:monolithic_singlemat}--\ref{fig:DMD_singlemat}, respectively, as well as plots of the profiles of these solutions along the interface in  Figures \ref{fig:interface_consistent_singlemat}--\ref{fig:interface_DMD_singlemat}.

\paragraph{Discussion of results}
\REV{First, we recall that for $\kappa_1=\kappa_2$ the IVR(C) solution of a hybrid monolithic discretization of the model problem should coincide with its \emph{non-hybrid} monolithic solution on the same mesh; see Section \ref{sec:closure}. The relative IVR(C) solution errors in Tables \ref{tab:singlemat_patch}-\ref{tab:singlemat_combo} verify this property.}

\REV{Insofar as the accuracy and the efficiency of the DMD-FS scheme are concerned}, our results in the single material case essentially mirror those in the multi-material setting. \REV{Specifically, the data in Tables \ref{tab:singlemat_patch}-\ref{tab:singlemat_combo}  confirm that} DMD-FS consistently outperforms IVR(L) not only in terms of computational cost but also in terms of accuracy.
The accuracy distinction is particularly strong for the patch test where on the finest mesh the relative errors of the DMD-FS solution are four orders of magnitude better than those of the IVR(L) solution; see Table \ref{tab:singlemat_patch}. Likewise, data in Table \ref{tab:singlemat_combo} shows that, for the combination test, DMD-FS also reliably delivers more accurate solutions than IVR(L) while being several times faster than the latter. 
Most notably, we see double digit speedups relative to IVR(C) for both test problems on all meshes, further confirming the potential of the DMD-FS approach to produce accurate solutions  at a fraction of the cost of the \REV{hybridized} IVR(C) scheme.

\REV{It is also worth pointing out that} for the combination test in the single material case, the distinctions between IVR(L) and the other schemes are already perceptible in the surface solution plots shown in Figures \ref{fig:monolithic_singlemat}--\ref{fig:DMD_singlemat}, which reveal the more diffusive nature of this \REV{first-order}   scheme. The higher rate of dissipation in IVR(L) is also visible in the plots of the solution profiles along the interface in  Figures \ref{fig:interface_consistent_singlemat}--\ref{fig:interface_DMD_singlemat}. In contrast, the profile of the DMD-FS solution is indistinguishable from that of the \REV{reference non-hybrid monolithic solution}.

%
\begin{table}[!t]
\renewcommand{\arraystretch}{1.15}
\centering
\textsf{
\begin{tabular}{p{0.2\linewidth}p{0.2\textwidth}p{0.2\textwidth}p{0.2\textwidth}}
 \multicolumn{4}{c}{Single Material Patch Test} \\
 \hline
 \multicolumn{4}{c}{IVR(C)} \\
 \hline
Grid & $\mathcal{E}^{0}_{X}$ & $\mathcal{E}^{1}_{X}$ & Speedup \\
\hline
 $16\times 16$ & 5.95E-15 & 2.19E-13 & N/A \\ 
 $32 \times 32$ &  9.07E-15   & 7.51E-13 & N/A  \\
$64 \times 64$ & 9.64E-15  & 1.41E-12 & N/A  \\ 
$128\times 128$ & 1.50E-14 & 4.84E-12 & N/A \\ 
\end{tabular}
\begin{tabular}{p{0.2\linewidth}p{0.2\textwidth}p{0.2\textwidth}p{0.2\textwidth}}
 \hline
 \multicolumn{4}{c}{IVR(L)} \\
 \hline
Grid & $\mathcal{E}^{0}_{X}$ & $\mathcal{E}^{1}_{X}$ & Speedup \\
 $16 \times 16$ & 1.16E-3 & 1.28E-2 & $\times$ 1.30 \\ 
 $32 \times 32$ & 4.17E-4   & 6.49E-3 & $\times$ 1.88 \\ 
 $64 \times 64$ & 1.49E-4  & 2.91E-3 & $\times$ 3.08 \\ 
 $128 \times 128$ & 4.74E-5 & 1.05E-3 & $\times$ 2.82 \\ 
\end{tabular}
\begin{tabular}{p{0.2\linewidth}p{0.2\textwidth}p{0.2\textwidth}p{0.2\textwidth}}
 \hline
 \multicolumn{4}{c}{DMD-FS} \\
 \hline
Grid & $\mathcal{E}^{0}_{X}$ & $\mathcal{E}^{1}_{X}$ & Speedup \\
 \hline
 $16 \times 16$ & 4.15E-5 & 1.22E-3 & $\times$ 13.72 \\
 $32 \times 32$ &  1.04E-6   & 5.19E-5 & $\times$ 21.56 \\ 
 $64 \times 64$ & 9.65E-8  & 8.12E-6 & $\times$ 37.36 \\ 
 $128 \times 128$ & 4.74E-9 & 6.53E-7 & $\times$ 17.58 \\ 
 \hline
\end{tabular}
}
\caption{Relative $L^2$ and $H^1$ errors of the single material patch test solutions computed by the IVR(C), IVR(L), and DMD-FS schemes at the final time $T=2\pi$, along with average speedup of the latter two relative to IVR(C).}
\label{tab:singlemat_patch}
\end{table}


\begin{table}[!t]
\renewcommand{\arraystretch}{1.15}
\centering
\textsf{
\begin{tabular}{p{0.2\linewidth}p{0.2\textwidth}p{0.2\textwidth}p{0.2\textwidth}}
 \multicolumn{4}{c}{Single Material Combination Test} \\
 \hline
 \multicolumn{4}{c}{IVR(C)} \\
 \hline
Grid & $\mathcal{E}^{0}_{X}$ & $\mathcal{E}^{1}_{X}$ & Speedup \\
 \hline
 $16 \times 16$ & 9.18E-15 & 4.97E-14 & N/A \\ 
 \hline
 $32 \times 32$ &  2.99E-14   & 4.05E-13 & N/A \\ 
 \hline
 $64 \times 64$ & 2.05E-14  & 5.22E-13 & N/A \\ 
 \hline
 $128 \times 128$ & 3.26E-14 & 1.83E-12 & N/A \\ 
 \hline
\end{tabular}
\begin{tabular}{p{0.2\linewidth}p{0.2\textwidth}p{0.2\textwidth}p{0.2\textwidth}}
 \multicolumn{4}{c}{IVR(L)} \\
 \hline
Grid & $\mathcal{E}^{0}_{X}$ & $\mathcal{E}^{1}_{X}$ & Speedup \\
 \hline
 $16 \times 16$ & 3.37E-1 & 6.05E-1 & $\times$ 1.23 \\
 \hline
 $32 \times 32$ &  2.78E-1  & 5.60E-1 & $\times$ 1.74 \\
 \hline
 $64 \times 64$ & 1.69E-1  & 3.71E-1 & $\times$ 3.16 \\ 
 \hline
 $128 \times 128$ & 7.16E-2 & 1.63E-1& $\times$ 3.25 \\ 
 \hline
\end{tabular}
\begin{tabular}{p{0.2\linewidth}p{0.2\textwidth}p{0.2\textwidth}p{0.2\textwidth}}
 \multicolumn{4}{c}{DMD-FS} \\
 \hline
Grid & $\mathcal{E}^{0}_{X}$ & $\mathcal{E}^{1}_{X}$ & Speedup \\
 \hline
 $16 \times 16$ & 6.20E-2 & 1.42E-1 & $\times$ 11.48 \\ 
 \hline
 $32 \times 32$ &  2.62E-3  & 6.54E-3 & $\times$ 19.63 \\ 
 \hline
 $64 \times 64$ & 6.82E-4  &1.69E-3 & $\times$ 39.56 \\ 
 \hline
 $128 \times 128$ & \REV{6.08E-4} & \REV{1.44E-3} & $\times$ 26.60 \\ 
 \hline
\end{tabular}
}
\caption{Relative $L^2$ and $H^1$ errors of the single material combination test solutions computed by the IVR(C), IVR(L), and DMD-FS schemes at the final time $T=2\pi$, along with average speedup of the latter two relative to IVR(C).}
\label{tab:singlemat_combo}
\end{table}


%% file: Conclusion.tex
\REV{In this paper we formulated and demonstrated a new class of partitioned schemes for parameterized interface problems that employ dynamic data-driven surrogates to perform the synchronization between the subdomain problems. Our approach uses a non-standard dynamic mode decomposition operator acting on a hybrid state comprising the interface flux and small patches of subdomain states near the interface. We provided analysis establishing the stability of the hybrid monolithic formulation of the coupled model  problem and then related the stability of the data-driven surrogate-based scheme to that of  the hybrid monolithic formulation. In particular, we showed that if the flux surrogate operator is sufficiently accurate it does not introduce additional stability restrictions on the time step.}

\REV{To train this operator we invoke parallels with the identification of linear time-invariant systems through their impulse response and consider a family of Gaussian initial conditions that provide information about the dynamic response of the interface to generic stimuli.
In so doing we obtain  accurate, efficient and predictive surrogates for the dynamics of the interface flux that can predict the latter for initial conditions not seen during the training process.}

We tested parametric ($\mu$DMD-FS) and non-parametric (DMD-FS) versions of the new partitioned scheme using a multi-material and a single material configuration of a transmission problem \REV{modeling the diffusive transport of a scalar quantity across an interface}. The tests were carried out for a manufactured solution and a ``combination'' solution commonly used in the literature. 
\REV{These tests revealed that, compared to a hybridized monolithic scheme, the new data-driven partitioned methods realize significant double-digit speedups. Notably, $\mu$DMD-FS and DMD-FS also outperformed a proxy for a traditional loosely-coupled scheme, while exceeding its accuracy by at least an order of magnitude.}

\REV{We also briefly sketched extensions of the new approach to more general time integration settings and imperfect interfaces. Problems having such interfaces appear to be a particularly promising application area for partitioned schemes employing the new dynamic surrogates because the complex dynamic response of imperfect interfaces is challenging for surrogates regressing the static Poincare-Steklov maps between the subdomains.
Future work will focus on the development of the new approach for non-standard coupling conditions arising in, e.g., the coupled ocean-atmosphere system, and more complex asynchronous and multi-rate time integration schemes that require multiple data transfers within a coupling window. Such schemes  stand to gain significantly from the availability of dynamic surrogates that can perform these transfers in a fast and accurate manner.}
%

%% file: main.bbl
\begin{thebibliography}{10}
\expandafter\ifx\csname url\endcsname\relax
  \def\url#1{\texttt{#1}}\fi
\expandafter\ifx\csname urlprefix\endcsname\relax\def\urlprefix{URL }\fi
\expandafter\ifx\csname href\endcsname\relax
  \def\href#1#2{#2} \def\path#1{#1}\fi

\bibitem{Felippa_01_CMAME}
C.~A. Felippa, K.~Park, C.~Farhat,
  \href{https://www.sciencedirect.com/science/article/pii/S0045782500003911}{Partitioned
  analysis of coupled mechanical systems}, Computer Methods in Applied
  Mechanics and Engineering 190~(24) (2001) 3247--3270, advances in
  Computational Methods for Fluid-Structure Interaction.
\newblock \href {https://doi.org/https://doi.org/10.1016/S0045-7825(00)00391-1}
  {\path{doi:https://doi.org/10.1016/S0045-7825(00)00391-1}}.
\newline\urlprefix\url{https://www.sciencedirect.com/science/article/pii/S0045782500003911}

\bibitem{Newman_13_SISC}
C.~Newman, D.~A. Knoll, \href{https://doi.org/10.1137/120881397}{Physics-based
  preconditioners for ocean simulation}, SIAM Journal on Scientific Computing
  35~(5) (2013) S445--S464.
\newblock \href {http://arxiv.org/abs/https://doi.org/10.1137/120881397}
  {\path{arXiv:https://doi.org/10.1137/120881397}}, \href
  {https://doi.org/10.1137/120881397} {\path{doi:10.1137/120881397}}.
\newline\urlprefix\url{https://doi.org/10.1137/120881397}

\bibitem{Knoll_04_JCP}
D.~Knoll, D.~Keyes,
  \href{https://www.sciencedirect.com/science/article/pii/S0021999103004340}{Jacobian-free
  newton--krylov methods: a survey of approaches and applications}, Journal of
  Computational Physics 193~(2) (2004) 357--397.
\newblock \href {https://doi.org/https://doi.org/10.1016/j.jcp.2003.08.010}
  {\path{doi:https://doi.org/10.1016/j.jcp.2003.08.010}}.
\newline\urlprefix\url{https://www.sciencedirect.com/science/article/pii/S0021999103004340}

\bibitem{Connors_22_SINUM}
J.~M. Connors, K.~C. Sockwell, \href{https://doi.org/10.1137/21M1461149}{A
  multirate discontinuous-{G}alerkin-in-time framework for interface-coupled
  problems}, SIAM Journal on Numerical Analysis 60~(5) (2022) 2373--2404.
\newblock \href {http://arxiv.org/abs/https://doi.org/10.1137/21M1461149}
  {\path{arXiv:https://doi.org/10.1137/21M1461149}}, \href
  {https://doi.org/10.1137/21M1461149} {\path{doi:10.1137/21M1461149}}.
\newline\urlprefix\url{https://doi.org/10.1137/21M1461149}

\bibitem{Caldwell_19_JAMES}
P.~M. Caldwell, A.~Mametjanov, Q.~Tang, L.~P. Van~Roekel, J.-C. Golaz, W.~Lin,
  D.~C. Bader, N.~D. Keen, Y.~Feng, R.~Jacob, M.~E. Maltrud, A.~F. Roberts,
  M.~A. Taylor, M.~Veneziani, H.~Wang, J.~D. Wolfe, K.~Balaguru,
  P.~Cameron-Smith, L.~Dong, S.~A. Klein, L.~R. Leung, H.-Y. Li, Q.~Li, X.~Liu,
  R.~B. Neale, M.~Pinheiro, Y.~Qian, P.~A. Ullrich, S.~Xie, Y.~Yang, Y.~Zhang,
  K.~Zhang, T.~Zhou,
  \href{https://agupubs.onlinelibrary.wiley.com/doi/abs/10.1029/2019MS001870}{The
  {DOE} {E3SM} coupled model version 1: Description and results at high
  resolution}, Journal of Advances in Modeling Earth Systems n/a~(n/a) (2019).
\newblock \href
  {http://arxiv.org/abs/https://agupubs.onlinelibrary.wiley.com/doi/pdf/10.1029/2019MS001870}
  {\path{arXiv:https://agupubs.onlinelibrary.wiley.com/doi/pdf/10.1029/2019MS001870}},
  \href {https://doi.org/10.1029/2019MS001870}
  {\path{doi:10.1029/2019MS001870}}.
\newline\urlprefix\url{https://agupubs.onlinelibrary.wiley.com/doi/abs/10.1029/2019MS001870}

\bibitem{Lemarie_15_PCS}
F.~Lemari{\'e}, E.~Blayo, L.~Debreu,
  \href{http://www.sciencedirect.com/science/article/pii/S1877050915012818}{Analysis
  of ocean-atmosphere coupling algorithms: Consistency and stability}, Procedia
  Computer Science 51 (2015) 2066 -- 2075, international Conference On
  Computational Science, \{ICCS\} 2015 Computational Science at the Gates of
  Nature.
\newblock \href {https://doi.org/http://dx.doi.org/10.1016/j.procs.2015.05.473}
  {\path{doi:http://dx.doi.org/10.1016/j.procs.2015.05.473}}.
\newline\urlprefix\url{http://www.sciencedirect.com/science/article/pii/S1877050915012818}

\bibitem{Foerster_07_CMAME}
C.~F{\"o}rster, W.~A. Wall, E.~Ramm,
  \href{http://www.sciencedirect.com/science/article/pii/S0045782506002544}{Artificial
  added mass instabilities in sequential staggered coupling of nonlinear
  structures and incompressible viscous flows}, Computer Methods in Applied
  Mechanics and Engineering 196~(7) (2007) 1278 -- 1293.
\newblock \href {https://doi.org/http://dx.doi.org/10.1016/j.cma.2006.09.002}
  {\path{doi:http://dx.doi.org/10.1016/j.cma.2006.09.002}}.
\newline\urlprefix\url{http://www.sciencedirect.com/science/article/pii/S0045782506002544}

\bibitem{Jiao_04_IJNME}
X.~Jiao, M.~T. Heath,
  \href{http://dx.doi.org/10.1002/nme.1147}{Common-refinement-based data
  transfer between non-matching meshes in multiphysics simulations},
  International Journal for Numerical Methods in Engineering 61~(14) (2004)
  2402--2427.
\newblock \href {https://doi.org/10.1002/nme.1147}
  {\path{doi:10.1002/nme.1147}}.
\newline\urlprefix\url{http://dx.doi.org/10.1002/nme.1147}

\bibitem{Ullrich_15_MWR}
P.~A. Ullrich, M.~A. Taylor, Arbitrary-order conservative and consistent
  remapping and a theory of linear maps: Part i,, Mon. Weather Rev. 143 (2015.)
  2419--2440,.

\bibitem{Gatzhammer_14_THESIS}
B.~Gatzhammer, Efficient and flexible partitioned simulation of fluid-structure
  interactions, Doctoral thesis, Technische Universitaet Muenchen, Fakultaet
  fuer Informatik. Informatik 5 -- Lehrstuhl fuer Wissenschaftliches Rechnen
  (September 2014).

\bibitem{Slattery_16_JCP}
S.~R. Slattery,
  \href{http://www.sciencedirect.com/science/article/pii/S0021999115008037}{Mesh-free
  data transfer algorithms for partitioned multiphysics problems: Conservation,
  accuracy, and parallelism}, Journal of Computational Physics 307 (2016) 164
  -- 188.
\newblock \href {https://doi.org/http://dx.doi.org/10.1016/j.jcp.2015.11.055}
  {\path{doi:http://dx.doi.org/10.1016/j.jcp.2015.11.055}}.
\newline\urlprefix\url{http://www.sciencedirect.com/science/article/pii/S0021999115008037}

\bibitem{Bungartz_16_CF}
H.-J. Bungartz, F.~Lindner, B.~Gatzhammer, M.~Mehl, K.~Scheufele, A.~Shukaev,
  B.~Uekermann,
  \href{http://www.sciencedirect.com/science/article/pii/S0045793016300974}{{preCICE}
  -- a fully parallel library for multi-physics surface coupling}, Computers \&
  Fluids 141 (2016) 250 -- 258, advances in Fluid-Structure Interaction.
\newblock \href
  {https://doi.org/https://doi.org/10.1016/j.compfluid.2016.04.003}
  {\path{doi:https://doi.org/10.1016/j.compfluid.2016.04.003}}.
\newline\urlprefix\url{http://www.sciencedirect.com/science/article/pii/S0045793016300974}

\bibitem{Kuberry_19_MISC}
P.~Kuberry, P.~Bosler, N.~Trask,
  \href{https://doi.org/10.5281/zenodo.2560287}{Compadre toolkit} (Feb. 2019).
\newblock \href {https://doi.org/10.5281/zenodo.2560287}
  {\path{doi:10.5281/zenodo.2560287}}.
\newline\urlprefix\url{https://doi.org/10.5281/zenodo.2560287}

\bibitem{Rowley_09_JFM}
C.~W. Rowley, I.~Mezi{\'c}, S.~Bagheri, P.~Schlatter, D.~S. Henningson,
  Spectral analysis of nonlinear flows, Journal of Fluid Mechanics 641 (2009)
  115--127.
\newblock \href {https://doi.org/10.1017/S0022112009992059}
  {\path{doi:10.1017/S0022112009992059}}.

\bibitem{Mezic_05_ND}
I.~Mezi{\'c}, \href{https://doi.org/10.1007/s11071-005-2824-x}{Spectral
  properties of dynamical systems, model reduction and decompositions},
  Nonlinear Dynamics 41~(1) (2005) 309--325.
\newblock \href {https://doi.org/10.1007/s11071-005-2824-x}
  {\path{doi:10.1007/s11071-005-2824-x}}.
\newline\urlprefix\url{https://doi.org/10.1007/s11071-005-2824-x}

\bibitem{Aletti_17_IJNME}
M.~Aletti, D.~Lombardi,
  \href{https://onlinelibrary.wiley.com/doi/abs/10.1002/nme.5490}{A
  reduced-order representation of the {P}oincar{\'e}--{S}teklov operator: an
  application to coupled multi-physics problems}, International Journal for
  Numerical Methods in Engineering 111~(6) (2017) 581--600.
\newblock \href
  {http://arxiv.org/abs/https://onlinelibrary.wiley.com/doi/pdf/10.1002/nme.5490}
  {\path{arXiv:https://onlinelibrary.wiley.com/doi/pdf/10.1002/nme.5490}},
  \href {https://doi.org/https://doi.org/10.1002/nme.5490}
  {\path{doi:https://doi.org/10.1002/nme.5490}}.
\newline\urlprefix\url{https://onlinelibrary.wiley.com/doi/abs/10.1002/nme.5490}

\bibitem{Chen_22_arXiv}
S.~Chen, Z.~Ding, Q.~Li, S.~J. Wright, A reduced order schwarz method for
  nonlinear multiscale elliptic equations based on two-layer neural networks
  (2022).
\newblock \href {http://arxiv.org/abs/2111.02280} {\path{arXiv:2111.02280}}.

\bibitem{Discacciati_23_UNPUB}
N.~Discacciati, J.~S. Hesthaven, Model reduction of coupled systems based on
  non-intrusive approximations of the boundary response maps, Tech. rep.,
  Institute of Mathematics {\'E}cole Polytechnique F{\'e}d{\'e}rale de Lausanne
  (EPFL) CH-1015 Lausanne, Switzerland (2023).

\bibitem{Parish_23_arXiv}
E.~Parish, P.~Lindsay, T.~Shelton, J.~Mersch, Embedded symmetric positive
  semi-definite machine-learned elements for reduced-order modeling in
  finite-element simulations with application to threaded fasteners (2023).
\newblock \href {http://arxiv.org/abs/2307.05434} {\path{arXiv:2307.05434}}.

\bibitem{Javili_14_CMAME}
A.~Javili, S.~Kaessmair, P.~Steinmann,
  \href{http://www.sciencedirect.com/science/article/pii/S0045782514000802}{General
  imperfect interfaces}, Computer Methods in Applied Mechanics and Engineering
  275 (2014) 76 -- 97.
\newblock \href {https://doi.org/https://doi.org/10.1016/j.cma.2014.02.022}
  {\path{doi:https://doi.org/10.1016/j.cma.2014.02.022}}.
\newline\urlprefix\url{http://www.sciencedirect.com/science/article/pii/S0045782514000802}

\bibitem{Huhn_23_JCP}
Q.~A. Huhn, M.~E. Tano, J.~C. Ragusa, Y.~Choi,
  \href{https://www.sciencedirect.com/science/article/pii/S0021999122009159}{Parametric
  dynamic mode decomposition for reduced order modeling}, Journal of
  Computational Physics 475 (2023) 111852.
\newblock \href {https://doi.org/https://doi.org/10.1016/j.jcp.2022.111852}
  {\path{doi:https://doi.org/10.1016/j.jcp.2022.111852}}.
\newline\urlprefix\url{https://www.sciencedirect.com/science/article/pii/S0021999122009159}

\bibitem{Ciarlet_02_BOOK}
P.~Ciarlet, The Finite Element Method for Elliptic Problems, SIAM Classics in
  Applied Mathematics, SIAM, Philadelphia, 2002.

\bibitem{Taraneh_15_PF}
T.~Sayadi, P.~J. Schmid, F.~Richecoeur, D.~Durox,
  \href{https://doi.org/10.1063/1.4913868}{{Parametrized data-driven
  decomposition for bifurcation analysis, with application to
  thermo-acoustically unstable systems}}, Physics of Fluids 27~(3) (2015)
  037102.
\newblock \href
  {http://arxiv.org/abs/https://pubs.aip.org/aip/pof/article-pdf/doi/10.1063/1.4913868/15922234/037102\_1\_online.pdf}
  {\path{arXiv:https://pubs.aip.org/aip/pof/article-pdf/doi/10.1063/1.4913868/15922234/037102\_1\_online.pdf}},
  \href {https://doi.org/10.1063/1.4913868} {\path{doi:10.1063/1.4913868}}.
\newline\urlprefix\url{https://doi.org/10.1063/1.4913868}

\bibitem{Mahan_09_PRB}
G.~D. Mahan, \href{https://link.aps.org/doi/10.1103/PhysRevB.79.075408}{Kapitza
  thermal resistance between a metal and a nonmetal}, Phys. Rev. B 79 (2009)
  075408.
\newblock \href {https://doi.org/10.1103/PhysRevB.79.075408}
  {\path{doi:10.1103/PhysRevB.79.075408}}.
\newline\urlprefix\url{https://link.aps.org/doi/10.1103/PhysRevB.79.075408}

\bibitem{Bochev_19_CAMWA}
K.~Peterson, P.~Bochev, P.~Kuberry,
  \href{https://www.sciencedirect.com/science/article/pii/S0898122118305637}{Explicit
  synchronous partitioned algorithms for interface problems based on lagrange
  multipliers}, Computers \& Mathematics with Applications 78~(2) (2019)
  459--482, proceedings of the Eight International Conference on Numerical
  Methods for Multi-Material Fluid Flows (MULTIMAT 2017).
\newblock \href {https://doi.org/https://doi.org/10.1016/j.camwa.2018.09.045}
  {\path{doi:https://doi.org/10.1016/j.camwa.2018.09.045}}.
\newline\urlprefix\url{https://www.sciencedirect.com/science/article/pii/S0898122118305637}

\bibitem{Ascher_98_BOOK}
U.~M. Ascher, L.~R. Petzold,
  \href{http://dl.acm.org/citation.cfm?id=551054}{Computer Methods for Ordinary
  Differential Equations and Differential-Algebraic Equations}, 1st Edition,
  Society for Industrial and Applied Mathematics, Philadelphia, PA, USA, 1998.
\newline\urlprefix\url{http://dl.acm.org/citation.cfm?id=551054}

\bibitem{Brezzi_91_BOOK}
F.~Brezzi, M.~Fortin, Mixed and Hybrid Finite Element Methods, Springer,
  Berlin, 1991.

\bibitem{Farhat_94_IJNME}
C.~Farhat, L.~Crivelli, F.-X. Roux,
  \href{http://dx.doi.org/10.1002/nme.1620371111}{A transient {FETI}
  methodology for large-scale parallel implicit computations in structural
  mechanics}, International Journal for Numerical Methods in Engineering
  37~(11) (1994) 1945--1975.
\newblock \href {https://doi.org/10.1002/nme.1620371111}
  {\path{doi:10.1002/nme.1620371111}}.
\newline\urlprefix\url{http://dx.doi.org/10.1002/nme.1620371111}

\bibitem{Park_01_CMAME}
K.~C. Park, C.~A. Felippa, R.~Ohayon,
  \href{http://www.sciencedirect.com/science/article/pii/S0045782500003789}{Partitioned
  formulation of internal fluid--structure interaction problems by localized
  {L}agrange multipliers}, Computer Methods in Applied Mechanics and
  Engineering 190~(24) (2001) 2989--3007.
\newblock \href {https://doi.org/https://doi.org/10.1016/S0045-7825(00)00378-9}
  {\path{doi:https://doi.org/10.1016/S0045-7825(00)00378-9}}.
\newline\urlprefix\url{http://www.sciencedirect.com/science/article/pii/S0045782500003789}

\bibitem{Ross_09_CMAME}
M.~R. Ross, M.~A. Sprague, C.~A. Felippa, K.~Park,
  \href{http://www.sciencedirect.com/science/article/pii/S0045782508004076}{Treatment
  of acoustic fluid--structure interaction by localized {L}agrange multipliers
  and comparison to alternative interface-coupling methods}, Computer Methods
  in Applied Mechanics and Engineering 198~(9--12) (2009) 986 -- 1005.
\newblock \href {https://doi.org/http://dx.doi.org/10.1016/j.cma.2008.11.006}
  {\path{doi:http://dx.doi.org/10.1016/j.cma.2008.11.006}}.
\newline\urlprefix\url{http://www.sciencedirect.com/science/article/pii/S0045782508004076}

\bibitem{Ross_08_CMAME}
M.~R. Ross, C.~A. Felippa, K.~Park, M.~A. Sprague,
  \href{http://www.sciencedirect.com/science/article/pii/S0045782508000625}{Treatment
  of acoustic fluid--structure interaction by localized {L}agrange multipliers:
  Formulation}, Computer Methods in Applied Mechanics and Engineering
  197~(33--40) (2008) 3057 -- 3079.
\newblock \href {https://doi.org/http://dx.doi.org/10.1016/j.cma.2008.02.017}
  {\path{doi:http://dx.doi.org/10.1016/j.cma.2008.02.017}}.
\newline\urlprefix\url{http://www.sciencedirect.com/science/article/pii/S0045782508000625}

\bibitem{Hughes_00_BOOK}
T.~Hughes, The Finite Element Method: Linear Static and Dynamic Finite Element
  Analysis, Dover Civil and Mechanical Engineering, 2000.

\bibitem{Smith_23_NSE}
I.~V. Ethan~Smith, R.~McClarren,
  \href{https://doi.org/10.1080/00295639.2022.2142025}{Variable dynamic mode
  decomposition for estimating time eigenvalues in nuclear systems}, Nuclear
  Science and Engineering 197~(8) (2023) 1769--1778.
\newblock \href
  {http://arxiv.org/abs/https://doi.org/10.1080/00295639.2022.2142025}
  {\path{arXiv:https://doi.org/10.1080/00295639.2022.2142025}}, \href
  {https://doi.org/10.1080/00295639.2022.2142025}
  {\path{doi:10.1080/00295639.2022.2142025}}.
\newline\urlprefix\url{https://doi.org/10.1080/00295639.2022.2142025}

\bibitem{Lemarie_13_ETNA}
F.~Lemarie, L.~Debreu, E.~Blayo, Toward an optimized global-in-time {S}chwarz
  algorithm for diffusion equations with discontinuous and spatially variable
  coefficients. part 1: The constant coefficients case, ETNA 40 (2013)
  148--169.

\bibitem{Johnson_92_BOOK}
C.~Johnson, Numerical {S}olution of {P}artial {D}ifferential {E}quations by the
  {F}inite {E}lement {M}ethod, Cambridge University Press, 1992.

\bibitem{Brenner_02_BOOK}
S.~Brenner, R.~Scott, The Mathematical Theory of Finite Element Methods, no.~15
  in Texts in Applied Mathematics, Springer Verlag, New York, 2002.

\bibitem{Lu_19_SISC}
H.~Lu, D.~M. Tartakovsky, \href{https://doi.org/10.1137/19M1259948}{Prediction
  accuracy of dynamic mode decomposition}, SIAM Journal on Scientific Computing
  42~(3) (2020) A1639--A1662.
\newblock \href {http://arxiv.org/abs/https://doi.org/10.1137/19M1259948}
  {\path{arXiv:https://doi.org/10.1137/19M1259948}}, \href
  {https://doi.org/10.1137/19M1259948} {\path{doi:10.1137/19M1259948}}.
\newline\urlprefix\url{https://doi.org/10.1137/19M1259948}

\bibitem{Bochev_20_CMAME}
P.~Bochev, D.~Ridzal, M.~D'Elia, M.~Perego, K.~Peterson,
  \href{https://www.sciencedirect.com/science/article/pii/S0045782520301651}{Optimization-based,
  property-preserving finite element methods for scalar advection equations and
  their connection to algebraic flux correction}, Computer Methods in Applied
  Mechanics and Engineering 367 (2020) 112982.
\newblock \href {https://doi.org/https://doi.org/10.1016/j.cma.2020.112982}
  {\path{doi:https://doi.org/10.1016/j.cma.2020.112982}}.
\newline\urlprefix\url{https://www.sciencedirect.com/science/article/pii/S0045782520301651}

\bibitem{Leveque_96_SINUM}
R.~J. LeVeque, \href{https://doi.org/10.1137/0733033}{High-resolution
  conservative algorithms for advection in incompressible flow}, SIAM Journal
  on Numerical Analysis 33~(2) (1996) 627--665.
\newblock \href {http://arxiv.org/abs/https://doi.org/10.1137/0733033}
  {\path{arXiv:https://doi.org/10.1137/0733033}}, \href
  {https://doi.org/10.1137/0733033} {\path{doi:10.1137/0733033}}.
\newline\urlprefix\url{https://doi.org/10.1137/0733033}

\end{thebibliography}
